\documentclass[preprint]{ptephy_v1}

\preprintnumber{XXXX-XXXX} 

\usepackage{graphicx}
\usepackage{subfigure}
\usepackage{slashbox}
\usepackage{amsmath,amssymb,ascmac,color}
\usepackage{type1cm}
\usepackage{bm}

\begin{document}

\title{Kaon-baryon coupling schemes and kaon condensation in hyperon-mixed matter}


\author{Takumi Muto}
\affil{Department of Physics, Chiba Institute of Technology, 2-1-1 Shibazono, Narashino, Chiba 275-0023, Japan  \email{takumi.muto@it-chiba.ac.jp}}

\author{Toshiki Maruyama}
\affil{Advanced Science Research Center, Japan Atomic Energy Agency , Ibaraki 319-1195, Japan}

\author{Toshitaka Tatsumi}
\affil{Kitashirakawa Kamiikeda-Cho, 52-4, Kyoto 606-8287,Japan}



\begin{abstract}%
Possible coexistence of kaon condensation and hyperons in highly dense matter [the ($Y+K$) phase] is investigated on the basis of the relativistic mean-field theory combined with the effective chiral Lagrangian. 
Two coupling schemes for the $s$-wave kaon-baryon interaction are compared regarding the onset density of kaon condensation in the hyperon-mixed matter and equation of state for the developed ($Y+K$) phase: One is the contact interaction scheme related to the nonlinear effective chiral Lagrangian. 
The other is the meson-exchange scheme, where the interaction vertices between the kaon field and baryons are described by exchange of mesons ($\sigma$, $\sigma^\ast$  mesons for scalar coupling, and $\omega$, $\rho$, $\phi$ mesons for vector coupling). 
It is shown that in the meson exchange scheme, the contribution 
from the nonlinear scalar self-interaction gives rise to 
a repulsive effect for kaon effective energy, pushing up the onset density of kaon condensation as compared with the case of the contact interaction scheme. 
In general, the difference of kaon-baryon dynamics between the contact interaction scheme and the meson-exchange scheme relies on the specific forms of the nonlinear self-interacting meson terms. 
They generate many-baryon forces through the equations of motion for the meson mean fields. However,  they should have a definite role on the ground state properties of nuclear matter only around the saturation density. It is shown that the nonlinear self-interacting term is not relevant to repulsive energy leading to stiffening of the equation of state at high densities and that it cannot be compensated with large attractive energy due to the appearance of the ($Y$+$K$) phase in the case of the contact interaction scheme.
We also discuss in the contact interaction scheme what effects are necessary so as to make the equation of state with (Y+K) phase stiff enough to be consistent with recent observations of massive neutron stars. 
\end{abstract}

\subjectindex{D33}

\maketitle

\section{Introduction}
\label{sec:intro}

Multi-strangeness system in dense hadronic matter has been investigated extensively from nuclear and astrophysical points of view. As a possible form in neutron stars, Bose-Einstein condensation of antikaons ($K^-$) (kaon condensation) has been attracting much interest as a macroscopic form of strangeness degree of freedom\cite{kn86,t88,mt92,m93,mtt93,fmtt94,tpl94,kvk95,lbm95,fmmt96,tstw98,ty1998}, and its implications for astrophysical phenomena related with compact stars have been discussed: It has a large impact on structure and thermal evolution of neutron stars through softening of the equation of state (EOS)\cite{tpl94,fmmt96} and enhancement of neutrino emissions\cite{t88,mtt93,fmtt94}. 

It has been shown that the $s$-wave kaon condensation can be discussed model-independently on the basis of current algebra and partial conservation of axial-vector current (PCAC)\cite{t88,mt92,mtt93}. A series of works based on chiral symmetry have shown that kaon condensation arises at baryon number density $\rho_B$=(3$-$4) $\rho_0$ with $\rho_0$ being the normal nuclear density, for the value of the $KN$ sigma term, $\Sigma_{KN}$=(300$-$400) MeV, which simulates explicit chiral symmetry breaking\cite{mt92,mtt93,tpl94,lbm95,fmmt96}. 
In these approaches, the $s$-wave kaon-baryon ($K$-$B$) interaction is represented within the contact interaction (CI) scheme, where the structure of $K$-$B$ and kaon-kaon ($K$-$K$) interaction vertices becomes nonlinear and is inherently determined from chiral symmetry. 

As another hadron phase with multi-strangeness, it has been suggested that hyperons ($Y$=$\Lambda$, $\Sigma^-$, $\Xi^-$ , $\dots$) are mixed ($Y$-mixing) in the ground state of neutron-star matter by the use of the relativistic mean-field (RMF) theories\cite{g85,ekp95,kpe95,sm96,g01,phz99,h00,s00,tolos2017}, many-body approaches based on the reaction matrix theory, variational methods, and so forth\cite{bg97,h98,bbs98,v00,nyt02,t04,t2016}. 
The mixing of hyperons as well leads to softening of the EOS\cite{g85,phz99,h00,s00,bg97,h98,bbs98,v00,nyt02,t2016} and posing another candidate for rapid cooling of neutron stars\cite{pplp92,t98}.  
The onset density of hyperons, $\rho_{\rm B}^c(Y)$, has been estimated to be $\rho_B$ = (2$-$4) $\rho_0$. Thus it may be plausible that kaon condensates and hyperons may coexist in dense neutron-star matter. 

One of the authors (T.~Muto) has considered possible coexistence of kaon-condensed phase and hyperon-mixed matter [abbreviated to the ($Y+K$) phase in this paper] in neutron stars by the use of the effective chiral Lagrangian for the $s$-wave $K$-$B$ interaction and a phenomenological potential model for the baryon-baryon ($B$-$B$) interaction\cite{m08}. One of the serious problems resulting from the existence of the ($Y+K$) phase is that both kaon condensates and $Y$-mixing in dense matter lead to significant softening of the EOS, which is not compatible with recent observations of massive neutron stars with (1.928$\pm$0.017)$M_\odot$ for PSR J1614-2230 and (2.01$\pm$0.04)$M_\odot$ for PSR J0348+0432, where $M_\odot$ is the solar mass\cite{demo10,f16,ant13}. Pulsars definitely exceeding 2$M_\odot$ have been detected such as (2.08$\pm$0.07)$M_\odot$ for the 2.8~ms radio pulsar PSR~J0740+6620~\cite{Cromartie2020,Fonseca2021} and (2.13$\pm$0.04)$M_\odot$ for PSR~J1810+1744~\cite{Romani2021}. 
Coexistence of kaon condensation and hyperons in dense matter has also been discussed by several authors in terms of the RMF theories\cite{ekp95,kpe95,sm96,g01,bb01,sl2010}, effective chiral models\cite{pbg00,mk2010}, and quark-meson coupling models\cite{mpp05,rhhk07}. 
Most of the models including the ($Y+K$) phase predict maximum neutron-star masses less than 1.85 $M_\odot$ except for recent works with the density-dependent meson-baryon coupling strengths in the RMF~\cite{cb2014,ts2020,mbb2021}. 

Recent progress in observational facilities and satellites has brought about new information on the structure of compact stars: 
The detection of gravitational waves from neutron-star mergers (GW170817) by LIGO-Virgo collaboration\cite{a2017} has shed light on constraining the EOS of dense matter by setting limits on the tidal deformabilities of compact stars\cite{yy2017,a2018}. The precise measurement of masses and radii of neutron stars has become possible from the $X$-ray observation by Neutron star Interior Composition ExploreR (NICER)\cite{Riley2019,Miller2019}. The mass and radius for the above mentioned pulsar PSR~J0740+6620 has been detected as $R$ = (12.35$\pm$0.75) km for $M$ = 2.08~$M_\odot$~\cite{Miller2021} and $R$ = (12.39$^{+1.30}_{-0.98}$)~km for $M$ = (2.072~$^{+0.067}_{-0.066}$) $M_\odot$~\cite{Riley2021}. The EOS including the ($Y+K$) phase should be in conformity with stringent constraints deduced from these observations. 

Studies of kaon condensation in neutron stars triggered researches on deeply bound kaonic states which may be formed in terrestrial experiments\cite{ay02,yda04,ynoh2005,mmt09,gfgm2009,zs2013,bbg12,ghm16}, and associated kaon dynamics in nuclear medium has been extensively studied theoretically and experimentally\cite{bbg12,ghm16,hj2016}. Recently, formation of basic kaonic clusters, $K^- p p $, has been reported in the E27 and E15 experiments at J-PARC\cite{ichikawa2015,sada2016,a2019,yamaga2020}. 

In order to find out the connection between deeply bound kaonic states in nuclei and kaon condensation in dense matter, we have studied multi-antikaonic bound states in nuclei based upon the RMF theory coupled to the effective chiral Lagrangian \cite{mmt09}. In this framework, we have adopted the meson-exchange (ME) scheme for the $s$-wave $K$-$B$ interaction, where the CI vertices between the nonlinear kaon field and baryons in the original effective chiral Lagrangian are replaced by exchange of mesons (scalar mesons $\sigma$, $\sigma^\ast$ and vector mesons $\omega$, $\rho$, $\phi$). (We called this interaction model the ``chiral model'' in Ref.~\cite{mmt09}.) 
Thus many-body effects on the $K$-$B$ interaction coming from the meson-exchange in a nuclear medium can be taken into account in the ME scheme. 
In a series of our works, we have considered not only possible bound states of kaons and hyperons in nuclei (kaon-condensed hypernuclei) but also the ($Y$+$K$) phase in neutron stars within the same interaction model based on the ME scheme and discussed interplay of kaons and hyperons in multi-strangeness systems in a unified way for both nuclei and neutron stars\cite{mmt14}. 
It is to be noted that the ``chiral model'' based on the ME scheme reduces essentially to the meson-exchange model (abbreviated to MEM in Ref.~\cite{mmt09}) in the limit of linear approximation for the nonlinear kaon fields~\cite{mmt09}. 
The MEM has been utilized by several authors for description of kaon condensation\cite{ekp95,kpe95,sm96,g01,pbg00,bb01} and multi-antikaonic nuclei\cite{gfgm2009} in the context of the RMF theories. 

Toward a description of the ($Y+K$) phase consistent with the recent observations of massive neutron stars or gravitational waves from the neutron-star merger, we should start with removing uncertainties stemming from the $s$-wave $K$-$B$ interaction, since it is a driving force for kaon condensation and may crucially affect the onset density and the EOS of the ($Y+K$) phase, depending upon the choice of the coupling schemes for the $s$-wave $K$-$B$ interaction. 

First, in this paper, we discuss in detail the validity of the CI and ME coupling schemes for the $K$-$B$ interaction vertices based upon the effective chiral Lagrangian coupled to the RMF model. We take into account the nonlinear self-interacting (NLSI) potential term of the $\sigma$ mesons, $U_\sigma$, which is usually introduced phenomenologically, in addition to the two-body $B$-$B$ interactions which are mediated by scalar and vector mesons in the RMF framework. [Throughout this paper, we call this two-body $B$-$B$ interaction part of the model the ``minimal RMF'' (MRMF). ] The results on the onset density and characteristic features of the ($Y+K$) phase in these two schemes based on the (MRMF+NLSI) model are compared. 
It is shown that the effect of the NLSI term is propagated to the kaon self-energy in the ME scheme as a derivative term, $dU_\sigma/d\sigma$, through the equation of motion for the $\sigma$ meson. (Similar results have been pointed out in Refs.~\cite{kpe95,sm96}.)
Such a model-dependent term is added as an extra repulsive contribution to the lowest kaon energy in the ME scheme beyond the scope of chiral symmetry. As a result, kaon condensation from hyperon-mixed matter does not occur in the case of the ME scheme unless $\Sigma_{KN}$ is taken to be very large. On the other hand, in the CI scheme, the $K$-$B$ and $K$-$K$ interactions are controlled model-independently within chiral symmetry. It will be shown that the onset density for kaon condensation realized from hyperon-mixed matter and the EOS with the ($Y$+$K$) phase are obtained with moderate values of the $\Sigma_{KN}$ [= (300$-$400) MeV ]. 

Second we consider a role of the NLSI term, which generates many-baryon forces through the equations of motion for the meson mean fields, as a possible origin of repulsive forces at high densities in view of a solution to the ``hyperon puzzle''. It will be shown that the NLSI term poses quite different aspects for saturation mechanisms of symmetric nuclear matter (SNM) from those in the conventional variational methods with the phenomenological three-baryon forces. Further the NLSI term becomes only a minor contribution to the repulsive energy at high densities, not being able to make the EOS stiff enough to be consistent with recent observations of massive neutron stars.  
We address that the universal three-baryon repulsive forces (UTBR), as introduced phenomenologically in place of the NLSI meson terms in our recent work~\cite{mmt2021}, can make the EOS with the ($Y+K$) phase stiff enough to reconcile theories with observations. It is also pointed out that the kaon self-energy in hyperon-mixed matter with such UTBR is formally equivalent between the CI and ME schemes. 

The paper is organized as follows: In Sec.~\ref{sec:overview}, the ($Y$+$K$) phase is overviewed in the context of chiral symmetry. The $s$-wave $K$-$B$ interaction in the CI scheme and the baryon-baryon ($B$-$B$) interaction in the (MRMF+NLSI) model are described in Sec.~\ref{sec:CI} and Sec.~\ref{sec:BB}, respectively. These sections are followed by Sec.~\ref{sec:CI-energy}, where the energy expression of the ($Y$+$K$) phase in the CI scheme is given. In Sec.~\ref{sec:ME}, the formulation in the ME scheme is explained, and the CI and ME schemes are compared with each other. 
The numerical results on the onset density of kaon condensation 
are presented in Sec.~\ref{sec:results}, where the effects of the NLSI term as many-baryon forces are discussed.
In Sec.~\ref{sec:NLSI} the roles of the NLSI term on saturation mechanisms in symmetric nuclear matter (SNM) and on the EOS of the ($Y$+$K$) phase in the CI scheme are figured out. 
In Sec.~\ref{sec:fraction}, properties of the ($Y$+$K$) phase, for instance, density-dependence of particle fractions and hyperon potentials in the CI scheme are addressed with the (MRMF+NLSI) model as common features in the presence of kaon condensates. The self-suppression effect of the $s$-wave $K$-$B$ attraction unique to the case of kaon condensation in the RMF framework is also discussed. In Sec.~\ref{sec:discussion}, our alternative model with the UTBR in the RMF in place of the NLSI term is remarked, and circumventing the problem caused by the NLSI term, which is connected with avoiding the extra many-body effect in the ME scheme, is discussed. 
 Summary and concluding remarks are devoted in Sec.~\ref{sec:summary}. In the Appendix A, the allowable value of the $\Sigma_{Kn}$ is evaluated. In the Appendix B, the $K^-$ optical potential depths are derived in both the CI and ME schemes and related to the $s$-wave scalar $K$-$N$ interaction, the $KN$ sigma terms. 

\section{Overview of the ($Y+K$) phase}
\label{sec:overview}

For the description of the ($Y$+$K$) phase, we use the SU(3)$_{\rm L}$$\times$SU(3)$_{\rm R}$ chiral effective Lagrangian, where the nonlinear representation of the kaon field is given as $U = \exp[2i(K^+T_{4+i5}+K^-T_{4-i5})/f]$ with $T_{4\pm i5} (\equiv T_4\pm iT_5)$ being the flavor SU(3) generators and $f$ the meson decay constant. The numerical value of $f$ is set to be the one of the pion decay constant ($f_\pi$ = 93 MeV) instead of the kaon decay constant ($f_K$=113 MeV) following our previous papers\cite{mt92,m93,mtt93,fmtt94,fmmt96}, which corresponds to taking the lowest-order value in chiral perturbation theory under the SU(3) flavor symmetry.  
On the basis of chiral symmetry, the $s$-wave kaon-condensed state, $|K\rangle$, is represented by chiral rotation from the  normal state, and the classical kaon field stands for an order parameter for kaon condensation\cite{t88,mt92,ty1998}. Here it is taken to be spatially uniform with momentum ${\bf k}=0$: 
 \begin{equation}
K^\pm =\frac{f}{\sqrt{2}}\theta\exp(\pm i\mu_K t) \ , 
\label{eq:kfield}
\end{equation}
where $\theta$ is the chiral angle, and $\mu_K$ is the $K^-$ chemical potential. 
$U$ is expressed explicitly in terms of (\ref{eq:kfield}) as 
\begin{equation}
U=1+iA\sin\theta+A^2 (\cos\theta-1)
\label{eq:U}
\end{equation}
with the matrix $A$ being defined by 
\begin{eqnarray}
A=\left(
\begin{array}{ccc}
 0         & 0   &  K^+/|K|  \\
0          & 0   &  0   \\
K^-/|K| & 0   &  0    \\
\end{array}\right) \ , 
\label{xmatrix}
\end{eqnarray}
where $|K| \equiv (K^+ K^-)^{1/2}=f\theta/\sqrt{2}$~\cite{kn86}.  

The ($Y+K$) phase consists of kaon condensates, degenerate baryons and leptons in beta equilibrium. We take into account only protons ($p$), neutrons ($n$), $\Lambda$, $\Sigma^-$, $\Xi^-$ for octet baryons, since degenerate leptons assist mixing of neutral or negatively charged hyperons through beta equilibrium conditions. For leptons, muons may appear at a density slightly higher than that of electrons under consideration. However, quantitative effects of muons on the properties of matter are expected to be small, so that we take into account only electrons ($e^-$) for simplicity. 
We impose the charge neutrality condition and baryon number conservation, and construct the effective energy density ${\cal E}_{\rm eff}$ by introducing the charge chemical potential $\mu$ and the baryon number chemical potential $\nu$, respectively, as Lagrange multipliers. The resulting effective energy density is then written in the form 
\begin{equation}
{\cal E}_{\rm eff}={\cal E}+\mu (\rho_p-\rho_{\Sigma^-}-\rho_{\Xi^-}-\rho_{K^-}-\rho_e) + \nu(\rho_p+ \rho_\Lambda+\rho_n+\rho_{\Sigma^-}+\rho_{\Xi^-}) \ , 
\label{eq:eff}
\end{equation}
where 
${\cal E}$ is the total energy density of the kaon-condensed phase, and 
$\rho_i$ ($i$= $p$, $n$, $\Lambda$, $\Sigma^-$, $\Xi^-$, $K^-$, $e^-$) is the number density of the particle $i$. 

The classical kaon field equation is given from $\partial{\cal E}_{\rm eff}/\partial\theta=0$. 
From the extremum conditions for ${\cal E}_{\rm eff}$ with respect to variation of $\rho_i$, one obtains the following relations, 
\begin{subequations}\label{eq:chemeq}
\begin{eqnarray}
\mu_K&=&\mu_e=\mu \ ,\label{eq:chemeq1} \\
\mu_p&=&-\mu-\nu \ , \label{eq:chemeq2} \\
\mu_{\Sigma^-}&=&\mu_{\Xi^-}=\mu-\nu \ , \label{eq:chemeq3} \\
\mu_\Lambda&=&\mu_n=-\nu \ , \label{eq:chemeq4}
\end{eqnarray}
\end{subequations}
where $\mu_i$ ($i$= $p$, $\Lambda$, $n$, $\Sigma^-$, $\Xi^-$, $K^-$, $e^-$) are the chemical potentials, which are given by $\mu_i=\partial{\cal E}/\partial\rho_i$. 
Obviously Eqs.~(\ref{eq:chemeq1}) $-$ (\ref{eq:chemeq4}) imply that the system is in chemical equilibrium for the weak interaction processes, $n\rightleftharpoons p + K^-$, $n\rightleftharpoons p + e^-(+\bar\nu_e)$, $n + e^-\rightleftharpoons \Sigma^-(+\nu_e)$, $\Lambda + e^-\rightleftharpoons \Xi^-(+\nu_e)$, and $n\rightleftharpoons \Lambda(+\nu_e\bar\nu_e)$. 

\section{$S$-wave $K$-$B$ interaction in the contact Interaction scheme }
\label{sec:CI} 

\subsection{Kaon and baryon parts of Lagrangian density}
\label{subsec:3-1-1}

The effective chiral Lagrangian for kaons and baryons in the CI scheme~\textcolor{blue}{\cite{kn86}} is given by 
\begin{eqnarray}
{\cal L}_{K,B}&=&\frac{1}{4}f^2 \ {\rm Tr} 
\partial^\mu U^\dagger\partial_\mu U 
+\frac{1}{2}f^2\Lambda_{\chi{\rm SB}}({\rm Tr}M(U-1)+{\rm h.c.}) 
+{\rm Tr}\overline{\Psi}(i{\gamma^\mu\partial_\mu}-M_B)\Psi \cr
&+&{\rm Tr}\overline{\Psi}i\gamma^\mu\lbrack V_\mu, \Psi\rbrack 
+ D {\rm Tr}\overline{\Psi}\gamma^\mu \gamma^5\lbrace A_\mu, \Psi\rbrace
+F {\rm Tr}\overline{\Psi}\gamma^\mu \gamma^5\lbrack A_\mu, \Psi\rbrack \cr
&+& a_1{\rm Tr}\overline{\Psi}(\xi M^\dagger\xi+{\rm h.c.})\Psi 
+a_2{\rm Tr}\overline{\Psi}\Psi(\xi M^\dagger\xi+{\rm h.c.}) 
+ a_3({\rm Tr}MU +{\rm h.c.}){\rm Tr}\overline{\Psi}\Psi \ , 
\label{eq:lagkb}
\end{eqnarray}
where the first and second terms on the r.~h.~s. of Eq.~(\ref{eq:lagkb}) are the kinetic and mass terms of kaons $U$ in the nonlinear representation, respectively, and $\Lambda_{\chi{\rm SB}}$ is the chiral symmetry breaking scale ($\sim$ 1 GeV), $M$ the quark mass matrix,  
$M\equiv {\rm diag}(m_u, m_d, m_s)$ with the quark masses $m_i$\cite{kn86}. The free kaon mass is identified with  $m_K=\lbrack\Lambda_{\chi{\rm SB}}(m_u+m_s)\rbrack^{1/2}$ and is set to be the empirical value (=494 MeV).  
The third term in Eq.~(\ref{eq:lagkb}) denotes the kinetic and mass terms for baryons, where $M_{\rm B}$ is a  baryon mass generated by spontaneous breaking of chiral symmetry.   
The fourth term in Eq.~(\ref{eq:lagkb}) gives the $s$-wave $K$-$B$ vector interaction corresponding to the Tomozawa-Weinberg term with $V_\mu$ being the mesonic vector current defined by $V_\mu\equiv 
1/2(\xi^\dagger\partial_\mu\xi+\xi\partial_\mu\xi^\dagger)$ with $\xi\equiv U^{1/2}$. 
The fifth and sixth terms are the $K$-$B$ axial-vector interaction with the mesonic axial-vector current defined by $A_\mu\equiv i/2(\xi^\dagger\partial_\mu\xi-\xi\partial_\mu\xi^\dagger)$. Throughout this paper, we simply omit these axial-vector coupling terms and retain only the $s$-wave $K$-$B$ vector interaction in order to figure out the consequences from $s$-wave kaon condensation.
 The last three terms in Eq.~(\ref{eq:lagkb}) give the 
$s$-wave $K$-$B$ scalar interaction, which explicitly breaks chiral symmetry. 

By the use of Eq.~(\ref{eq:kfield}), the  Lagrangian density (\ref{eq:lagkb}) is separated into the baryon part ${\cal L}_B$ and kaon part ${\cal L}_K$ in the mean-field approximation. For ${\cal L}_B$ one obtains
\begin{equation}
 {\cal L}_B = \sum_{b=p,n,\Lambda, \Sigma^-, \Xi^-}\overline{\psi_b} (i\gamma^\mu \partial_\mu-M_b^\ast ) \psi_b \ , 
 \label{eq:lagb}
\end{equation} 
where the $s$-wave $K$-$B$ scalar interaction is absorbed into the effective baryon mass $M_b^\ast$ : 
\begin{equation}
M_b^\ast=M_b -\Sigma_{Kb}(1-\cos\theta) 
\label{eq:effbm}
\end{equation}
with $M_b$ ($b$=$p,n,\Lambda, \Sigma^-, \Xi^-$) being the baryon mass defined by 
\begin{eqnarray}
M_p &=& M_B-2(a_1m_u+a_2m_s)-2a_3(m_u+m_d+m_s) \ , \cr
M_n &=& M_B-2(a_1m_d+a_2m_s)-2a_3(m_u+m_d+m_s)  \ , \cr
M_\Lambda &=& M_B-1/3\cdot (a_1+a_2)(m_u+m_d+4m_s)-2a_3(m_u+m_d+m_s)  \ , \cr 
M_{\Sigma^-} &=& M_B-2(a_1m_d+a_2m_u)-2a_3(m_u+m_d+m_s)  \ , \cr
M_{\Xi^-} &=& M_B-2(a_1m_s+a_2m_u)-2a_3(m_u+m_d+m_s)  \ , 
\label{eq:fbmass}
\end{eqnarray}
and $\Sigma_{Kb}$ being the ``$K$-baryon sigma term'' which simulates the $K$-$B$ attractive 
interaction in the scalar channel\cite{m93,m08}: 
\begin{subequations}\label{eq:ckbsigma}
\begin{eqnarray}
\Sigma_{Kp}&=&\Sigma_{K\Xi^-}=-(a_1+a_2+2a_3)(m_u+m_s) \ , \label{eq:ckpsigma} \\
\Sigma_{Kn}&=&\Sigma_{K\Sigma^-}=-(a_2+2a_3)(m_u+m_s) \ , \label{eq:cknsigma} \\
\Sigma_{K\Lambda}&=&-\left(\frac{5}{6}a_1+\frac{5}{6}a_2+2a_3\right)(m_u+m_s) \ . \label{eq:cklsigma}
\end{eqnarray}
\end{subequations}
(See Appendix~\ref{subsec:appendix1}.)
In the following, $M_b$ ($b = p, n, \Lambda, \Sigma^-, \Xi^-$) [Eq.~(\ref{eq:fbmass})]  
are identified with the empirical baryon masses, i.~e.~, $M_p$=938.27 MeV, $M_n$=939.57 MeV, $M_\Lambda$=1115.68 MeV, $M_{\Sigma^-}$=1197.45 MeV, and $M_{\Xi^-}$=1321.71 MeV. 

Following Ref.~\cite{kn86}, the quark masses $m_i$ are chosen to be $m_u$ = 6 MeV, $m_d$ = 12 MeV, and $m_s$ = 240 MeV. Together with these values, the parameters $a_1$ and $a_2$ are fixed to be $a_1$ = $-$0.28, $a_2$ = 0.56 so as to reproduce the empirical octet baryon mass splittings\cite{kn86}. The remaining parameter $a_3$ is fixed so as to give the $KN$ sigma term, $\Sigma_{KN}$. Throughout this paper, we consider two cases as allowable values of $\Sigma_{Kn}$:  $\Sigma_{Kn}$ = 300 MeV and 400 MeV, for which $a_3 = -0.89$ and $-1.1$, respectively. In these cases, one also obtains, from Eqs.~(\ref{eq:ckpsigma})$-$(\ref{eq:cklsigma}), $\Sigma_{K\Sigma^-}$ = 300 MeV, $\Sigma_{Kp}$ = $\Sigma_{K\Xi^-}$ = 369 MeV, and $\Sigma_{K\Lambda}$ = 380 MeV for $a_3 = -0.89$, 
and $\Sigma_{K\Sigma^-}$ = 400 MeV, $\Sigma_{Kp}$ = $\Sigma_{K\Xi^-}$ = 469 MeV, and $\Sigma_{K\Lambda}$ = 480 MeV for $a_3 = -1.1$. 
According to the lattice QCD result, the current quark masses have been fixed to be rather smaller values, $(m_u, m_d, m_s)$ = (2.2, 4.7, 95) MeV~\cite{PDG2020}, than adopted in this paper. However, it can be shown that the $\Sigma_{KN}$ is little altered by the use of different quark masses as far as $(m_u+m_s)/\hat m$ ($\sim$14) is almost the same in both set of quark masses~\cite{mmt2021}. 

For ${\cal L}_K$ one obtains\cite{te97}
\begin{eqnarray}
{\cal L}_K&=& \frac{1}{2}\Bigg\lbrace 1+\left(\frac{\sin\theta}{\theta}\right)^2\Bigg\rbrace\partial^\mu K^+\partial_\mu K^- +\frac{1-\left(\frac{\sin\theta}{\theta}\right)^2}{2f^2\theta^2}\Big\lbrace (K^+\partial_\mu K^-)^2+(K^- \partial_\mu K^+)^2\Big\rbrace \cr
&-&  m_K^2\left( \frac{\sin(\theta/2)}{\theta/2}\right)^2 K^+ K^-
+ iX_0\left( \frac{\sin(\theta/2)}{\theta/2}\right)^2 \left(K^+\partial_0 K^- -\partial_0 K^+ K^-\right)  \cr
&=&\frac{1}{2}(f\mu_K\sin\theta)^2-f^2m_K^2(1-\cos\theta) 
+ 2 \mu_K X_0 f^2(1-\cos\theta) \ . 
\label{eq:lagk}
\end{eqnarray}
The last term on the r.h.s. of Eq.~(\ref{eq:lagk}) stands for the $s$-wave $K$-$B$ vector interaction with $X_0$ being given by 
\begin{equation}
X_0\equiv\frac{1}{2f^2}\sum_{b=p,n,\Lambda, \Sigma^-, \Xi^-} Q_V^b\rho_b = \frac{1}{2f^2}\left(\rho_p+\frac{1}{2}\rho_n-\frac{1}{2}\rho_{\Sigma^-}-\rho_{\Xi^-} \right)  \ , 
\label{eq:x0}
\end{equation}
where $Q_V^b$ $\equiv \frac{1}{2}\left(I_3^{(b)}+\frac{3}{2}Y^{(b)}\right)$ is the V-spin charge with $I_3^{(b)}$ and $Y^{(b)}$ being the third component of the isospin and hypercharge for baryon species  $b$, respectively. The form of Eq.~(\ref{eq:x0}) for $X_0$ is specified model-independently within chiral symmetry. 
From Eqs.~(\ref{eq:lagk}) and (\ref{eq:x0}), one can see that the $s$-wave $K$-$B$ vector interaction works attractively for protons and neutrons, while repulsively for $\Sigma^-$ and $\Xi^-$ hyperons, as far as $\mu_K > 0$. 

In Ref.~\cite{mmt2015}, we improved our model for the $s$-wave $K$-$B$ interaction by introduction of the range terms and a pole contribution from $\Lambda$(1405) (denoted as $\Lambda^\ast$) so as to reproduce the on-shell $s$-wave $KN$ scattering lengths\cite{m81}. 
It has been shown that the range terms work repulsively and that the onset density of kaon condensation  realized from hyperon-mixed matter is slightly pushed up to a higher density as compared with the case without the range terms. Nevertheless, the repulsive effect from the range terms ($\propto \mu_K^2$) on the EOS of the ($Y+K$) phase is tiny over the relevant densities as far as $\mu_K\lesssim O(m_\pi)$. 
The contribution from the $\Lambda^\ast$ pole to the energy is also negligible over the baryon density $\rho_B\gtrsim \rho_0$, since the kaon energy in matter lies well below the pole position of $M_{\Lambda^\ast}-M_N$. Therefore, these effects of the range terms and the pole contribution from $\Lambda^\ast$ are omitted in this paper. 

\section{$B$-$B$ interaction in the RMF}
\label{sec:BB}

\subsection{Lagrangian density for baryons and mesons}
\label{subsec:1}

In the RMF framework, the $B$-$B$ interaction is mediated by scalar ($\sigma$, $\sigma^\ast$) and vector ($\omega$, $\rho$, $\phi$) mesons. The scalar meson $\sigma^\ast$ ($\sim \bar s s$)  and the vector meson $\phi$ ($\sim\bar s\gamma^\mu s$), both of which carry the strangeness and couple only to hyperons ($Y$), are introduced according to the extension of baryons to include hyperons. The quark structure of the $\phi$ meson  comes from the assumption of an ideal mixing between $\omega$ and $\phi$ mesons. 
The scalar meson-baryon couplings lead to  modification of the effective baryon masses (\ref{eq:effbm}) to
\begin{equation}
\widetilde M_b^\ast=M_b-g_{\sigma b}\sigma-g_{\sigma^\ast b}\sigma^\ast-\Sigma_{Kb}(1-\cos\theta)  
\label{eq:effbmci}
\end{equation}
with 
$g_{\sigma b}$ ($b=p,n,\Lambda, \Sigma^-, \Xi^-$) and $g_{\sigma^\ast Y}$ ($Y=\Lambda, \Sigma^-, \Xi^-$) being the scalar meson-baryon coupling constants. 
The vector meson-baryon couplings are introduced by the covariant derivatives at the baryon kinetic terms as $\partial_\mu$$\rightarrow$$D^{(b)}_\mu\equiv \partial_\mu+i g_{\omega b}\omega_\mu+i \widetilde g_{\rho b}{\vec I}^{\ (b)}\cdot{\vec R}_\mu +ig_{\phi b}\phi_\mu$, 
where $\omega^\mu$,  $\vec R^\mu$, and $\phi^\mu$ denote the vector meson fields for $\omega$, $\rho$, and $\phi$ mesons, respectively. The arrow attached to the $\rho$-meson field $\vec R_\mu$ refers to an isovector with the isospin operator $\vec I^{\ (b)}$ for baryon $b$. 
The $g_{\omega b}$, $\widetilde g_{\rho b}$, and $g_{\phi b}$ in the covariant derivative are the vector meson-baryon coupling constants.
It is to be noted that the $\rho$-meson-baryon coupling constant, $g_{\rho b}$, is factorized as $g_{\rho b}\equiv\widetilde g_{\rho b}\cdot |I_3^{(b)}|$ with the third component of the isospin for the baryon $b$. The $\rho$-meson-baryon coupling constants utilized in the other RMF models mostly correspond to 
$\widetilde g_{\rho b}$ in our model. 

The baryon part of the Lagrangian density ${\cal L}_B$ is then modified from Eq.~(\ref{eq:lagb}) to
 \begin{equation}
{\cal L}_B=\sum_{b=p,n,\Lambda, \Sigma^-, \Xi^-}\overline{\psi_b} (i\gamma^\mu D^{(b)}_\mu-\widetilde M_b^\ast ) \psi_b \ . 
\label{eq:lagbci}
\end{equation}
In addition, the meson part of the Lagrangian density, ${\cal L}_M$, including the $\sigma$ self-interaction is introduced: 
\begin{eqnarray}
{\cal L}_{M}&=&\frac{1}{2}\left(\partial^\mu\sigma\partial_\mu\sigma-m_\sigma^2\sigma^2\right)-U_\sigma(\sigma) + \frac{1}{2}\left(\partial^\mu\sigma^\ast\partial_\mu\sigma^\ast-m_{\sigma^\ast}^2\sigma^{\ast 2}\right) \cr\vspace{0.1cm}~
&-&\frac{1}{4}\omega^{\mu\nu}\omega_{\mu\nu}+\frac{1}{2}m_\omega^2\omega^\mu\omega_\mu -\frac{1}{4}{\vec R}^{\mu\nu}\cdot{\vec R}_{\mu\nu}+\frac{1}{2}m_\rho^2 {\vec R}^\mu\cdot {\vec R}_\mu -\frac{1}{4}\phi^{\mu\nu}\phi_{\mu\nu}+\frac{1}{2}m_\phi^2\phi^\mu\phi_\mu \ . 
\label{eq:lagm}
\end{eqnarray}
The second term on the  r.~h.~s. of Eq.~(\ref{eq:lagm}) is the scalar self-interaction potential given by $U_\sigma(\sigma)$=$bM_N(g_{\sigma N}\sigma)^3/3+c(g_{\sigma N}\sigma)^4/4$ with $b$=0.008659 and $c$=$-$0.002421\cite{g85,mmt09}. $U_\sigma(\sigma)$ is introduced so as to set the incompressibility at the nuclear saturation density to be 240 MeV, which is consistent with an empirical value. 
In this paper, the $U_\sigma(\sigma)$ is solely considered as the NLSI term. 
The kinetic terms of the vector mesons are given in terms of 
$\omega^{\mu\nu}\equiv \partial^\mu \omega^\nu-\partial^\nu \omega^\mu$, $\vec R^{\mu\nu}\equiv \partial^\mu {\vec R}^\nu-\partial^\nu {\vec R}^\mu$, $\phi^{\mu\nu}\equiv \partial^\mu \phi^\nu-\partial^\nu \phi^\mu$. Throughout this paper, only time components of the vector meson mean fields and the third  component of the isovector $\rho$ mean field are considered for description of the ground state of the system. We simply denote these components as $\omega_0$, $R_0$, and $\phi_0$. Then the $\rho$-$B$ coupling term in the covariant derivative is rewritten as $i\widetilde g_{\rho b}I_3^{(b)}R_0=i g_{\rho b}\hat I_3^{(b)}R_0$ with $g_{\rho b}$ (=$\widetilde g_{\rho b}\cdot|I_3^{(b)}|$) and $\hat I_3^{(b)}\equiv I_3^{(b)}/|I_3^{(b)}|$, where $\hat I_3^{(b)}$ is assigned as $\hat I_3^{(p)}$=+1, $\hat I_3^{(n)}=\hat I_3^{(\Sigma^-)}=\hat I_3^{(\Xi^-)}=-1$. 

In the CI scheme, there is no direct kaon-meson ($m$) coupling ($m$ = $\sigma$, $\sigma^\ast$, $\omega$, $\rho$, $\phi$) [see Eqs.~(\ref{eq:lagk}) and (\ref{eq:lagm})].

\subsection{Choice of meson-baryon coupling constants}
\label{subsec:mbcc}

The values of $g_{\sigma N}$, $g_{\omega N}$, $g_{\rho N}$, which are related to the $N$-$N$ interaction,  are determined so as to reproduce not only the properties of normal nuclear matter with saturation density $\rho_0$ = 0.153~fm$^{-3}$, the binding energy (=16.3 MeV), and the symmetry energy (=32.8 MeV), but also proton-mixing ratio and density distributions of proton and neutron for normal nuclei\cite{mmt09}. One obtains $g_{\sigma N}$ = 6.39, $g_{\omega N}$ = 8.72, and $g_{\rho N}$ = 4.27. 
The $\sigma^\ast$-$N$ and $\phi$-$N$ couplings are not taken into account since they should be suppressed due to the OZI rule.

The vector meson couplings for the hyperon $Y$ are obtained from the relations in the SU(6) symmetry\cite{sdg94}: 
\begin{subequations}\label{eq:vy}
\begin{eqnarray}
g_{\omega\Lambda}&=&g_{\omega\Sigma^-}=2g_{\omega \Xi^-}=\frac{2}{3} g_{\omega N} \ , \label{eq:vy1} \\
g_{\rho \Lambda}&=&0 \ , \ g_{\rho\Sigma^-}=2g_{\rho\Xi^-}=2g_{\rho N} \ ,\label{eq:vy2} \\
g_{\phi\Lambda}&=&g_{\phi\Sigma^-}=\frac{1}{2} g_{\phi\Xi^-}=-\frac{\sqrt{2}}{3} g_{\omega N} \ . \label{eq:vy3}
\end{eqnarray}
\end{subequations} 

The scalar ($\sigma$, $\sigma^\ast$) mesons-hyperon couplings are determined from the phenomenological analyses of recent hypernuclear experiments as much as possible. The scalar $\sigma$ meson coupling for the hyperon $Y$, $g_{\sigma Y}$, is related with 
the potential depth of the hyperon $Y$ ($Y=\Lambda$, $\Sigma^-$, $\Xi^-$) at $\rho_{\rm B}=\rho_0$ in SNM, $V_Y^N$, which is written in the RMF as  
\begin{equation}
V_Y^N=-g_{\sigma Y}\langle\sigma\rangle_0 +g_{\omega Y}\langle\omega_0\rangle_0 
\label{eq:ypot}
\end{equation}
with $\langle\sigma\rangle_0$ and $\langle\omega_0\rangle_0$ being the meson mean fields at $\rho_B=\rho_0$ in SNM. 
For $Y=\Lambda$, the single $\Lambda$ orbital energies in ordinary hypernuclei are fitted well with the $\Lambda$-nucleus single particle potential with the depth $\sim$ $-27$ MeV\cite{mdg88}. Based on  this result,  $V_\Lambda^N$ is set to be $-27$ MeV. One then obtains $g_{\sigma\Lambda}$ = 3.84 from Eq.~(\ref{eq:ypot}). 

The depth of the $\Sigma^-$ potential $V_{\Sigma^-}^N$ has been shown to be repulsive, according to the recent theoretical calculations\cite{kf00,fk01} and phenomenological analyses on the ($K^-$, $\pi^\pm $) reactions at BNL\cite{b99,d99}, ($\pi^-$, $K^+$) reactions at KEK\cite{n02,dr04,hh05}, and the $\Sigma^-$ atom data\cite{mfgj95}.  Following Ref.~\cite{d99}, the $V_{\Sigma^-}^N$ is set to be 23.5 MeV as a typical value, from which one obtains $g_{\sigma\Sigma^-}$ = 2.28 from Eq.~(\ref{eq:ypot}). 

The depth of the $\Xi^-$ potential in nuclear matter is set to be attractive, $V_{\Xi^-}^N$ = $-14$ MeV with reference to the experimental results deduced from the ($K^-$, $K^+$) reactions, ($-$14)$-$($-$20) MeV\cite{f98,k00}. One then obtains $g_{\sigma\Xi^-}$ = 1.94 from Eq.~(\ref{eq:ypot}). 

The $\sigma^\ast$ meson couplings for the hyperon $Y$, $g_{\sigma^\ast Y}$,  
are relevant to the $Y$-$Y$ interaction as well as binding energy of hypernuclei.  From recent detection of the double $\Lambda$ hypernuclei in the KEK-E176-E373 experiments, separation energies for two $\Lambda$'s, $B_{\Lambda\Lambda}$, have been obtained for several double $\Lambda$ hypernuclei\cite{ghm16}.  
For the ``Hida event'', the experimental value of the separation energy for the ground state of the $^{\ 11}_{\Lambda\Lambda}{\rm Be}$ has been estimated to be 20.83$\pm$1.27 MeV\cite{ghm16,h2010}. 
We determined $g_{\sigma^\ast \Lambda}$ so as to reproduce the empirical values of $B_{\Lambda\Lambda}$($^{\ \ 11}_{\Lambda\Lambda}$Be) 
with use of the present  $B$-$B$ interaction model in the RMF extended to 
finite nuclei\cite{mmt09,mmt14}. One finds $g_{\sigma^\ast \Lambda}$ = 7.2, for which $B^{\rm th}_{\Lambda\Lambda}(^{\ 11}_{\Lambda\Lambda}{\rm Be})\equiv
B^{\rm th}(^{\ 11}_{\Lambda\Lambda}{\rm Be})-B^{\rm exp}(^{9}{\rm Be})$ = 20.7 MeV with $B^{\rm exp}(^{9}{\rm Be})$ = 58.16 MeV\cite{awt03}. [Throughout this paper, the superscript, ``exp" (``th''), denotes an experimental value (a theoretical value obtained in our $B$-$B$ interaction model).] 
In this case, the $\Lambda$-separation energy for $^{10}_{\ \Lambda}{\rm Be}$ is estimated to be 
$B^{\rm th}_\Lambda(^{10}_{\ \Lambda}{\rm Be}) \equiv B^{\rm th}(^{10}_{\ \Lambda}{\rm Be}) - B^{\rm exp}(^9{\rm Be})$ = 9.95 MeV, whereas the experimental value for the ground state peak (mixture of 1$^-$ and 2$^-$ states) at J-Lab is reported as 8.55 MeV\cite{g2016}.
It is to be noted that our $B$-$B$ interaction model applied to finite nuclei assumes local density approximation and spherical symmetry for profiles of baryon density distributions, whereas the  $^{\ 11}_{\Lambda\Lambda}{\rm Be}$ nucleus has a clustering structure\cite{h2010}. Furthermore a quantitatively accurate estimation  in our model may not be expected for a few-body system such as the $^{\ \ 6}_{\Lambda\Lambda}$He (the ``Nagara event''), although a precise extraction of the $\Lambda\Lambda$ binding energy $B_{\Lambda\Lambda}$($^{\ \ 6}_{\Lambda\Lambda}$He) and the bond energy $\Delta B_{\Lambda\Lambda}$($^{\ \ 6}_{\Lambda\Lambda}$He)$\equiv B_{\Lambda\Lambda}$($^{\ \ 6}_{\Lambda\Lambda}$He)$-$2$B_\Lambda$($^5_{\Lambda}$He) has been done experimentally\cite{ht01,a13}.

For the $\sigma^\ast$-$\Xi^-$ coupling, the bound $\Xi^-$ hypernucleus was detected through the ``Kiso'' event, $\Xi^-$ + $^{14}$N $\rightarrow$ $^{15}_{\ \Xi}$C $\rightarrow$ $^{10}_{\ \Lambda}$Be + $^5_\Lambda$He\cite{n15}. In Ref.~\cite{sh16}, $^{15}_{\ \Xi}$C was assumed to be an excited state with the $\Xi^-$ being in the 1$p$ state, and the estimated separation energies of $\Xi^-$ for both $^{\ \ 15}_{\ \Xi(p)}$C and the ground state of $^{\ \ 12}_{\ \Xi(s)}$Be were shown to be consistent with the empirical values, i.e., $B^{\rm exp}_{\Xi}(^{\ \ 15}_{\ \Xi(p)}$C) = 1.11$\pm$0.25 MeV and $B^{\rm exp}_{\Xi}(^{\ \ 12}_{\ \Xi(s)}$Be) $\approx$ 5 MeV.  In this case, the $B_{\Xi}(^{\ \ 15}_{\ \Xi(s)}$C) for the ground state of the $^{15}_{\ \Xi}$C was estimated to be (8.0 $-$ 9.4) MeV within the RMF calculation\cite{sh16}. This  interpretation concerning the energy-level structure and the binding energy of the $\Xi^-$-$^{14}$N system has been confirmed by the recent observation of the twin-$\Lambda$ hypernuclei (the ``IBUKI'' event) at the J-PARC E07 experiment\cite{hayakawa2021,yoshimoto2021}. In our $B$-$B$ interaction model, $g_{\sigma^\ast\Xi^-}$ is taken to be 4.0, for which one obtains $B^{\rm th}_{\Xi}(^{ \ \ 15}_{\ \Xi(s)}$C) = 8.1 MeV and $B^{\rm th}_{\Xi}(^{\ \ 12}_{\ \Xi(s)}$Be) = 5.1 MeV, which are consistent with the empirical values. 

 The remaining parameter for the $\sigma^\ast$-$\Sigma^-$ coupling, $g_{\sigma^\ast \Sigma^-}$, is simply set to be zero since there is little experimental information on $\Sigma$ hypernuclei. 
As is seen in the numerical result (Sec.~\ref{sec:eos}), the $\Sigma^-$ hyperons are not mixed over the relevant densities due to the strong repulsion of the $V_{\Sigma^-}^N$. Therefore it may safely be said that such simplification, $g_{\sigma^\ast \Sigma^-} = 0$, little affects the results in this paper. 
It is to be noted that $g_{\sigma^\ast Y}$ ($Y=\Lambda, \Sigma^-, \Xi^-$) is also related with the $s$-wave $K$-$B$ scalar attraction in the ME scheme [see Eq.~(\ref{eq:mekbsigma}) in Sec.~\ref{subsec:ME-CI}].  

Together with these coupling constants, the meson masses are taken to be $m_\sigma$ = 400 MeV, $m_{\sigma^\ast}$ = 975 MeV, $m_\omega$ = 783 MeV, $m_\rho$ = 769 MeV, and $m_\phi$ = 1020 MeV. 
The parameters relevant to the meson-baryon interaction used in our RMF model are listed in Table~\ref{tab:para1} and \ref{tab:para2}.
\begin{table}[!]
\caption{The coefficients $b$ and $c$ in the $\sigma$ self-interaction potential $U_\sigma(\sigma)$ and  the meson masses $m_a$ ($a=\sigma, \sigma^\ast, \omega, \rho, \phi$) used in our RMF model. See the text for details. }
\begin{center}
\begin{tabular}{c c c c c c c}
\hline
 $b$ & $c$ & $m_\sigma$ & $m_{\sigma^\ast} $ & $m_\omega$ & $m_\rho$ & $m_\phi $  \\
     & & (MeV) & (MeV) & (MeV) & (MeV) & (MeV) \\ \hline
 0.008659 & \ $-$0.002421 & 400 & 975 & 783 & 769 & 1020  \\\hline
\end{tabular}
\label{tab:para1}
\end{center}
\end{table}

\begin{table}[!]
\caption{The meson-baryon coupling constants $g_{aB}$ ($a=\sigma, \sigma^\ast, \omega, \rho, \phi$ and $B=$N$, \Lambda, \Sigma^-, \Xi^-$) used in our RMF model.}
\begin{center}
\begin{tabular}{ c || c | c | c | c }
\hline
\backslashbox{$a$}{$B$} &  $N$ & $\Lambda$ & $\Sigma^-$ & $\Xi^-$ \\\hline\hline
 $\sigma$         &  6.39 & 3.84 & 2.28 & 1.94 \\
 $\sigma^\ast$ &  0      & 7.2   & 0      & 4.0 \\\hline
$\omega$       & 8.72 &  $\displaystyle\frac{2}{3}g_{\omega N}$ & $\displaystyle\frac{2}{3}g_{\omega N}$ & $\displaystyle\frac{1}{3}g_{\omega N}$ \\
 $\rho$       & 4.27 &  0 & $2g_{\rho N}$ & $g_{\rho N}$ \\
$\phi$              & 0 &  $\displaystyle -\frac{\sqrt{2}}{3}g_{\omega N}$ & $\displaystyle -\frac{\sqrt{2}}{3}g_{\omega N}$ &$\displaystyle -\frac{2\sqrt{2}}{3}g_{\omega N}$ \\[0.2cm]
 \hline
\end{tabular}
\label{tab:para2}
\end{center}
\end{table} 
\vspace{0.5cm}


\section{Energy expression in the CI scheme}
\label{sec:CI-energy}

\subsection{Effective energy density}
\label{subsubsec:effen}

The total Lagrangian density ${\cal L}$ in the CI scheme for the description of the ($Y$+$K$) phase is given by  Eqs.~(\ref{eq:lagk}), (\ref{eq:lagbci}), (\ref{eq:lagm}), together with the electron part ${\cal L}_e$ : 
${\cal L}$=${\cal L}_B$+${\cal L}_M$+${\cal L}_K$+${\cal L}_e$. 
The energy density of the ($Y+K$) phase, ${\cal E}$ (=${\cal E}_B+{\cal E}_M+{\cal E}_K+{\cal E}_e$), is obtained from the ground-state expectation value of the Hamiltonian density, ${\cal H}={\cal H}_B+{\cal H}_M+{\cal H}_K+{\cal H}_e$ with
\begin{subequations}\label{eq:h}
\begin{eqnarray}
{\cal H}_B&=&\sum_b(\partial{\cal L}_B/\partial\dot \psi_b)\dot \psi_b-{\cal L}_B \ , \label{eq:hb} \\
{\cal H}_M&=&\sum_m (\partial{\cal L}_M/\partial\dot \varphi_m)\dot \varphi_m-{\cal L}_M \ , \label{eq:hm} \\
{\cal H}_K&=&(\partial{\cal L}_K/\partial\dot K^-)\dot K^- +(\partial{\cal L}_K/\partial\dot K^+)\dot K^+-{\cal L}_K \ , \label{eq:hk} \\
{\cal H}_e&=& (\partial{\cal L}_e/\partial\dot \psi_e)\dot \psi_e -{\cal L}_e \ ,\label{eq:he}
\end{eqnarray}
\end{subequations}
where $\varphi_m$ ($=\sigma,\sigma^\ast,\omega_0,R_0, \phi_0$) is the meson field mediating the $B$-$B$ interaction, and $\psi_e$ the electron field. 
From Eqs.~(\ref{eq:hb}) and (\ref{eq:lagb}),  one obtains
\begin{equation}
{\cal E}_B=\sum_{b=p,n,\Lambda,\Sigma^-, \Xi^-} \Bigg\lbrace\frac{2}{(2\pi)^3}\int_{|{\bf p}|\leq p_F(b)} d^3|{\bf p}|(|{\bf p}|^2+\widetilde M_b^{\ast 2})^{1/2} 
+\rho_b\left(g_{\omega b}\omega_0 +g_{\rho b} \hat I_3^{(b)} R_0 +g_{\phi b}\phi_0\right) \Bigg\rbrace \ ,
\label{eq:eb}
\end{equation}
where $p_F(b)$ is the Fermi momentum of baryon $b$. 
From Eqs.~(\ref{eq:hm}) and (\ref{eq:lagm}), one obtains
\begin{equation}
{\cal E}_M=\frac{1}{2}m_\sigma^2\sigma^2+U(\sigma)+\frac{1}{2}m_{\sigma^\ast}^2\sigma^{\ast 2}
-\frac{1}{2}m_\omega^2\omega_0^2-\frac{1}{2}m_\rho^2 R_0^2-\frac{1}{2}m_\phi^2\phi_0^2 \ . 
\label{eq:em}
\end{equation}
The kaon part of the energy density, ${\cal E}_K$, is expressed from Eq.~(\ref{eq:hk}) as 
\begin{equation}
{\cal E}_K=\mu_K\rho_{K^-}-{\cal L}_K \ , 
\label{eq:ek}
\end{equation}
where the first term is obtained by rewriting the first two terms on the r.~h.~s. of Eq.~(\ref{eq:hk}) by the use of the time dependence of the classical kaon field, 
$\dot K^\pm=\pm i\mu_K K^\pm$, which follows from Eq.~(\ref{eq:kfield})\cite{kn86,bc79},  
and the number density of kaon condensates,  
\begin{equation}
\rho_{K^-}=-iK^-(\partial{\cal L}_K/\partial\dot{K^-})+iK^+(\partial{\cal L}_K/\partial\dot{K^+})\ . 
\label{eq:rhok}
\end{equation}

Substituting Eqs.~(\ref{eq:kfield}) and (\ref{eq:lagk}) into Eq.~(\ref{eq:rhok}), one obtains 
\begin{equation}
\rho_{K^-}=\mu_K f^2\sin^2\theta+2f^2X_0(1-\cos\theta) \ ,
\label{eq:rhokci}
\end{equation}
where the first term is the kaon kinetic part and the second term comes from the $s$-wave $K$-$B$ vector interaction.
With Eqs.~(\ref{eq:rhokci}) and (\ref{eq:lagk}), Eq.~(\ref{eq:ek}) reads
\begin{equation}
{\cal E}_K=\frac{1}{2}(\mu_K f\sin\theta)^2+f^2m_K^2(1-\cos\theta) \ .
\label{eq:ekfinal}
\end{equation}
The electron part, Eq.~(\ref{eq:he}), is simply written as
\begin{equation}
{\cal E}_e\simeq\mu_e^4/(4\pi^2) 
\label{eq:ee}
\end{equation}
for the ultra-relativistic electrons. 

With Eqs.~(\ref{eq:chemeq1})$-$(\ref{eq:chemeq4}) the effective energy density ${\cal E}_{\rm eff}$ (the thermodynamic potential density, which is equal to the sign-reversed total pressure, $-P$)  is separated  into 
\begin{equation}
{\cal E}_{\rm eff}={\cal E}_{{\rm eff},B}+{\cal E}_{{\rm eff},M}+{\cal E}_{{\rm eff},K}+{\cal E}_{{\rm eff},e} \ , 
\label{eq:eff2}
\end{equation}
where
\begin{subequations}\label{eq:effmod}
\begin{eqnarray}
{\cal E}_{{\rm eff},B}&=&{\cal E}_B-\sum_{b=p,n, \Lambda, \Sigma^-, \Xi^-} \mu_b\rho_b \ , \label{eq:effmodb} \\
{\cal E}_{{\rm eff},M}&=&{\cal E}_M \ , \label{eq:effmodm} \\
{\cal E}_{{\rm eff},K}&=&{\cal E}_K-\mu_K\rho_{K^-}=-{\cal L}_K \ , \label{eq:effmodk} \\
{\cal E}_{{\rm eff},e}&=&{\cal E}_e-\mu_e\rho_e=\mu_e^4/(4\pi^2)-\mu_e\cdot \mu_e^3/(3\pi^2)=-\mu_e^4/(12\pi^2)  \ . \label{eq:effmode}
\end{eqnarray}
\end{subequations}

\subsection{Classical kaon field equation and equations of motion for meson mean-fields}
\label{subsec:2}

The classical kaon field equation is obtained as $\partial{\cal E}_{\rm eff}/\partial\theta=0$, which reads
\begin{equation}
\mu_K^2\cos\theta+2X_0\mu_K-m_K^{\ast 2}=0 
\label{eq:keom2}
\end{equation} 
with
\begin{equation}
m_K^{\ast 2}\equiv m_K^2-\frac{1}{f^2}\sum_{b=p,n,\Lambda, \Sigma^-, \Xi^-}\rho_b^s\Sigma_{Kb} \ , 
\label{eq:ekm2}
\end{equation}
where $\rho_b^s$ is a scalar density for baryon $b$: 
\begin{equation}
 \rho_b^s=\frac{2}{(2\pi)^3} \int_{|{\bf p}|\leq p_F(b)} d^3{\bf p}\frac{\widetilde M_b^\ast}{(|{\bf p}|^2
+\widetilde M_b^{\ast 2})^{1/2}} \ . 
\label{eq:rhobs}
 \end{equation}

\begin{figure}[t]
\begin{minipage}[l]{0.50\textwidth}~\vspace{-3.5cm}~
\begin{center}
\includegraphics[height=.25\textheight]{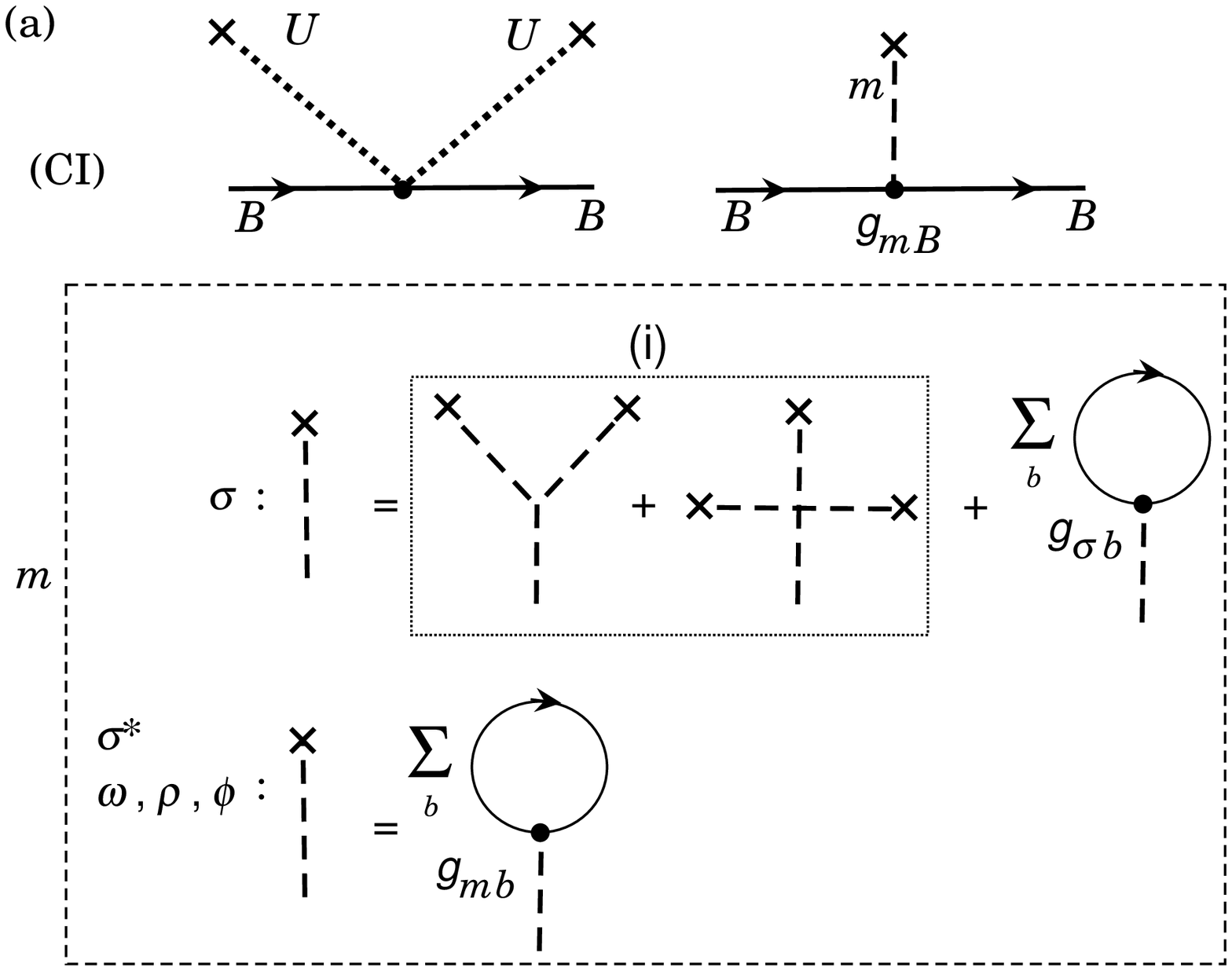}
\end{center}~
\end{minipage}~
\begin{minipage}[r]{0.50\textwidth}
\begin{center}
\includegraphics[height=.38\textheight]{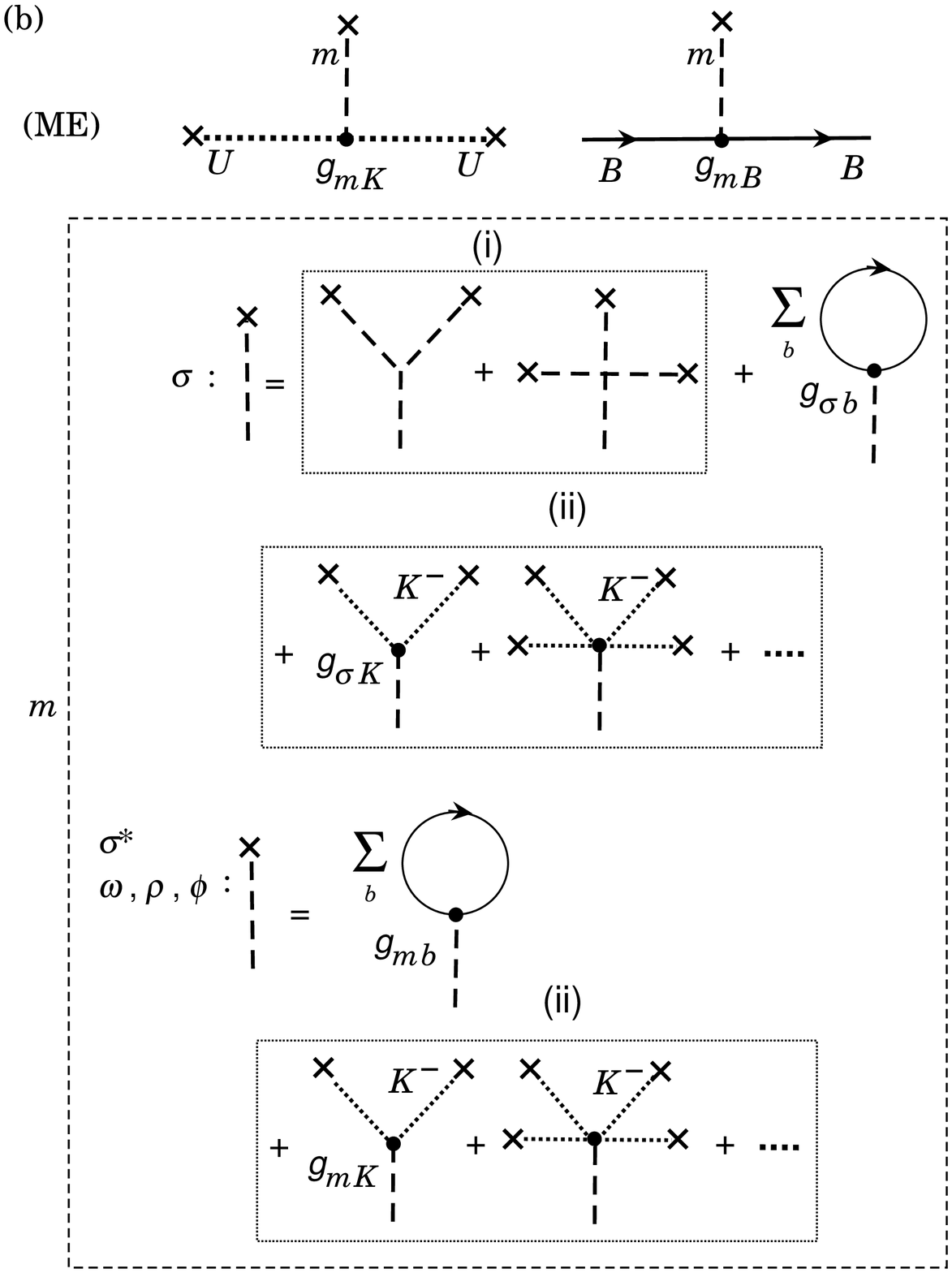}
\end{center}~
\end{minipage}
\caption{(a) The diagrams of the  interaction vertices in the CI scheme for the nonlinear $K$-$B$ interaction and the $B$-$m$ ($m$=$\sigma$, $\sigma^\ast$, $\omega$, $\rho$, $\phi$) interaction. The bold dotted line with a cross ($\times$) stands for the classical nonlinear $K^-$ field $U$,  the solid line for baryon $B$, and the dashed line with a cross ($\times$) for the mean field of meson $m$ ($m$ = $\sigma, \sigma^\ast$ for the scalar mesons, $\omega$, $\rho$, $\phi$ for the vector mesons). The dashed line with no cross at both ends, when connected to the interaction vertices, gives a propagator (=$1/{m_m}^2$) for the meson $m$ with mass $m_m$ in the Hartree approximation. In the dotted box, the equations of motion for the $\sigma$ [Eq.~(\ref{eq:cieom1})] and those for the other meson fields ($\sigma^\ast$, $\omega$, $\rho$, $\phi$) [Eqs.~(\ref{eq:cieom2}) $-$ (\ref{eq:cieom5})] are depicted by the diagrams. The diagram (i) comes from the self-interaction term of the multi-$\sigma$ mesons ($\propto dU_\sigma/d\sigma$).  (b) The same as (a) but in the ME scheme. In the dotted box in the ME scheme, the diagram (ii) stands for the expansion of the kaon source terms in Eqs.~(\ref{eq:meeom1}) $-$ (\ref{eq:meeom5}) in Sec.~\ref{subsec:ME-CI} in powers of the kaon field ($\propto \theta$) with the dotted lines. \break}
\label{fig1}
\end{figure}

 The equations of motion for the meson mean fields in the CI scheme are given from $\partial{\cal E}_{\rm eff}/\partial \varphi_m=~0$ ($\varphi_m =\sigma,\sigma^\ast,\omega_0,R_0, \phi_0$). With the use of  Eqs.~(\ref{eq:eb}), (\ref{eq:em}), and (\ref{eq:effbmci}) one obtains 
\begin{subequations}\label{eq:cieom}
\begin{eqnarray}
m_\sigma^2\sigma&=&-\frac{dU_\sigma}{d\sigma}+\sum_{b=p,n,\Lambda,\Sigma^-,\Xi^-}g_{\sigma b}\rho_b^s  \ , \label{eq:cieom1} \\
m_\sigma^{\ast 2}\sigma^\ast&=&\sum_{Y=\Lambda,\Sigma^-,\Xi^-}g_{\sigma^\ast Y}\rho_Y^s \ , \label{eq:cieom2}\\
m_\omega^2\omega_0&=&\sum_{b=p,n,\Lambda,\Sigma^-,\Xi^-} g_{\omega b}\rho_b \ ,  \label{eq:cieom3}\\
m_\rho^2 R_0&=&\sum_{b=p,n,\Lambda,\Sigma^-,\Xi^-} g_{\rho b}{\hat I}_3^{(b)} \rho_b \ , \label{eq:cieom4} \\
m_\phi^2\phi_0 &=&\sum_{Y=\Lambda,\Sigma^-,\Xi^-}g_{\phi Y}\rho_Y \ . \label{eq:cieom5}
\end{eqnarray}
\end{subequations}
 The diagrams of the interaction vertices in the CI scheme are depicted in Fig.~\ref{fig1}~(a) for the nonlinear $K$-$B$ interaction and the $B$-$m$ interaction. In the CI scheme, the structure of the $s$-wave $K$-$B$ and $K$-$K$ interactions is uniquely determined from chiral symmetry, and the mesons $m$ do not directly couple to kaons but only to baryons. There appears a many-body effect at the $\sigma$ - $B$ vertex through coupling of $B$ to the multi-$\sigma$ mesons ($\propto dU_\sigma/d\sigma$ being denoted as (i) in Fig.~\ref{fig1}~(a)). 

The ground state energy for the ($Y+K$) phase is obtained so as to satisfy Eqs.~(\ref{eq:keom2}), (\ref{eq:cieom}), under the charge neutrality condition, $\partial{\cal E_{\rm eff}}/\partial \mu=0$, and chemical equilibrium conditions for weak processes, $\partial{\cal E_{\rm eff}}/\partial \rho_i=0$ ($i=K^-, e^-, p, n, \Lambda, \Sigma^-, \Xi^-$) with the total baryon number density $\rho_B$ being fixed. 
 From the last conditions the relations between the chemical potentials, $\mu=\mu_K=\mu_e=\mu_n-\mu_p$, $\mu_\Lambda=\mu_n$, $\mu_{\Sigma^-}=\mu_{\Xi^-}=\mu_n+\mu_e$ [(\ref{eq:chemeq1}) $-$ (\ref{eq:chemeq4})] are assured.
Here the baryon chemical potential $\mu_b$ (for $b=p,n,\Lambda,\Sigma^-,\Xi^-$) is obtained from  Eqs.~(\ref{eq:eb}), (\ref{eq:em}), (\ref{eq:ekfinal}) with the help of Eqs.~(\ref{eq:x0}), (\ref{eq:effbmci}), (\ref{eq:rhokci}) and Eqs.~(\ref{eq:keom2}) $\sim$ (\ref{eq:cieom}) as 
\begin{equation}
\mu_b=\partial{\cal E}/\partial\rho_b=\left(p_F(b)^2+\widetilde M_b^{\ast 2}\right)^{1/2}+g_{\omega b}\omega_0+g_{\rho b}{\hat I}_3^{(b)} R_0+g_{\phi b}\phi_0-\mu Q_V^b(1-\cos\theta) \ . 
\label{eq:mub}
\end{equation}
From Eq.~(\ref{eq:mub}), the baryon potential $V_b$ ($b$ = $p$, $n$, $\Lambda$, $\Xi^-$, $\Sigma^-$)  reads
\begin{equation}
V_b= -g_{\sigma b}\sigma-g_{\sigma^\ast b}\sigma^\ast+g_{\omega b}\omega_0+g_{\rho b}{\hat I}_3^{(b)} R_0+g_{\phi b}\phi_0-(\Sigma_{Kb}+\mu Q_V^b)(1-\cos\theta) \ . 
\label{eq:vb}
\end{equation}
By setting $\rho_Y$ = 0 ($Y=\Lambda, \Xi^-, \Sigma^-$), the description of the energy expression for the ($Y$+$K$) phase in the CI scheme reduces essentially to the one studied for kaon-condensed phase in neutron-star matter without hyperon-mixing\cite{fmmt96}. 

\section{Meson-exchange (ME) scheme}
\label{sec:ME}

\subsection{Correspondence between the ME scheme and the CI scheme}
\label{subsec:ME-CI}

We recapitulate here 
the meson-exchange (ME) scheme for the $s$-wave $K$-$B$ interaction\cite{mmt09,mmt14} in comparison with the CI scheme. In the CI scheme, the $s$-wave $K$-$B$ scalar interaction is absorbed into the effective baryon mass $\widetilde M_b^\ast$ (\ref{eq:effbmci}) in the baryon part of the Lagrangian ${\cal L}_B$ (\ref{eq:lagbci}), and the effect of the $s$-wave $K$-$B$ scalar interaction on kaons is involved in the effective kaon mass squared $m_K^{\ast 2}$ (\ref{eq:ekm2}) through the classical kaon field equation (\ref{eq:keom2}). 

In the ME scheme, not only baryons but also the kaon field couple only directly with the mesons $m$ ($m$=$\sigma$, $\sigma^\ast$, $\omega$, $\rho$, and $\phi$)\cite{mmt09,mmt14}, and the $K$-$B$ interaction is mediated by exchange of these mesons. 
The $s$-wave $K$-$B$ scalar interaction in the ME scheme is introduced accordingly as follows: 
The effective baryon mass (\ref{eq:effbmci}) appearing in the baryon part of the Lagrangian density ${\cal L}_B$ (\ref{eq:lagbci}) should be modified as 
\begin{equation}
\widetilde M_b^\ast\rightarrow M_b^\ast({\rm ME})=M_b-g_{\sigma b}\sigma-g_{\sigma^\ast b}\sigma^\ast \ , 
\label{eq:meebm}
\end{equation}
since there is no direct $K$-$B$ coupling in the ME scheme. 
Also the kaon mass squared in the kaon part of the Lagrangian density ${\cal L}_K$ (\ref{eq:lagk}) is replaced by the effective kaon mass squared $m_K^{\ast 2}({\rm ME})$, i.~e., 
\begin{equation}
m_K^2\rightarrow 
m_K^{\ast 2}({\rm ME})=m_K^2-2m_K(g_{\sigma K}\sigma+g_{\sigma^\ast K}\sigma^\ast)  \ , 
\label{eq:meekm2}
\end{equation}
which should be compared with the $m_K^{\ast 2}$ in the CI scheme (\ref{eq:ekm2}). [See also Eq.~(\ref{eq:meekm3}) and subsequent discussion.] 
 
For the $s$-wave $K$-$B$ vector interaction, $X_0$ defined by Eq.~(\ref{eq:x0}) appearing in ${\cal L}_K$ (\ref{eq:lagk}) should be replaced as 
\begin{equation}
X_0\rightarrow X_0({\rm ME})=g_{\omega K}\omega_0+g_{\rho K}R_0+g_{\phi K}\phi_0  \ . 
\label{eq:mex0}
\end{equation}
With the above modification for the $K$-$B$ scalar and vector interactions, one obtains the Lagrangian density in the ME scheme as
${\cal L}({\rm ME})$=${\cal L}_B$(ME)+${\cal L}_M$+${\cal L}_K$(ME)+${\cal L}_e$ with
 \begin{equation}
{\cal L}_B({\rm ME})=\sum_{b=p,n,\Lambda, \Sigma^-, \Xi^-}\overline{\psi_b} (i\gamma^\mu D^{(b)}_\mu-M_b^\ast ({\rm ME}) ) \psi_b \ , 
\label{eq:lagbme}
\end{equation}
\begin{equation}
{\cal L}_K({\rm ME}) =\frac{1}{2}(f\mu_K\sin\theta)^2-f^2m_K^{\ast 2}({\rm ME})(1-\cos\theta) 
+ 2 \mu_K X_0({\rm ME}) f^2(1-\cos\theta) \ .
\label{eq:lagkme}
\end{equation}
The energy density in the ME scheme naturally follows from ${\cal L}$(ME): ${\cal E}$(ME) = ${\cal E}_B$(ME)+${\cal E}_M$+${\cal E}_K$(ME)+${\cal E}_e$ with
\begin{equation}
{\cal E}_B({\rm ME})=\sum_{b=p,n,\Lambda,\Sigma^-, \Xi^-} \Bigg\lbrace\frac{2}{(2\pi)^3}\int_{|{\bf p}|\leq p_F(b)} d^3|{\bf p}|\left(|{\bf p}|^2+(M_b^\ast({\rm ME}))^2\right)^{1/2} 
+\rho_b\left(g_{\omega b}\omega_0 +g_{\rho b} \hat I_3^{(b)} R_0 +g_{\phi b}\phi_0\right) \Bigg\rbrace \ ,
\label{eq:ebme}
\end{equation}
\begin{equation}
{\cal E}_K({\rm ME})=\mu_K\rho_{K^-}({\rm ME})-{\cal L}_K({\rm ME})=\frac{1}{2}(\mu_Kf\sin\theta)^2+f^2m_K^{\ast 2}({\rm ME}) (1-\cos\theta)  \ , 
\label{eq:ekme}
\end{equation}
where the number density of kaon condensates, $ \rho_{K^-}({\rm ME})$, is given from Eq.~(\ref{eq:rhok}) by the replacement of $m_K^2\rightarrow m_K^{\ast 2}$(ME) [(\ref{eq:meekm2})] and $X_0\rightarrow X_0$(ME) [(\ref{eq:mex0})] in ${\cal L}_K$ [(\ref{eq:lagk})] : 
\begin{equation}
 \rho_{K^-}({\rm ME})=\mu_K f^2\sin^2\theta+2f^2X_0({\rm ME}) (1-\cos\theta) \ . 
 \label{eq:rhokme}
 \end{equation}
The effective energy density in the ME scheme ${\cal E}_{\rm eff}$(ME) is given as 
\begin{equation}
{\cal E}_{\rm eff}({\rm ME})={\cal E}_{{\rm eff},B}({\rm ME})+{\cal E}_M+{\cal E}_{{\rm eff},K}({\rm ME})+{\cal E}_{{\rm eff},e} \ , 
\label{eq:eff2me}
\end{equation}
where
\begin{subequations}\label{eq:effmodme}
\begin{eqnarray}
{\cal E}_{{\rm eff},B}({\rm ME})&=&{\cal E}_B({\rm ME})-\sum_{b=p,n, \Lambda, \Sigma^-, \Xi^-} \mu_b({\rm ME})\rho_b \ , \label{eq:effmodmeb} \\
{\cal E}_{{\rm eff},K}({\rm ME})&=&{\cal E}_K({\rm ME})-\mu_K\rho_{K^-}({\rm ME})=-{\cal L}_K({\rm ME}) \ . \label{eq:effmodmek}  
\end{eqnarray}
\end{subequations}
In Eq.~(\ref{eq:effmodmeb}), $\mu_{\rm b}$(ME) is the baryon chemical potential in the ME scheme:
\begin{equation}
\mu_b({\rm ME}) =\partial {\cal E}({\rm ME})/\partial\rho_b=\left(p_F(b)^2+M_b^{\ast 2}({\rm ME})\right)^{1/2}+g_{\omega b}\omega_0+g_{\rho b}{\hat I}_3^{(b)} R_0+g_{\phi b}\phi_0 \ .
\label{eq:memub}
\end{equation}
Note that the meson part, ${\cal E}_{{\rm eff},M}$ (= ${\cal E}_M$), and the electron part, ${\cal E}_{{\rm eff},e}$, are the same as those in the CI scheme. 

In accordance with the above modification, the classical kaon field equation reads
\begin{equation}
\mu_K^2\cos\theta+2X_0({\rm ME}) \mu_K-m_K^{\ast 2}({\rm ME})=0  \ , 
\label{eq:mekeom}
\end{equation} 
which should be compared with the one in the CI scheme (\ref{eq:keom2}).
The equations of motion for the meson mean-fields in the ME scheme are modified to
\begin{subequations}\label{eq:meeom}
\begin{eqnarray}
m_\sigma^2\sigma&=&-\frac{dU_\sigma}{d\sigma}+\sum_{b=p,n,\Lambda,\Sigma^-,\Xi^-}g_{\sigma b}
\rho_b^s({\rm ME}) + 2f^2 g_{\sigma K}m_K(1-\cos\theta) \ , \label{eq:meeom1} \\
m_{\sigma^\ast}^2\sigma^\ast &=&\sum_{Y=\Lambda,\Sigma^-,\Xi^-}g_{\sigma^\ast Y}\rho_Y^s({\rm ME}) 
+ 2f^2 g_{\sigma^\ast K}m_K(1-\cos\theta) \ , \label{eq:meeom2}\\
m_\omega^2\omega_0&=&\sum_{b=p,n,\Lambda,\Sigma^-,\Xi^-} g_{\omega b}\rho_b-2f^2g_{\omega K}\mu_K (1-\cos\theta) \ ,  \label{eq:meeom3}\\
m_\rho^2 R_0&=&\sum_{b=p,n,\Lambda,\Sigma^-,\Xi^-} g_{\rho b} \hat I_3^{(b)}\rho_b 
- 2f^2g_{\rho K}\mu_K (1-\cos\theta) \ , \label{eq:meeom4} \\
m_\phi^2\phi_0 &=&\sum_{Y=\Lambda,\Sigma^-,\Xi^-}g_{\phi Y}\rho_Y-2f^2g_{\phi K}\mu_K(1-\cos\theta) \label{eq:meeom5}
\end{eqnarray}
\end{subequations}
with 
\begin{equation}
 \rho_b^s({\rm ME})=\frac{2}{(2\pi)^3} \int_{|{\bf p}|\leq p_F(b)} d^3 {\bf p}\frac{M_b^\ast({\rm ME})}{(|{\bf p}|^2
+M_b^{\ast 2}\left({\rm ME})\right)^{1/2}} 
\label{eq:rhobs}
 \end{equation}
 being the scalar density for baryon $b$ in the ME scheme.
There appear source terms in the equations of motion for meson mean-fields in the presence of kaon condensates ($\theta>0$) (the last terms on the r.~h.~s. of the equations of motion (\ref{eq:meeom1})$-$(\ref{eq:meeom5}) ), originating from the kaon-meson ($K$-$m$) couplings in the ME scheme. 
In Fig.~\ref{fig1}~(b),  the diagrams of the interaction vertices in the ME scheme for the nonlinear $K$-$m$ interaction and the $B$-$m$ interaction are depicted. In the dotted box in the ME scheme, the diagram (ii) stands for the expansion of the  kaon source terms in Eqs.~(\ref{eq:meeom1}) $-$ (\ref{eq:meeom5}) in powers of the kaon field ($\propto \theta$) with the dotted lines. 

In the ME scheme, the mesons ($m$) couple not only to the baryons ($B$) but also to the kaon field $U$ [Fig.~\ref{fig1}~(b)], while they couple only to baryons in the CI scheme [Fig.~\ref{fig1}~(a)]. As a result, there appear two kinds of many-body effects on the $K$-$m$ couplings in the ME scheme: (i) the derivative term of the $\sigma$ self-interaction potential $U_\sigma(\sigma)$ with respect to the $\sigma$ [denoted as (i) in Fig.~\ref{fig1}~(b)] and (ii) the kaon source terms in the equations of motion for the mesons $m$ [denoted as (ii) in Fig.~\ref{fig1}~(b)]. 
Both effects (i) and (ii) lead to difference for the quantities associated with kaon properties between the CI and ME schemes. On the other hand, the nonlinear kaon ($U$)-$B$ interactions in the ME scheme are generated from the components of mesons ($m$) connected to baryons in the $U$-$m$ couplings [the second term on the r.~h.~s. of the $\sigma$ mean field diagram and the first term of the other meson mean fields diagram in Fig.~\ref{fig1}~(b)], which are identified with the $U$-$B$ contact interactions in the CI scheme in Fig.~\ref{fig1}~(a). 
For instance, the expression of the effective kaon mass squared, $m_K^{\ast 2}({\rm ME})$ [ (\ref{eq:meekm2}) ], is rewritten   
by the help of Eqs.~(\ref{eq:meeom1}) and (\ref{eq:meeom2}) as
\begin{eqnarray}
m_K^{\ast 2}({\rm ME})&=&m_K^2-\frac{1}{f^2}\sum_b \rho^s_b \Sigma_{Kb}({\rm ME}) +2g_{\sigma K}\frac{m_K}{m_\sigma^2}\frac{dU_\sigma}{d\sigma} \cr
&-&(2fm_K)^2\Bigg\lbrace \left(\frac{g_{\sigma K}}{m_\sigma}\right)^2+\left(\frac{g_{\sigma^\ast K}}{m_{\sigma^\ast}}\right)^2 \Bigg\rbrace (1-\cos\theta) \ , 
\label{eq:meekm3}
\end{eqnarray}
 where $\Sigma_{Kb}({\rm ME}) $ ($b=p,n,\Lambda,\Sigma^-, \Xi^-$) is defined as
  \begin{equation}
 \Sigma_{Kb}({\rm ME})\equiv 2f^2m_K\left(\frac{g_{\sigma K}g_{\sigma b}}{m_\sigma^2}
 +\frac{g_{\sigma^\ast K}g_{\sigma^\ast b}}{m_{\sigma^\ast}^2} \right)  \ , 
 \label{eq:mekbsigma}
 \end{equation}
which is generated from the $U$-$m$ couplings through exchange of scalar mesons [the second term on the r.~h.~s. of the $\sigma$ meson diagram and the first term on the r.~h.~s. of the $\sigma^\ast$ meson diagram in Fig.~\ref{fig1}~(b)], 
and is identified with the kaon-baryon sigma terms [Eq.~(\ref{eq:ckbsigma})]. 
As compared with the effective kaon mass squared in the CI scheme [(\ref{eq:ekm2})], there are two additional terms in the third and last terms on the r.~h.~s. of Eq.~(\ref{eq:meekm3}) in the ME scheme: The derivative term proportional to the $dU_\sigma/d\sigma$ (=$bM_Ng_{\sigma N}^3\sigma^2+cg_{\sigma N}^4\sigma^3$), coming from the E.O.M. of the $\sigma$ mean field [Eq.~(\ref{eq:meeom1})], represents the many-body effect (i) and has a repulsive effect on the effective kaon mass squared, pushing up the lowest kaon energy $\omega_K$  
as compared with that in the CI scheme. [see Sec.~\ref{subsec:onset}.] 
This many-body effect (i) entails a difference between the CI and ME schemes for quantities associated with kaon properties such as the effective kaon mass even in the normal phase ($\theta$ = 0).
On the other hand, the last term, coming from the nonlinear kaon field-scalar meson couplings through the source terms of Eqs.~(\ref{eq:meeom1}) and (\ref{eq:meeom2}), represents the many-body effect (ii), and it works attractively in the presence of kaon condensates, leading to reduction of $m_K^{\ast 2}$(ME). 

To compare the $X_0$, representing the $s$-wave $K$-$B$ vector interaction, between the CI and ME schemes, the expression of $X_0 ({\rm ME})$ [ (\ref{eq:mex0}) ] is rewritten  
by the help of Eqs.~(\ref{eq:meeom3})$-$(\ref{eq:meeom5}) as 
\begin{equation}
X_0({\rm ME})=\frac{1}{2f^2}\sum_b Q_V^b\rho_b
-2f^2\mu_K
\Bigg\lbrace \left(\frac{g_{\omega K}}{m_\omega}\right)^2+\left(\frac{g_{\rho K}}{m_\rho}\right)^2+\left(\frac{g_{\phi K}}{m_\phi}\right)^2 \Bigg\rbrace (1-\cos\theta) \ , 
\label{eq:mex01}
\end{equation}
where the first term on the r.h.s.of Eq.~(\ref{eq:mex01}) results from the following constraints among the vector meson-$K$ and vector meson-$B$ coupling constants:
\begin{subequations}\label{eq:metw}
\begin{eqnarray} 
&2f^2&\left(\frac{g_{\omega K} g_{\omega N}}{m_\omega^2} +\frac{g_{\rho K} g_{\rho N}}{m_\rho^2}\right)=1 \ , \label{eq:metw1} \\
&2f^2&\left(\frac{g_{\omega K} g_{\omega N}}{m_\omega^2}-\frac{g_{\rho K} g_{\rho N}}{m_\rho^2}\right)=\frac{1}{2} \ , \label{eq:metw2} \\
&2f^2&\left(\frac{g_{\omega K} g_{\omega \Lambda}}{m_\omega^2}+\frac{g_{\phi K} g_{\phi \Lambda}}{m_\phi^2}\right)=0 \ , \label{eq:metw3} \\
&2f^2&\left(\frac{g_{\omega K} g_{\omega \Sigma^-}}{m_\omega^2}-\frac{g_{\rho K} g_{\rho \Sigma^-}}{m_\rho^2}+\frac{g_{\phi K}g_{\phi \Sigma^-}}{m_\phi^2}\right) = -\frac{1}{2} \ , \label{eq:metw4} \\
&2f^2&\left(\frac{g_{\omega K} g_{\omega \Xi^-}}{m_\omega^2} - \frac{g_{\rho K} g_{\rho \Xi^-}}{m_\rho^2}+\frac{g_{\phi K}g_{\phi \Xi^-}}{m_\phi^2}\right)
= -1 \ . \label{eq:metw5} 
\end{eqnarray}
\end{subequations}
These constraints are imposed in order that the terms depending on the number densities of baryons
 in (\ref{eq:mex01}) correspond to the Tomozawa-Weinberg terms prescribed by chiral symmetry.
From Eqs.~(\ref{eq:metw1})$-$(\ref{eq:metw3}), one obtains the kaon-vector meson coupling constants as 
\begin{eqnarray}
g_{\omega K}&=&3 m_\omega^2/(8f^2g_{\omega N})=3.05 \ ,  \cr
g_{\rho K}&=&m_\rho^2/(8f^2g_{\rho N})=2.01 \ ,  \cr
g_{\phi K}&=& 3 \sqrt{2}m_\phi^2/(8f^2g_{\omega N})=7.33 \ , 
\label{eq:vk}
\end{eqnarray}
where the SU(6) relations for the vector meson-baryon coupling constants, $g_{\omega\Lambda}=(2/3)g_{\omega N}$, $g_{\phi \Lambda}=(-\sqrt{2}/3)g_{\omega N}$, have been used [see Sec.~\ref{subsec:mbcc}]. 
The values in (\ref{eq:vk}) should be compared with those obtained with the quark and isospin counting rule, $g_{\omega K}=g_{\omega N}/3$=2.90, $g_{\rho K}=g_{\rho N}$=4.27, and $g_{\phi K}$=$g_{\rho\pi\pi}/\sqrt{2}$=4.27 from SU(6) relation with $g_{\rho\pi\pi}$=6.04\cite{mmt14}. 
 It should be noted that the remaining constraints (\ref{eq:metw4}), (\ref{eq:metw5}) are  shown to be automatically fulfilled by the use of the relations (\ref{eq:vk}) for the vector meson-kaon couplings together with the SU(6) relations for the vector meson - $\Sigma^-$ and $\Xi^-$ coupling constants in (\ref{eq:vy}). 
  
 As compared with the $X_0$ [Eq.~(\ref{eq:x0})] in the CI scheme, the first term in (\ref{eq:mex01}), which is generated from the $U$-$m$ couplings through exchange of vector mesons [the first term on the r.~h.~s. of the vector meson diagram in Fig.~\ref{fig1}~(b)], is identified with the one in (\ref{eq:x0}) corresponding to the Tomozawa-Weinberg term in the CI scheme. In addition, there appears a term on the r.~h.~s. of (\ref{eq:mex01}) coming from the nonlinear kaon field-vector meson couplings through the source terms of Eqs.~(\ref{eq:meeom3})$-$(\ref{eq:meeom5}) as a many-body effect (ii) in the ME scheme. This term works repulsively  in the presence of kaon condensates to weaken the $s$-wave $K$-$B$ vector attraction as far as $\mu_K > 0$. 
 
It is to be noted that the $K$-$K$ interaction is inherent in the nonlinear kaon field in the CI scheme.  The resulting $K$-$K$ scattering length agrees with the current algebra prediction, $-m_K/(16\pi f_K^2)$\cite{w66}. On the other hand, in the ME scheme, there is an additional contribution to the $K$-$K$ scattering length arising from the kaon-scalar meson couplings as positive contribution and the kaon-vector meson couplings as negative contribution, leading to negatively overestimated value of the $K$-$K$ scattering length. Nevertheless, there is still uncertainty about the experimental value of the $K$-$K$ scattering length. Therefore throughout this paper, we leave the problem associated with the $K$-$K$ interaction in case of the ME scheme as it is, until consistent description of the nonlinear kaon field with meson-exchange picture becomes possible in future study. 
 
As for the baryon ($B$)-meson ($m$) couplings in the ME scheme, the first three terms on the r.~h.~s. of the $\sigma$ mean field diagram and the first term on the r.~h.~s. of the other meson mean fields diagram in Fig.~\ref{fig1}~(b) are common to those in Fig.~\ref{fig1}~(a) in the CI scheme. Further the $B$-$m$ couplings through the many-body effect (ii) arising from the kaon source term in the ME scheme [the fourth term on the r.~h.~s. of the $\sigma$ mean field diagram and the second term on the r.~h.~s. of the other meson mean fields diagram in Fig.~\ref{fig1}~(b)] are identified with the nonlinear kaon field ($U$)-$B$ contact interactions in the CI scheme in Fig.~\ref{fig1}~(a). Therefore, 
in contrast to the quantities associated with kaon properties, there is no difference for quantities associated with baryons between the CI and ME schemes. For example, the effective baryon mass $M_b^\ast$(ME) [(\ref{eq:meebm})] is shown to be equal to the $\widetilde M_b^\ast$ [(\ref{eq:effbmci})] in the CI scheme, after being rewritten by the use of Eqs.~(\ref{eq:meeom1}) and (\ref{eq:meeom2}) for $M_b^\ast$(ME) and by the use of Eqs.~(\ref{eq:cieom1}) and (\ref{eq:cieom2}) for $\widetilde M_b^\ast$ together with the ``$Kb$ sigma term'' $\Sigma_{Kb}$ (ME) [(\ref{eq:mekbsigma})]. 

In a similar way,  the baryon chemical potential $\mu_b({\rm ME})$ (for $b=p,n,\Lambda,\Sigma^-,\Xi^-$) in the ME scheme, which is given by Eq.~(\ref{eq:memub}), 
is shown to be equal to the $\mu_b$ [(\ref{eq:mub})] in the CI scheme. In order to reach this result, 
the former is rewritten by the use of Eqs.~(\ref{eq:meeom3}) $-$ (\ref{eq:meeom5}) for the vector mean fields, together with Eq.~(\ref{eq:metw}) to identify the $Q_V^b$ in $\mu_b$ [(\ref{eq:mub})], and the latter is rewritten by the use of Eqs.~(\ref{eq:cieom3}) $-$ (\ref{eq:cieom5}). 

In terms of $\mu_b$(ME), the chemical equilibrium conditions for the weak processes are imposed as $\mu=\mu_K=\mu_e=\mu_n({\rm ME})-\mu_p({\rm ME})$, $\mu_\Lambda({\rm ME})=\mu_n({\rm ME})$, $\mu_{\Sigma^-}({\rm ME})=\mu_{\Xi^-}({\rm ME})=\mu_n({\rm ME})+\mu_e$. 
 The charge neutrality condition is written as
 $\rho_p-\rho_{\Sigma^-}-\rho_{\Xi^-}-\rho_{K^-}({\rm ME})=0$ with the number density of kaon condensates $\rho_{K^-}({\rm ME})$ (\ref{eq:rhokme}). 
 
\subsection{Meson-kaon coupling constants in the ME scheme}
\label{subsec:mk-coupling}

In the ME scheme, 
there remain unknown parameters, the scalar meson-kaon coupling constants, $g_{\sigma K}$ and $g_{\sigma^\ast K}$. 
As seen from Eq.~(\ref{eq:mekbsigma}), the $g_{\sigma K}$ is related to the $Kn$ sigma term as $g_{\sigma K}$=$\Sigma_{Kn}({\rm ME})m_\sigma^2/(2f^2 m_K g_{\sigma N})$ with $g_{\sigma^\ast N}$=0. Here the value of $\Sigma_{Kn}({\rm ME})$ is adjusted to be (300$-$400) MeV, the value adopted in the CI scheme. 
Then $g_{\sigma K}$ is determined to be $g_{\sigma K}$=0.88 (1.17) for $\Sigma_{Kn}({\rm ME})$=300 MeV (400 MeV). 
The scalar $\sigma^\ast$ meson coupling to kaons is chosen to be $g_{\sigma^\ast K}$=2.65/2 from decay of $f_0$(975)\cite{sm96,bb01}. 

The scale of the $s$-wave $K$-$N$ attractive interaction can be measured by the $K^-$ optical potential $U_K$ defined at $\rho_B$=$\rho_0$ in SNM. 
The $U_K$ in the CI scheme and that in the ME scheme are related by 
$\displaystyle U_K({\rm ME})=U_K({\rm CI})+\frac{g_{\sigma K}}{m_\sigma^2}\left(dU_\sigma/d\sigma\right)_{\sigma=\langle\sigma\rangle} $, as shown Eq.~(\ref{eq:ukme}) in Appendix~B. The $U_K({\rm ME})$ is pushed up to a larger value than the $U_K({\rm CI})$ due to the repulsive contribution from the kaon-multi-$\sigma$ meson coupling. 
The parameters relevant to the meson-kaon interaction in the ME scheme of our RMF model are listed in Table~\ref{tab:para3}. Also the kaon-baryon sigma terms $\Sigma_{Kb}$ ($b=p, n, \Lambda, \Sigma^-, \Xi^-$) and $U_K$ adopted for both the CI and ME schemes are listed in Table.~\ref{tab:para4}. 
In this paper, two cases for $\Sigma_{Kn}$=300 MeV and 400 MeV are mainly considered in both CI and ME schemes.
Recently the inclusive missing-mass spectrum of $^{12}$C ($K^-$, $p$) reactions has been measured by the J-PARC E05 experiment\cite{ichikawa2020}. The measured spectrum shape has been reproduced with the real part of the $\bar K$-nucleus potential depth $U_K$=$-$80 MeV and with the imaginary part $W_0$=$-$40 MeV, while it is difficult to reproduce the spectrum with the very deep potential such as $|U_K|\sim $ 200 MeV. As seen in Table~\ref{tab:para4}, our deduced values of the potential depth $U_K$ correponding to each $\Sigma_{Kn}$ for the CI and ME schemes are consistent with these  experimental implications for the $K^-$ optical potential depth.
\begin{table}[h]
\caption{The meson-kaon coupling constants $g_{m K}$ ($m=\sigma, \sigma^\ast, \omega, \rho, \phi$)  used in the ME scheme. The $K^-$ optical potential $U_K$ for symmetric nuclear matter at saturation density is given as $U_K$(ME)$\simeq$$-(g_{\sigma K}\langle\sigma\rangle+g_{\omega K}\langle\omega_0\rangle)$, where $\langle\sigma\rangle$ and $\langle\omega_0\rangle$ are the meson mean-fields at $\rho_p=\rho_n=\rho_0/2$. See the text for details. }
\begin{center}
\begin{tabular}{c c | c c c c c }
\hline
$\Sigma_{Kn}$ & $U_K({\rm ME})$ & $g_{\sigma K}$ & $g_{\sigma^\ast K}$ & $g_{\omega K}$ & $g_{\rho K}$ & $g_{\phi K}$   \\
(MeV) & (MeV) & & & &  & \\ \hline
300  & $-$77 & 0.88 & 2.65/2 & \ $\displaystyle 3 m_\omega^2/(8f^2g_{\omega N})$ \  & $\ \displaystyle m_\rho^2/(8f^2g_{\rho N})$ \ & $\displaystyle 3 \sqrt{2}m_\phi^2/(8f^2g_{\omega N})$  \\
400 & $-$87 &1.17 & 2.65/2 & (=3.05) & (=2.01) & (=7.33) \\
\hline
\end{tabular}
\label{tab:para3}
\end{center}
\end{table}

\begin{table}[h]
\caption{The ``K-baryon sigma term'' $\Sigma_{Kb}$ ($b=p, n, \Lambda, \Sigma^-, \Xi^-$)
and $K^-$ optical potential $U_K$ in symmetric nuclear matter at $\rho_B=\rho_0$ for both the CI and ME coupling schemes. Eq.~(\ref{eq:ckbsigma}) [Eq.~(\ref{eq:mekbsigma})] is used for $\Sigma_{Kb}$ in the CI scheme (in the ME scheme). Eq.~(\ref{eq:ukci}) [Eq.~(\ref{eq:ukme})] is used for $U_K$ in the CI scheme (in the ME scheme). }
\begin{center}
\begin{tabular}{c || c c c c c | c }
\hline
   & $\Sigma_{Kn}$ &  $\Sigma_{Kp}$ & $\Sigma_{K\Lambda}$ & $\Sigma_{K\Sigma^-}$ & $\Sigma_{K\Xi^-}$ & $U_K$ \\
    & (MeV)   & (MeV) & (MeV) & (MeV) & (MeV) & (MeV)  \\ \hline\hline
CI & 300 & 369 & 380 & 300 & 369 & $-$98   \\
    & 400 & 469 & 480 & 400 & 469 & $-$118  \\\hline
ME & 300 & 300 & 266 & 107 & 139 & $-$77   \\
    & 400 & 400 & 326 & 143 & 169 & $-$87  \\\hline 
\end{tabular} 
\label{tab:para4}
\end{center}
\end{table}

\section{Onset density of kaon condensation}
\label{sec:results}

We discuss the onset of kaon condensation realized from hyperon-mixed matter in both CI and ME schemes based on our model interaction and compare the results of the two schemes. 

\subsection{Lowest $K^-$ energy in hyperon-mixed matter in the CI and ME schemes}
\label{subsec:onset}

\begin{figure}[h]
\begin{minipage}[l]{0.60\textwidth}~
\vspace{-2.0cm}~
\begin{center}
\includegraphics[height=.28\textheight]{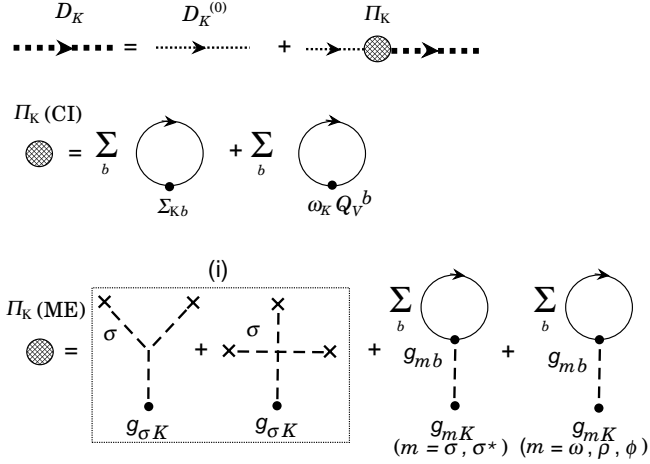}
\end{center}~
\end{minipage}~
\begin{minipage}[r]{0.40\textwidth}~\vspace{2.0cm}~
\caption{The diagrams for the kaon propagator $D_K$ (the bold dotted line) and the kaon self-energy $\Pi_K(\omega_K; \rho_{\rm B})$ in the CI and ME schemes (the shaded circle). The thin dotted line denotes a free kaon propagator $D_K^{(0)}$ [=$1/(\omega_K^2-m_K^2)$]. The diagrams for the self-energy in the CI scheme, $\Pi_K$(CI), correspond to Eq.~(\ref{eq:selfk}) and those for the ME scheme, $\Pi_K$(ME), to Eq.~(\ref{eq:selfkme}). See the text for details.  }
\label{fig2}
\end{minipage}
\end{figure}

We will show, in the later section~\ref{subsec:result-onset}, that hyperon-mixing precedes kaon condensation at lower densities for the allowable range of $\Sigma_{Kn}$=(300$-$400) MeV. Therefore we consider a continuous phase transition from pure hyperon-mixed matter to the ($Y+K$) phase. 
At the onset of kaon condensation, the lowest kaon energy $\omega_K(\rho_B)$ at $\rho_B$ meets the kaon chemical potential $\mu_K$, which is equal to the charge chemical potential $\mu$ due to chemical equilibrium for weak processes, $n\rightleftharpoons p+K^-$, $n\rightleftharpoons p+e^- (+\bar\nu_e)$\cite{mt92}. Therefore the onset density $\rho_B^c$ is given by
\begin{equation}
\omega_K (\rho_B^c)=\mu \ , 
\label{eq:onsetk}
\end{equation}
where $\omega_K(\rho_B)$ is given as a pole of the kaon propagator at $\rho_B$, i.e., $D_K^{-1}(\omega_K; \rho_{B})=0$. The kaon inverse propagator is obtained through expansion of the effective energy density ${\cal E}_{\rm eff}$ with respect to the classical kaon field, 
\begin{equation}
{\cal E}_{\rm eff}(\theta)={\cal E}_{\rm eff}(0)-\frac{f^2}{2}D_K^{-1}(\mu; \rho_{\rm B})\theta^2+O(\theta^4)  \ , 
\label{eq:dkinv1}
\end{equation}
which gives
\begin{equation}
D_K^{-1}(\omega_K; \rho_{B})=\omega_K^2-m_K^2-\Pi_K(\omega_K; \rho_B) 
\label{eq:dkinv}
\end{equation}
with $\Pi_K(\omega_K; \rho_B)$ being the self-energy of kaons. 
In the CI scheme, it is given by
\begin{equation}
\Pi_K(\omega_K; \rho_B)({\rm CI})=m_K^{\ast 2}-m_K^2-2X_0\omega_K=-\frac{1}{f^2}\sum_{b=p,n,\Lambda, \Sigma^-, \Xi^-}\left(\rho_b^s\Sigma_{Kb}+\omega_K\rho_bQ_V^b\right) \ , 
\label{eq:selfk}
\end{equation}
 which is read off from Eq.~(\ref{eq:keom2}) by setting $\mu_K\rightarrow \omega_K$, $\theta\rightarrow 0$ and by the use of Eqs.~(\ref{eq:x0}) and (\ref{eq:ekm2}). 
In the ME scheme, one has
\begin{eqnarray}
\Pi_K(\omega_K; \rho_B)({\rm ME})&=&\left(m_K^{\ast 2}({\rm ME})-m_K^2-2X_0 ({\rm ME})\omega_K\right)_{\theta\rightarrow 0} \cr
&=&-2m_K(g_{\sigma K}\sigma+g_{\sigma^\ast K}\sigma^\ast)-2\omega_K(g_{\omega K}\omega_0+g_{\rho K}R_0+g_{\phi K}\phi_0) \cr
&=& 2g_{\sigma K}\frac{m_K}{m_\sigma^2}\frac{dU_\sigma}{d\sigma}-\frac{1}{f^2}\sum_{b=p,n,\Lambda, \Sigma^-, \Xi^-}\left(\rho_b^s\Sigma_{Kb}({\rm ME})+\omega_K\rho_bQ_V^b\right) \ , 
\label{eq:selfkme}
\end{eqnarray}
where the 2nd line or the 3rd line on the r.~h.~s. is obtained from the 1st line by the use of Eqs.~(\ref{eq:meekm2}), (\ref{eq:mex0}) and Eqs.~(\ref{eq:meekm3}), (\ref{eq:mex01}), respectively.
The diagrams for the kaon propagators $D_K$ in the CI and ME schemes are shown in Fig.~\ref{fig2}. The dotted line stands for kaons and the solid line for the baryon $b$ (=$p$, $n$, $\Lambda$, $\Sigma^-$, $\Xi^-$). In the ME scheme, the dashed line stands for the $\sigma$ meson.   
Comparing Eqs.~(\ref{eq:selfk}) and (\ref{eq:selfkme}), one can see that the kaon-multi-$\sigma$-meson coupling, stemming from the $dU_\sigma/d\sigma$ in the ME scheme denoted as (i), induces additionally repulsive term to the kaon self-energy with nonlinear density-dependence, which is absent in the case of the CI scheme.  

\subsection{Numerical results for the onset density of kaon condensation realized from hyperon-mixed matter}
\label{subsec:result-onset}

\begin{figure}[h]
\begin{minipage}[r]{0.50\textwidth}
\begin{center}
\includegraphics[height=.34\textheight]{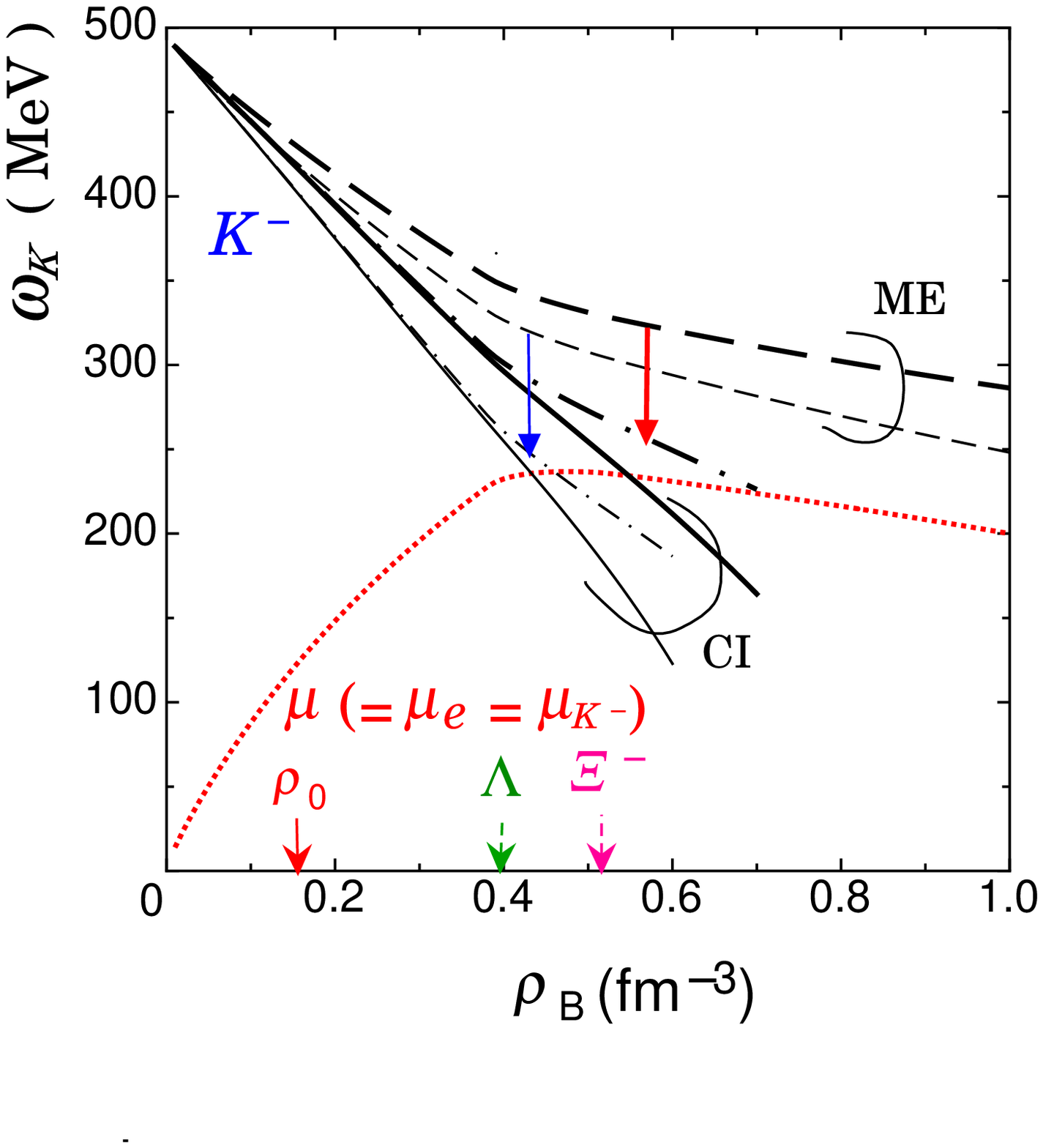}
\end{center}~
\caption{The lowest kaon energy $\omega_K$ as a function of baryon number density $\rho_{\rm B}$ for the CI case (solid lines) and the ME case (dashed lines). The bold lines (thin lines) are for the $\Sigma_{Kn}$=300 MeV ($\Sigma_{Kn}$=400 MeV). The dependence of the charge chemical potential $\mu$ (=$\mu_e=\mu_{K^-}$) on $\rho_{\rm B}$ is also shown by the dotted line. See the text for details. \break}
\label{fig3}
\end{minipage}~
\begin{minipage}[l]{0.50\textwidth}~\vspace{-2.5cm}~
\begin{center}
\includegraphics[height=.30\textheight]{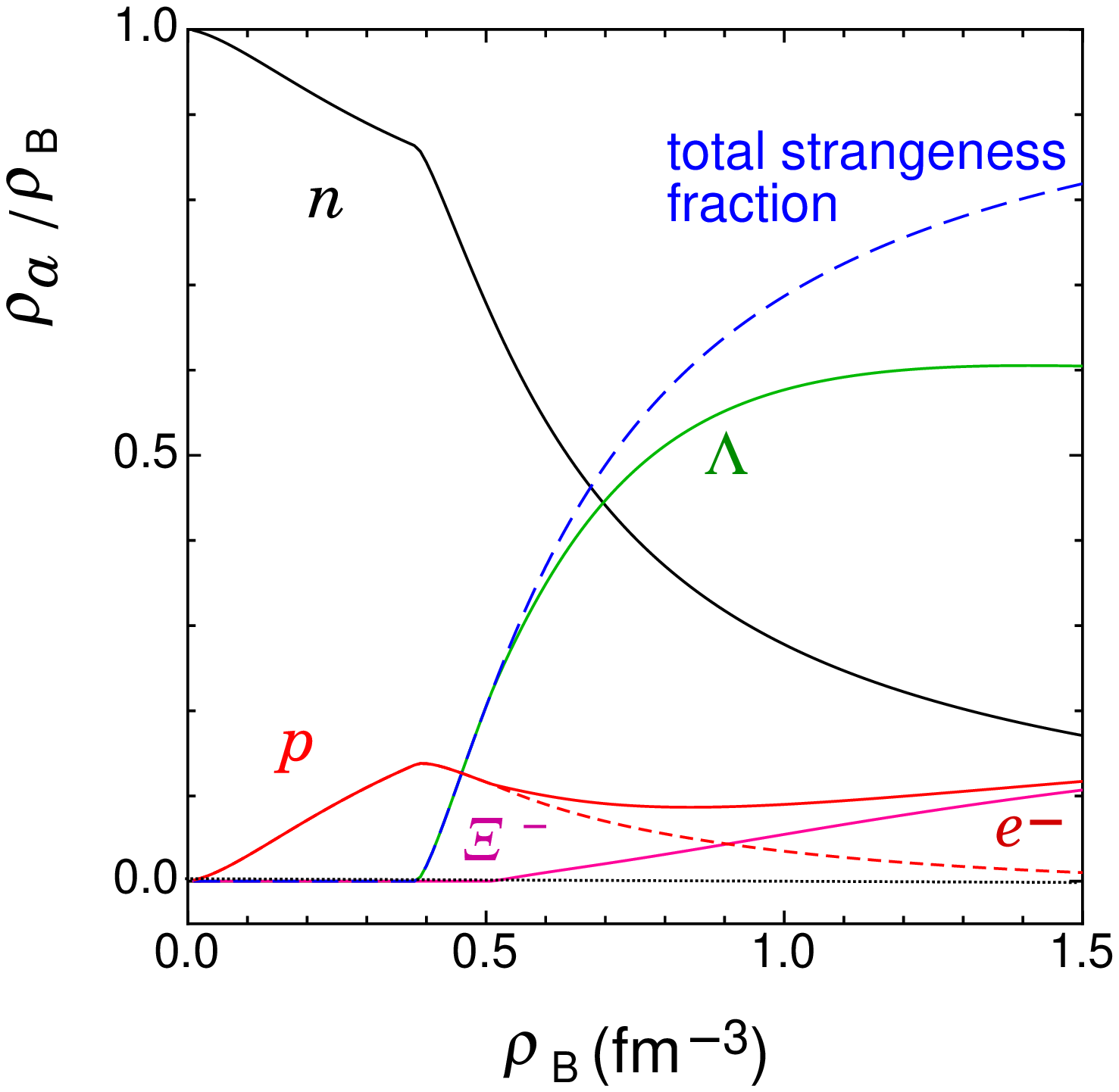}
\end{center}~\vspace{1.0cm}~
\caption{The particle fractions in pure hyperon-mixed matter with $\theta = 0$ as a function of the baryon number density $\rho_{\rm B}$. The total strangeness fraction (the dashed line)  is given by $(\rho_\Lambda+2\rho_{\Xi^-})/\rho_{\rm B}$. }
\label{fig4}
\end{minipage}
\end{figure}

In Fig.~\ref{fig3}, the lowest kaon energy $\omega_K$ is shown as a function of baryon number density $\rho_{\rm B}$  for the CI case (solid lines) and the ME case (dashed lines). The bold lines (thin lines) are for the $\Sigma_{Kn}$ = 300 MeV ($\Sigma_{Kn}$ = 400 MeV). The dependence of the charge chemical potential $\mu$ (=$\mu_e=\mu_{K^-}$) on $\rho_{\rm B}$ is also shown by the dotted line. 

In the CI scheme, the energy $\omega_K$ decreases almost linearly with $\rho_{\rm B}$ from the mass in the  vacuum to the value of $O$(200 MeV) at $\rho_{\rm B}$ = 0.6 fm$^{-3}$ (0.5 fm$^{-3}$) for $\Sigma_{Kn}$ = 300 MeV (400 MeV). The onset density of kaon condensates is read as $\rho_{\rm B}^c(K^-)$ = 0.548 fm$^{-3}$ (= 3.58 $\rho_0$) for $\Sigma_{Kn}$ = 300 MeV and $\rho_{\rm B}^c(K^-)$ = 0.433 fm$^{-3}$ (= 2.83 $\rho_0$) for $\Sigma_{Kn}$ = 400 MeV. 
For reference, we show particle fractions in pure hyperon-mixed matter ($\theta$ set to zero) as a function of $\rho_{\rm B}$ in Fig.~\ref{fig4}. 
In pure hyperon-mixed matter, $\Lambda$ hyperons start to be mixed in nucleon matter at $\rho_{\rm B}$ = $\rho_{\rm B}^c(\Lambda)$ = 0.384 fm$^{-3}$ (= 2.51$\rho_0$), and subsequently $\Xi^-$ hyperons appear at a higher density $\rho_{\rm B}$ =$\rho_{\rm B}^c(\Xi^-)$ = 0.508 fm$^{-3}$ (= 3.32 $\rho_0$). 
 It is to be noted that $\Sigma^-$ hyperons are not mixed over the relevant densities, since the potential $V_{\Sigma^-}^N(\rho_0)$ is taken to be strongly repulsive and $\Xi^-$ hyperons are mixed in place of $\Sigma^-$ hyperons. 

As the number density of $\Lambda$ hyperons increases with $\rho_{\rm B}$, the number densities of proton and electron decrease through the weak process, $p+e^-\rightarrow \Lambda +\nu_e$, keeping with $\rho_p=\rho_e$,  so that the charge chemical potential $\mu$ [=$\mu_e$=$(3\pi^2\rho_e)^{1/3}$] decreases as $\rho_{\rm B}$ increases after the onset density of the $\Lambda$ hyperons. In a way similar to the $\Lambda$-mixing case, mixing of $\Xi^-$ hyperons suppresses $\mu$ as $\rho_{\rm B}$ increases, through $\Lambda+e^- \rightarrow \Xi^- + \nu_e$, as seen in Fig.~\ref{fig4}. As a result, the onset density $\rho_{\rm B}^c(K^-)$ is pushed up to a higher density in hyperon-mixed matter as compared with that realized from  neutron-star matter without hyperon-mixing\cite{ekp95,sm96}. For $\Sigma_{Kn}$ = 300 MeV, kaon condensation occurs at a higher density than the onset density of the $\Xi^-$-mixing. For $\Sigma_{Kn}$ = 400 MeV,  the lowest kaon energy $\omega_K$ is smaller than that for the case of $\Sigma_{Kn}$ = 300 MeV at a given density due to stronger $K$-$B$ scalar attraction, and kaon condensation occurs at a density just after the onset density of the $\Lambda$-mixing and before the density at which the $\Xi^-$-mixing starts. 
 
In the ME scheme, the  lowest kaon energy $\omega_K$ (the dahed line) lies higher than that in the CI scheme (the solid line). The main difference of $\omega_K$ between the CI and ME schemes
stems from the kaon-multi-$\sigma$-meson coupling term from (i) [the first term in the third line on the r.~h.~s. of Eq.~(\ref{eq:selfkme})], which works to increase the energy $\omega_K$ as compared with the CI scheme case. 
For comparison, the lowest kaon energy obtained after this term is subtracted, denoted as $\omega'_K$ below, is shown as a function of $\rho_{\rm B}$ by the bold (thin) dashed-dotted lines for $\Sigma_{Kn}$ = 300 MeV (400 MeV) in Fig.~\ref{fig3}. The energy $\omega'_K$ is almost equal to the energy $\omega_K$ in the CI scheme for $\rho_{\rm B}\lesssim \rho_{\rm B}^c(\Lambda)$. 
For $\rho_{\rm B}\gtrsim \rho_{\rm B}^c(\Lambda)$, the mixing of $\Lambda$ leads to reduction of the $K$-$\Lambda$ scalar attraction in the ME scheme as compared with that in the CI scheme due to the relation, $\Sigma_{K\Lambda}$(ME) $<$ $\Sigma_{K\Lambda}$ (CI) [see Table~\ref{tab:para4}]. Thus the decrease in $\omega'_K$ with $\rho_{\rm B}$ becomes moderate in the presence of $\Lambda$ hyperons, so does the energy $\omega_K$ in the ME scheme. As a result, the energy $\omega_K$ in the ME scheme 
 does not cross the charge chemical potential $\mu$ over the relevant densities for the standard values of $\Sigma_{Kn}$ = (300$-$400) MeV, so that kaon condensation does not occur over the relevant densities. In our previous works based on the ME scheme, kaon condensation in the hyperon-mixed matter appears at $\rho_{\rm B} \sim 3.3 \rho_0$ for $U_K$= $-$ 120 MeV\cite{mmt14}, which, however, corresponds to a large $Kn$-sigma term, i.e., $\Sigma_{Kn}$(ME)  = 754 MeV, estimated from Eq.~(\ref{eq:mekbsigma}). In this case the repulsive effect from (i) is compensated by huge attraction given by the $s$-wave $K$-$B$ scalar interaction. In most of the other works in the ME scheme with the NLSI  scalar potential $U_\sigma$, kaon condensation in hyperon-mixed matter appears only for large kaon optical potential depth, $|U_K|\gtrsim$ 120 MeV. 
 
 The many-body effect (i) appearing solely in the ME scheme
 depends upon the specific form of the $U_\sigma(\sigma)$ which is phenomenologically introduced in order to reproduce the empirical value of the incompressibility of nuclear matter at saturation density ($K$=240 MeV). 

In another example of the many-body effects (i), the nonlinear potential of the $\omega$ meson, $c(\omega^\mu\omega_\mu)^2/4$, which is introduced to reproduce properties of stable and unstable nuclei systematically 
in the RMF models\cite{st94}, would modify the $K$-$B$ vector interaction, leading to 
an extra repulsive term, $-c g_{\omega K}\omega_0^3/m_\omega^2$, in the expression of $X_0$(ME) in Eq.~(\ref{eq:mex0}). Also inclusion of the $\omega$-$\rho$ meson coupling term $\lambda(\omega^\mu\omega_\mu)(\vec R^\mu\cdot\vec R_\mu)$\cite{Horowitz2001,fattoyev2020}, which affects the symmetry energy $S(\rho_0)$ and its slope $L$ at $\rho_{\rm B}=\rho_0$, would lead to both the repulsive and attractive contributions to the $X_0$, $-2\lambda g_{\omega K} R_0^2\omega_0 /m_\omega^2$ and $-2\lambda g_{\rho K} R_0\omega_0^2/m_\rho^2$ ($R_0 < 0$) , respectively. 
Because of arbitrariness of the NLSI terms as shown above, corresponding many-body terms entering into the kaon self-energy cannot be fixed uniquely. 

\section{Role of the nonlinear self-interacting term as many-baryon forces}
\label{sec:NLSI}

We have shown that the difference of kaon dynamics in dense matter between the CI and ME schemes is caused by many-body effects (i) derived from the NLSI term, $U_\sigma$ in this paper. The $U_\sigma$ itself is introduced commonly in both schemes in order to reproduce the empirical incompressibility in SNM. Here we reconsider a role of the NLSI term as many-baryon forces associated with the saturation mechanisms in SNM. We also examine for the NLSI term a possible origin of the many-baryon repulsion 
in the context of stiffening the EOS for the ($Y$+$K$) phase at high densities. 

\subsection{Effects of the NLSI term on saturation mechanisms in the SNM}
\label{subsec:NLSI}

\begin{figure}[!]
\begin{minipage}[l]{0.50\textwidth}~
\begin{center}
\includegraphics[height=.28\textheight]{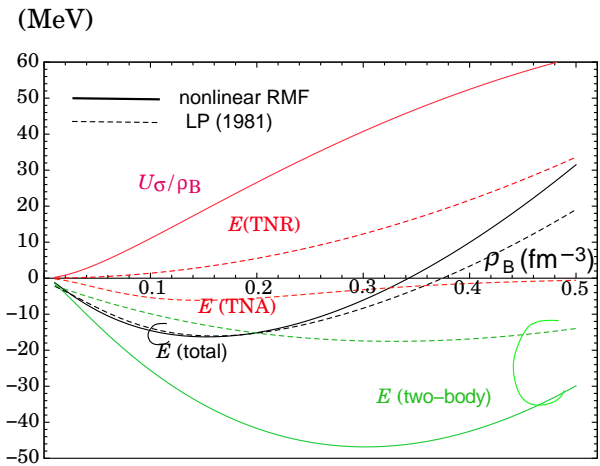}
\end{center}~
\end{minipage}~
\begin{minipage}[r]{0.50\textwidth}
\caption{The total energy per nucleon, $E$~(total), and energy contributions from the NLSI term, $E$~(NLSI) (=$U_\sigma/\rho_{\rm B}$), and the sum of kinetic and two-body interaction energies, $E$~(two-body), in SNM are shown as functions of $\rho_{\rm B}$ by the solid lines. For comparison, the $E$~(total), $E$~(two-body), and the energy contributions from the three-nucleon repulsion, $E$~(TNR), and three-nucleon attraction, $E$~(TNA), are read from Ref.~\cite{lp1981}[LP~(1981)] by the dotted lines. See the text for details.\break }
\label{fig5}
\end{minipage}
\end{figure}

The total energy per nucleon, $E$~(total), in SNM is separated as 
$E$~(total)=[${\cal E}$~(two-body)+${\cal E}$~(NLSI)]/$\rho_{\rm B}$ with
\begin{eqnarray}
{\cal E}~({\rm two-}{\rm body})&=&\sum_{N=p,n} \frac{2}{(2\pi)^3}\int_{|{\bf p}|\leq p_F} d^3|{\bf p}|(|{\bf p}|^2+M_N^{\ast 2})^{1/2}\cr
&+&\frac{1}{2}m_\sigma^2\sigma^2+\frac{1}{2}m_{\sigma^\ast}^2\sigma^{\ast 2}
+\frac{1}{2}m_\omega^2\omega_0^2+\frac{1}{2}m_\rho^2 R_0^2+\frac{1}{2}m_\phi^2\phi_0^2 \ , \cr
\cr
{\cal E}~({\rm NLSI})&=& U_\sigma \ , 
\label{eq:edSNM}
\end{eqnarray}
where $p_F$ is the Fermi momentum of the nucleon in SNM, $ p_F=\left(3\pi^2 \rho_{\rm B}/2\right)^{1/3}$, and $M_N^{\ast }$ (=$M_N-g_{\sigma N}\sigma$) the effective nucleon mass. 
The $B$-$B$ two-body interactions, stemming from $\sigma$, $\omega$, and $\rho$ mesons exchange in the RMF, are included in $E$~(two-body). In Fig.~\ref{fig5}, $E$~(total), $E$~(two-body) (=${\cal E}$~(two-body)/$\rho_{\rm B}$ ), and $E$~(NLSI) (=$U_\sigma/\rho_{\rm B}$) are shown as functions of $\rho_{\rm B}$ by the solid lines. Note that these curves are common to both CI and ME schemes. For comparison, the $E$~(total), $E$~(two-body), and the energy contributions from the three-nucleon repulsion, $E$~(TNR), and three-nucleon attraction, $E$~(TNA), which are read from the result by Lagaris and Pandharipande in Ref.~\cite{lp1981}[LP~(1981)], are shown by the dotted lines as a reference for the standard nuclear matter calculation with the variational method. In the case of LP~(1981), the three-nucleon forces, both TNR and TNA [$E$~(TNR)=3.5 MeV and $E$~(TNA)=$-$6.1 MeV] play an important role to shift the location of the saturation point due to the $E$~(two-body) to the empirical one, where $\rho_{\rm B}$ = 0.16~fm$^{-3}$ and the binding energy =16.3 MeV. On the other hand, in the present model, the NLSI term brings about large repulsion (20 MeV) at $\rho_0$ as compared with the TNR in LP~(1981). The $E$~(NLSI) monotonically increases with $\rho_{\rm B}$ for $0 < \rho_{\rm B}\lesssim$ 0.6~fm$^{-3}$. Further a large cancellation between the $E$~(NLSI) and $E$~(two-body) maintains saturation of SNM. Therefore the NLSI term shows quite different aspects quantitatively with respect to saturation mechanisms from those with the standard nuclear matter calculation. 

\subsection{Contribution of the NLSI term to the EOS for the ($Y+K$) phase in the CI scheme}
\label{subsec:eos}

Following the results on the onset of kaon condensation in Sec.~\ref{subsec:result-onset}, we concentrate on the CI scheme for discussion of the EOS including the ($Y+K$) phase and contribution of the NLSI term to the EOS as many-baryon forces.  

\begin{figure}[!]
\begin{minipage}[r]{0.50\textwidth}
\begin{center}~
\vspace{1.5cm}~
\includegraphics[height=.31\textheight]{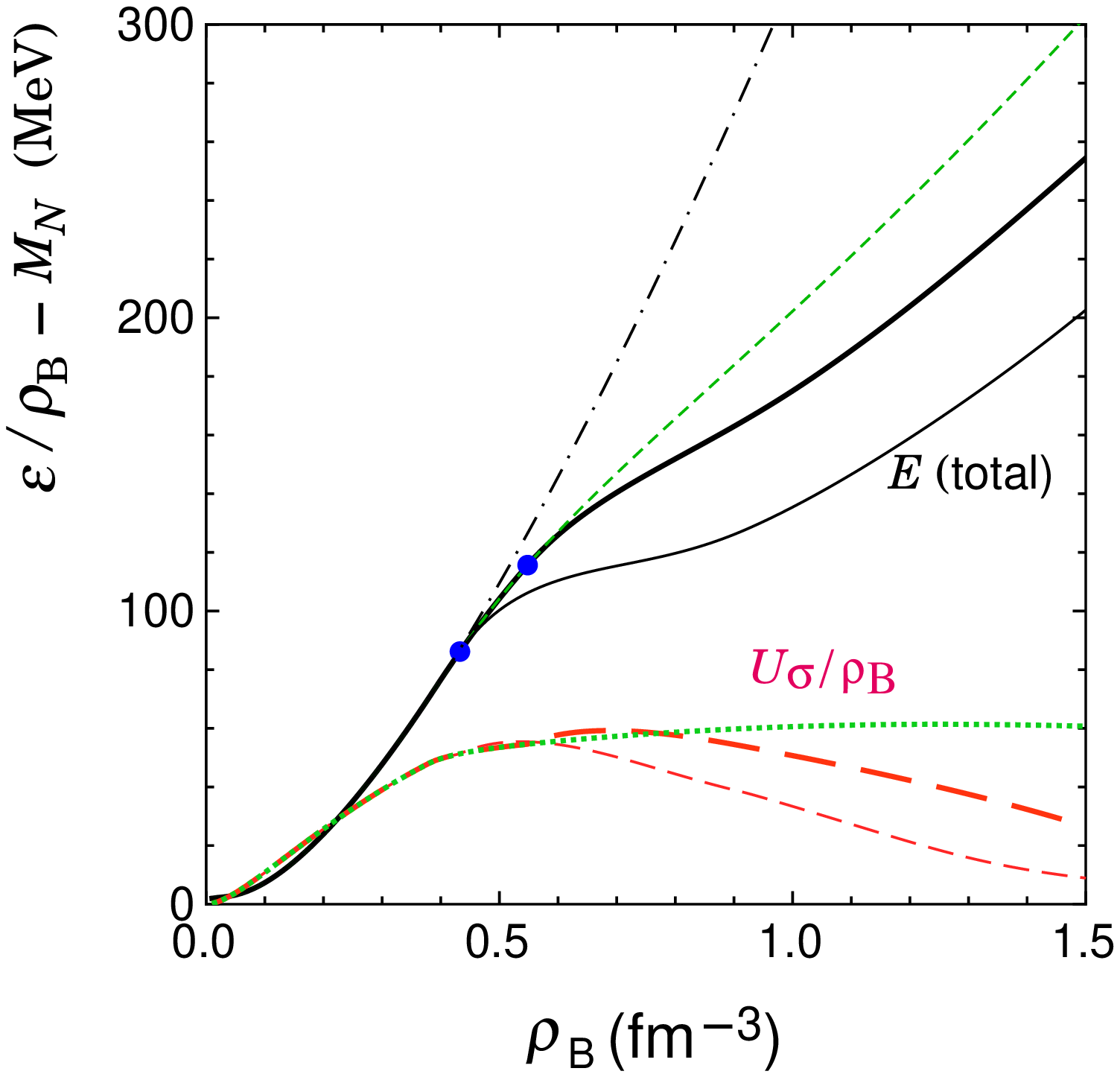}
\end{center}
\caption{The energy per baryon, ${\cal E}/\rho_{\rm B}$ (=$E$~(total)), for the ($Y+K$) phase as a function of $\rho_{\rm B}$ in the CI scheme for $\Sigma_{Kn}$=300 MeV and 400 MeV by the bold and thin solid lines, respectively. (The nucleon rest mass is subtracted.) 
Contribution from the NLSI term, $U_\sigma/\rho_{\rm B}$, is also shown for $\Sigma_{Kn}$=300 MeV and 400 MeV by the bold and thin red dashed lines, respectively. 
For comparison, the one for pure hyperon-mixed matter (set to be $\theta = 0$), which is equal to the energy per baryon in the ME scheme, and the one for pure nucleon matter (set to be $\theta$=0 and the hyperon-mixing ratio $\rho_Y/\rho_{\rm B}$=0) are shown by the green dotted line and the dash-dotted line, respectively. The energy contribution from the NLSI term in the case of the pure hyperon-mixed matter (i.~e., in the case of the ME scheme) is also shown in the lower green dotted line. \break }
\label{fig6}
\end{minipage}~
\begin{minipage}[l]{0.50\textwidth}~
\vspace{-7.8cm}~
\begin{center}
\includegraphics[height=.32\textheight]{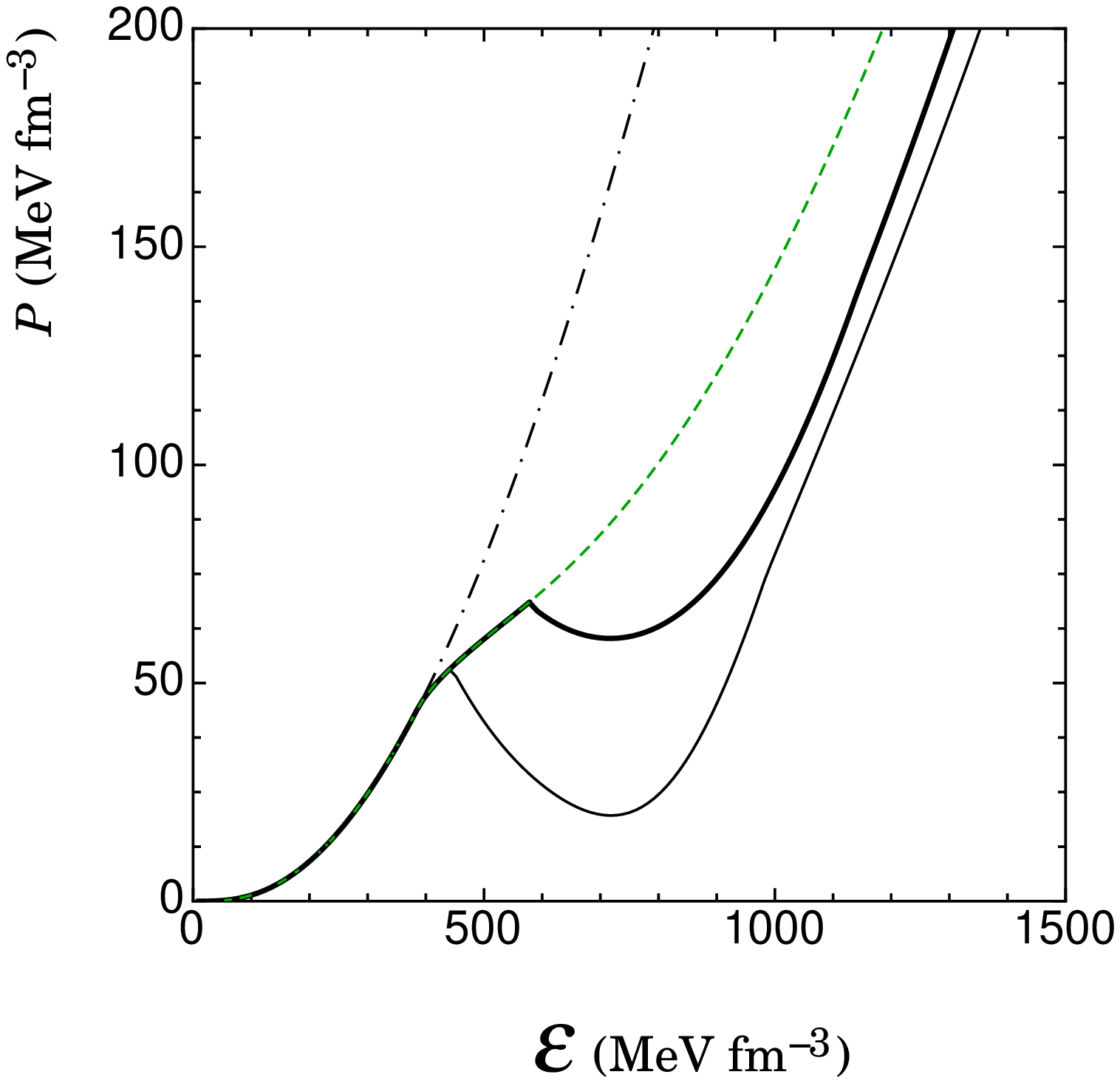}
\end{center}
\vspace{1.0cm}~
\caption{The pressure $P$ (=$-{\cal E}_{\rm eff}$ [(\ref{eq:eff2})]) in the ($Y+K$) phase as a function of the energy density ${\cal E}$ in the CI scheme, for $\Sigma_{Kn}$=300 MeV and 400 MeV by the bold and thin solid lines, respectively. The other curves denote the same cases as those for the energies per baryon in Fig.~\ref{fig6}.}
\label{fig7}
\end{minipage}
\end{figure}

In Fig.~\ref{fig6}, the energy per baryon, ${\cal E}/\rho_{\rm B}$, in the ($Y+K$) phase with the nucleon rest mass being subtracted is shown as a function of $\rho_{\rm B}$, which is obtained in the CI scheme for $\Sigma_{Kn}$=300 MeV and 400 MeV by the bold and thin solid lines, respectively. 
Contribution from the NLSI term, $U_\sigma/\rho_{\rm B}$, is also shown for $\Sigma_{Kn}$=300 MeV and 400 MeV by the bold and thin red dashed lines, respectively.
In Fig.~\ref{fig7}, the pressure $P$ (=$-{\cal E}_{\rm eff}$ [(\ref{eq:eff2})]) for the ($Y+K$) phase is shown as a function of the energy density ${\cal E}$ in the CI scheme, for $\Sigma_{Kn}$=300 MeV and 400 MeV by the bold and thin solid lines, respectively. 
For comparison, in both Figs.~\ref{fig6} and \ref{fig7}, the cases for pure hyperon-mixed matter without kaon condensation (set to be $\theta = 0$), which are equal to those obtained in the ME scheme, and for pure nucleon matter (set to be $\theta$=0 and the hyperon-mixing ratio $\rho_Y/\rho_{\rm B}$=0) are also shown by the green dotted line and the dash-dotted line, respectively. 
One can see from both figures that once kaon condensation occurs in hyperon-mixed matter, it leads to significant softening of the EOS from the one for the pure nucleon matter, since the attractive effect of the $s$-wave $K$-$B$ interaction is added as well as the effect of avoiding the $N$-$N$ repulsion by making the relative number densities of nucleons lower through mixing of hyperons\cite{nyt02}. 
As seen in Fig.~\ref{fig7}, there appears an unstable region, $dP /d\epsilon < 0$, as a result of large softening of the EOS in the presence of kaon condensates. 
In such a density region, the phase equilibrium between the hyperon-mixed phase and the ($Y+K$) phase should be taken into account under the Gibbs condition, which may lead to an inhomogeneous mixed phase consisting of these phases\cite{mtvt06}.  

It is to be noted that the EOS for the pure nucleon matter (the dash-dotted line) is slightly stiffer than the one according to the A18+$\delta v$+UIX$^\ast$ model by Akmal, Pandharipande, and Ravenhall\cite{apr1998}. 

In Fig.~\ref{fig6}, the energy contribution from the NLSI term, $E$~(NLSI) (=$U_\sigma/\rho_{\rm B}$), is also shown for $\Sigma_{Kn}$=300 MeV and 400 MeV by the bold and thin red dashed lines, respectively. The $E$~(NLSI) in the case of the pure hyperon-mixed matter (i.~e., in the case of the ME scheme) is also shown in the lower green dotted line. 
The $E$~(NLSI), which should be relevant to the properties of the SNM around $\rho_0$, increases with density up to $\rho_{\rm B} \sim$~0.6~fm$^{-3}$. In the case of the ($Y$+$K$) phase, however, the energy contribution from the NLSI term 
turns to decreasing beyond the density $\rho_{\rm B}$ = (0.6$-$0.7)~fm$^{-3}$, and the total energy per baryon is dominated by the two-body $B$-$B$ interaction. In the case of pure hyperon-mixed matter, 
the $E$~(NLSI) becomes saturated around $\rho_{\rm B}$ $\sim$1~fm$^{-3}$ and has a minor contribution to the total energy per baryon. Thus, in the context of stiffening the EOS at high densities, the NLSI term 
is not relevant to an origin of the extra repulsive energy at high densities leading to the solution to the hyperon puzzle. 
Since the validity of applying such NLSI terms to high densities cannot be assured beyond the phenomenological introduction around $\rho_0$ for the properties of SNM, 
 the many-body effects appearing in the kaon self-energy in the ME scheme should not be considered as physically solid. 

\section{Properties of the ($Y$+$K$) phase in the CI scheme}
\label{sec:fraction}

Here we summarize the main properties of the ($Y$+$K$) phase, density-dependence of particle fractions and hyperon potentials in the CI scheme with the (MRMF+NLSI) model. 
The self-suppression effect of the $s$-wave $K$-$B$ attraction unique to the case of kaon condensation in the RMF framework~\cite{fmmt96} is also discussed.
These features may be common features in the presence of kaon condensates in the relativistic models. 

\subsection{Particle fractions}
\label{subsec:fractions}

\begin{figure}[h]
\begin{minipage}[r]{0.50\textwidth}
\begin{center}
~\vspace{4.0cm}~
\includegraphics[height=.32\textheight]{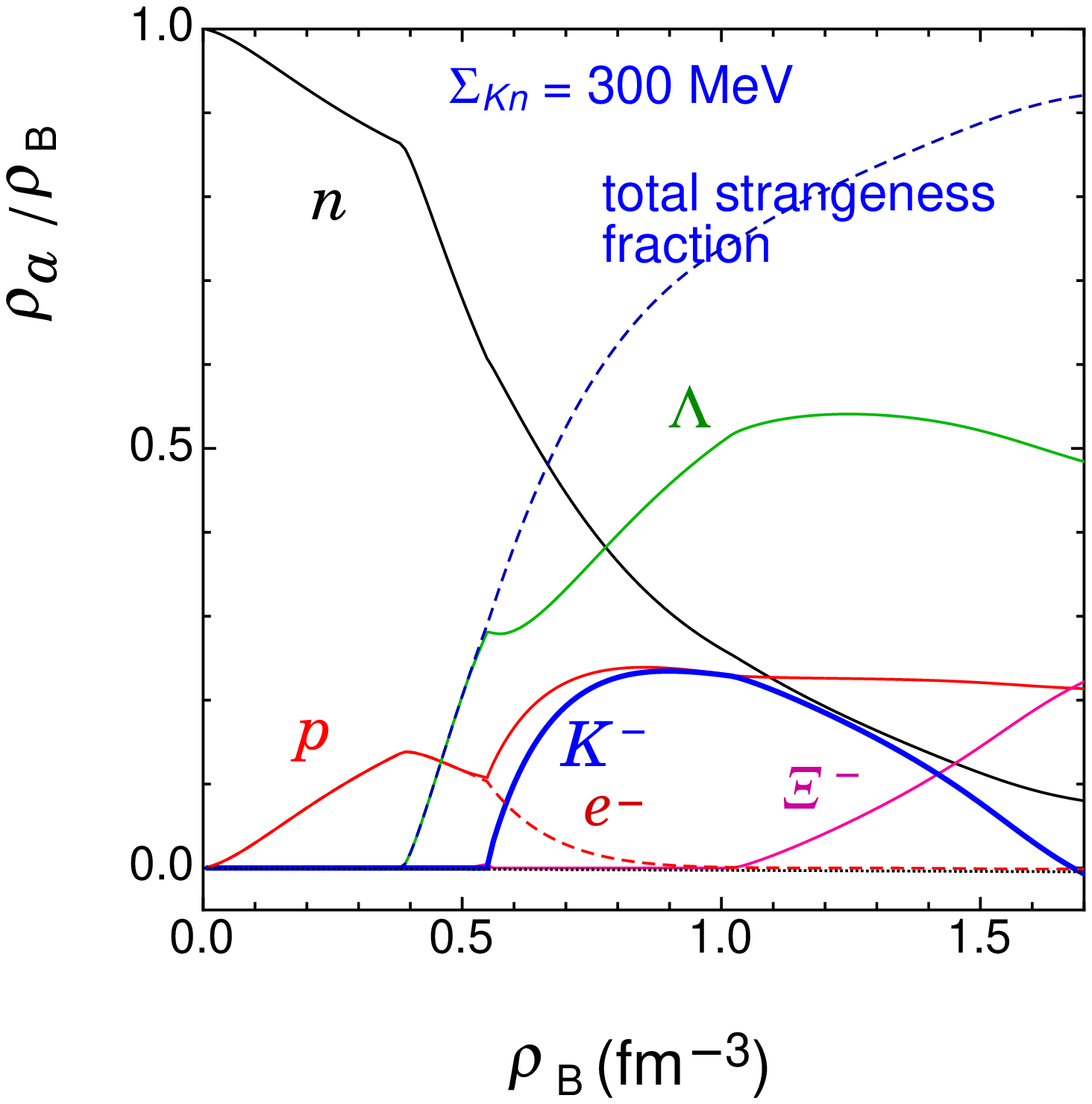}
\end{center}
~\vspace{-2.0cm}~
\caption{The particle fractions in the ($Y+K$) phase as functions of baryon number density $\rho_B$ for $\Sigma_{Kn} = 300$ MeV. The total strangeness fraction is defined by $(\rho_{K^-}+\rho_\Lambda+2\rho_{\Xi^-})/\rho_{\rm B}$. \break}
\label{fig8}
\end{minipage}~
\begin{minipage}[l]{0.50\textwidth}
~\vspace{-3.0cm}~
\begin{center}
\includegraphics[height=.32\textheight]{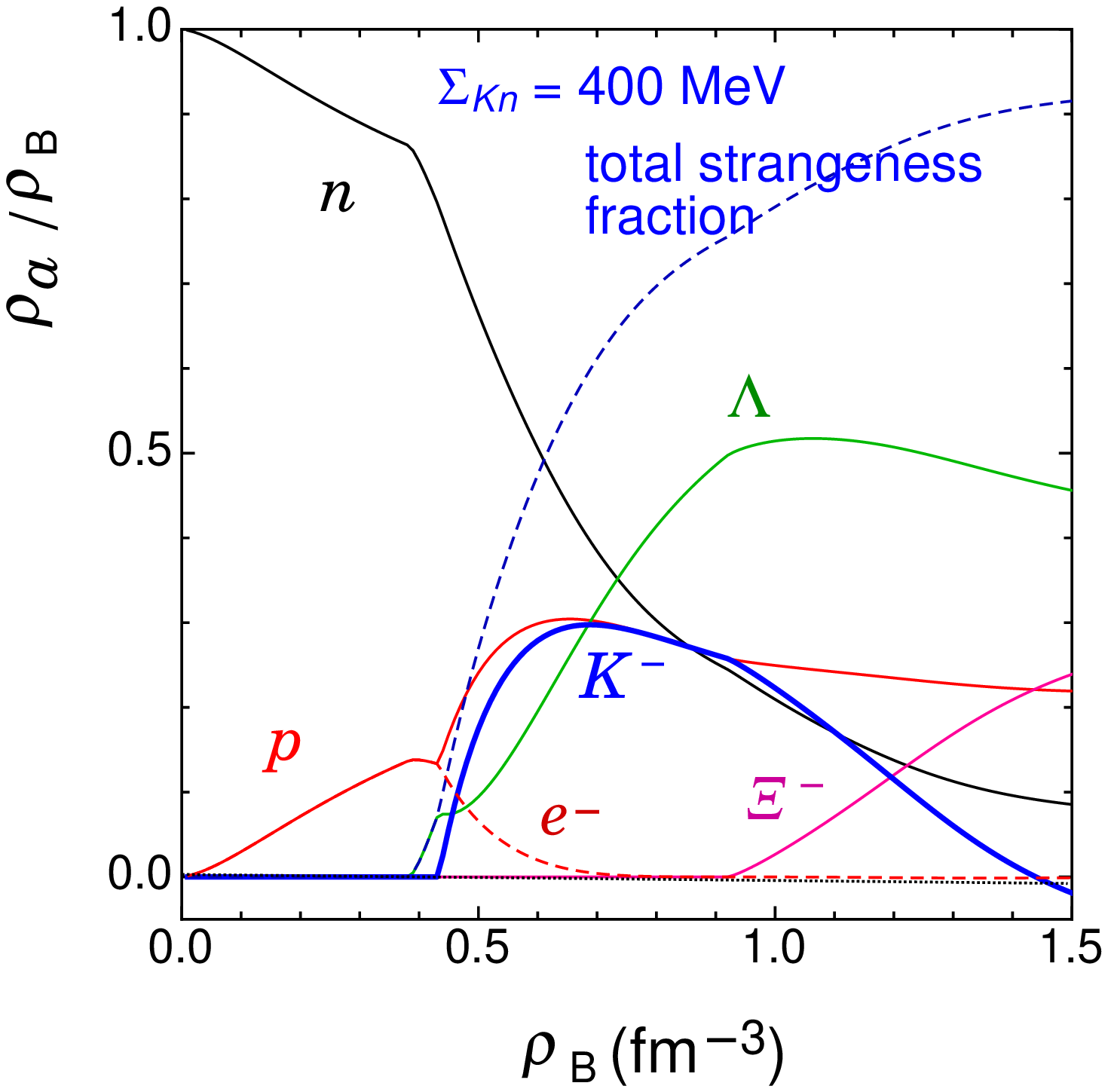}
\end{center}
~\vspace{1.5cm}~
\caption{The same as Fig.~\ref{fig8}, but for $\Sigma_{Kn} = 400$ MeV. }
\label{fig9}
\end{minipage}
\end{figure}

The particle fractions $\rho_a/\rho_{\rm B}$  ($a$ = $p$, $n$, $\Lambda$, $\Xi^-$, $K^-$, $e^-$) in the ($Y+K$) phase are shown as functions of $\rho_{\rm B}$ for $\Sigma_{Kn} = 300$ MeV and 400 MeV in Figs.~\ref{fig8} and \ref{fig9}, respectively. For reference, those for the pure hyperon-mixed matter (i.~e., in the case of the ME scheme) are shown in Fig.~\ref{fig4} in Sec.~\ref{subsec:result-onset}. 
One can see some common behaviors about the density dependence of particle fractions for both cases of $\Sigma_{Kn} = 300$ MeV and 400 MeV as follows: 
(I) The fraction of $\Lambda$ hyperons monotonically increases with density for both cases until kaon condensates appear at $\rho_{\rm B}^c(K^-)$. Just after the onset of kaon condensation, the growth rate of the $\Lambda$ fraction with density is slightly suppressed, but it soon recovers monotonic increase with density.  
On the other hand, $\Xi^-$ hyperons appear just before the onset of kaon condensation [$\rho_{\rm B}^c(\Xi^-)$ (= 0.508 fm$^{-3}$) $ < $ $\rho_{\rm B}^c(K^-)$ (= 0.548 fm$^{-3}$) ] for $\Sigma_{Kn}$ = 300 MeV, but the fraction is tiny, $\rho_{\Xi^-}/\rho_{\rm B}\lesssim 5\times 10^{-3}$, around the density $\rho_{\rm B} = \rho_{\rm B}^c(\Xi^-)$. For $\Sigma_{Kn}$ = 400 MeV, onset of kaon condensation precedes the $\Xi^-$-mixing. In both cases of $\Sigma_{Kn}$, the $\Xi^-$-mixing ratio vanishes once kaon condensates appear until the $\Xi^-$ appears again at higher densities $\rho_{\rm B}\geq \rho_{\rm B}^c(\Xi^-~{\rm in}~K^-)$, where $\rho_{\rm B}^c(\Xi^-~{\rm in}~K^-)$ is the onset density of the $\Xi^-$ hyperons in the presence of kaon condensates 
[$\rho_{\rm B}^c(\Xi^-~{\rm in}~K^-)$ = 1.03 fm$^{-3}$ (= 6.73 $\rho_0$) for $\Sigma_{Kn}$ = 300 MeV and $\rho_{\rm B}^c(\Xi^-~{\rm in}~K^-)$ = 0.920 fm$^{-3}$ (= 6.01 $\rho_0$) for $\Sigma_{Kn}$ = 400 MeV]. 

Once kaon condensates appear, they develop together with $\Lambda$ hyperons as $\rho_{\rm B}$ increases. However, both kaon condensates and the $\Lambda$-mixing ratio gradually decrease as the $\Xi^-$- mixing starts in the fully-developed ($Y+K$) phase at $\rho_{\rm B}^c(\Xi^-~{\rm in}~K^-)$ and further as the fraction of $\Xi^-$ hyperons increases  with density. Here one can see a competition between $\Xi^-$ hyperons and kaon condensates. This competitive effect results model-independently from the fact that the number density of kaon condensates, $\rho_{K^-}$ [(\ref{eq:rhokci})], decreases as the number density of $\Xi^-$ hyperons increases due to the negative factor $Q_V^b$ (=$-1$ for $\Xi^-$ hyperons) in the function $X_0$ [(\ref{eq:x0})], which is uniquely assigned as a consequence of chiral symmetry. 

(II) The electron fraction is suppressed after the appearance of $\Lambda$ hyperons or kaon condensates. In particular, the negative charge carried by electrons is taken over by that of kaon condensates avoiding a cost of degenerate energy of electrons and due to the $K^-$-$B$ attractive interaction. After the onset of kaon condensation, the charge chemical potential $\mu$ [=$(3\pi^2\rho_e)^{1/3}$ ] decreases as density increases and has the value of $\mu\lesssim O(m_\pi)$. At density $\rho_{\rm B}\gtrsim$ 1.22 fm$^{-3}$ (=7.97 $\rho_0$) for $\Sigma_{Kn}$ = 300 MeV [$\rho_{\rm B}\gtrsim$ 0.960 fm$^{-3}$ (=6.27 $\rho_0$) for $\Sigma_{Kn}$ = 400 MeV], $\mu$ becomes negative, where positrons ($e^+$) are present in place of electrons.  

(III) The proton fraction increases along with the growth of kaon condensates, so that the negative charge carried by kaon condensates is compensated by the positive charge of protons keeping the charge neutrality. The neutron fraction decreases with density following the appearance of protons and hyperons due to the baryon number conservation. 

\begin{figure}[!]
\begin{minipage}[r]{0.50\textwidth}
\begin{center}
~\vspace{1.5cm}~
\includegraphics[height=.32\textheight]{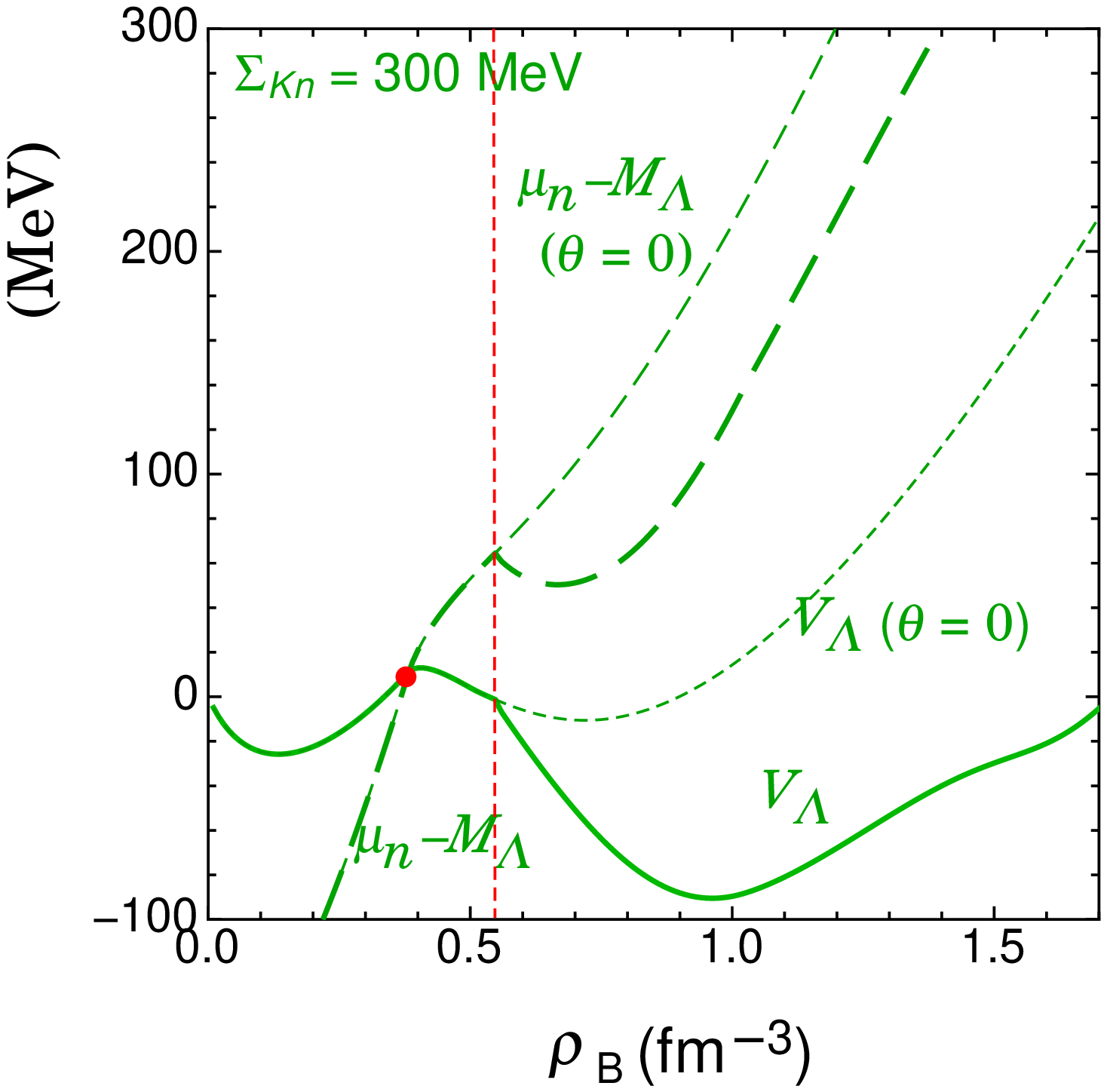}
\end{center}
\caption{The $\Lambda$ potential $V_\Lambda$ and ($\mu_n-M_\Lambda$) as functions of baryon number density $\rho_{\rm B}$ by the solid line and the long-dashed line, respectively for $\Sigma_{Kn}$ = 300 MeV. For reference, those for the pure hyperon-mixed matter (set to be $\theta=0$) are shown by the dotted line and the short dashed line, respectively. The vertical dotted line indicates the onset density for kaon condensation, $\rho_{\rm B}^c(K^-)$. See the text for details.\break }
\label{fig10}
\end{minipage}~
\begin{minipage}[l]{0.50\textwidth}
~\vspace{-5.5cm}~
\begin{center}
\includegraphics[height=.32\textheight]{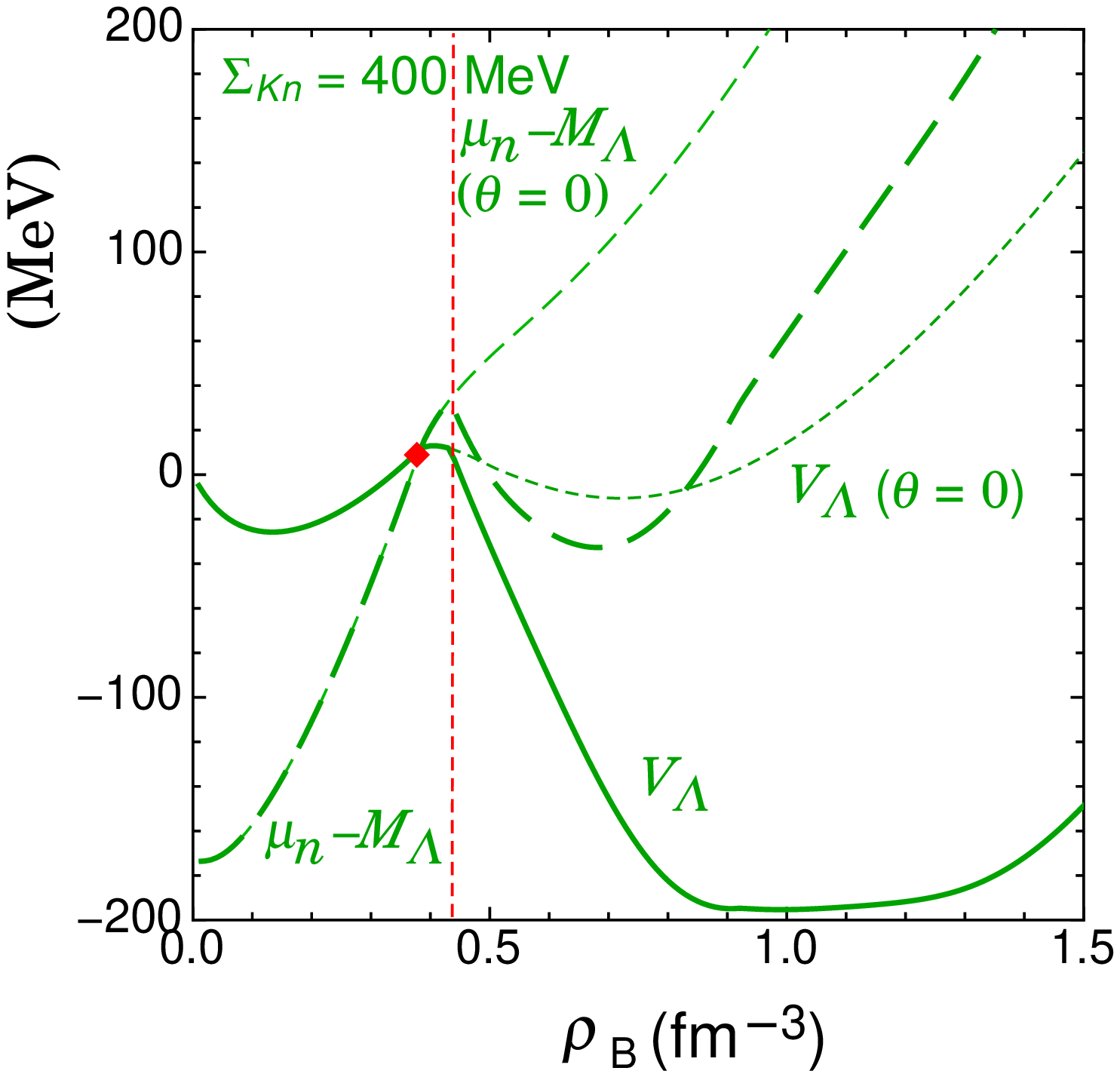}
\end{center}
~\vspace{1.0cm}~
\caption{The same as in Fig.~\ref{fig10}, but for $\Sigma_{Kn}$ = 400 MeV.}
\label{fig11}
\end{minipage}
\end{figure}

Here the density dependence of the $\Lambda$ and $\Xi^-$-mixing is reconsidered in terms of the hyperon potentials $V_\Lambda$ and $V_{\Xi^-}$. The potential $V_\Lambda$ is shown as a function of $\rho_{\rm B}$ by the solid line, 
together with the value of ($\mu_n-M_\Lambda$) by the long-dashed line, for $\Sigma_{Kn}$ = 300 MeV in Fig.~\ref{fig10} and for $\Sigma_{Kn}$ = 400 MeV in Fig.~\ref{fig11}. 
For reference, $V_\Lambda$ and the value of ($\mu_n-M_\Lambda$) for the pure hyperonic matter (set to be $\theta=0$) are shown by the dotted line and the short dashed line, respectively.
~
The $\Lambda$-mixing condition is given by $\mu_n-M_\Lambda >V_\Lambda$. The filled circle (crossing point of the solid line and the long-dashed line) corresponds to the onset density of the $\Lambda$-mixing, $\rho_{\rm B}^c(\Lambda)$, which is lower than the onset density of kaon condensation, $\rho_{\rm B}^c(K^-)$ (indicated by the vertical dotted line). In the vicinity of $\rho_{\rm B}^c (K^-)$, both the $V_\Lambda$ and ($\mu_n-M_\Lambda$) decrease with density as kaon condensation develops due to the $s$-wave $K$-$\Lambda$ and $K$-$n$ attractive interactions [the last terms on the r.~h.~s. of Eqs.~(\ref{eq:vb}) and (\ref{eq:mub})]. The reduction of the $V_\Lambda$ is more remarkable than that of the $\mu_n-M_\Lambda$, so that the $\Lambda$-mixing condition is always met for the relevant densities. Thus, beyond the density $\rho_{\rm B}^c (\Lambda)$, $\Lambda$ hyperons continue to be mixed before and after the onset of kaon condensation. 

\begin{figure}[!]
\begin{minipage}[r]{0.50\textwidth}
\begin{center}~\vspace{1.5cm}~
\includegraphics[height=.32\textheight]{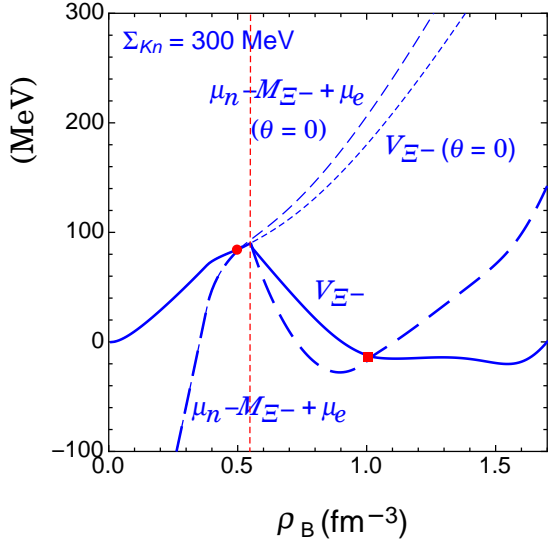}
\end{center}
\caption{The $\Xi^-$ potential $V_{\Xi^-}$ and ($\mu_n-M_{\Xi^-}+\mu_e$) as functions of baryon number density $\rho_{\rm B}$ by the solid line and the long-dashed line, respectively for $\Sigma_{Kn}$ = 300 MeV. For reference, those for the pure hyperon-mixed matter (set to be $\theta=0$) are shown by the dotted line and the short dashed line, respectively. The vertical dotted line indicates the onset density for kaon condensation, $\rho_{\rm B}^c(K^-)$. See the text for details.\break }
\label{fig12}
\end{minipage}~
\begin{minipage}[l]{0.50\textwidth}~
\vspace{-5.8cm}~
\begin{center}
\includegraphics[height=.32\textheight]{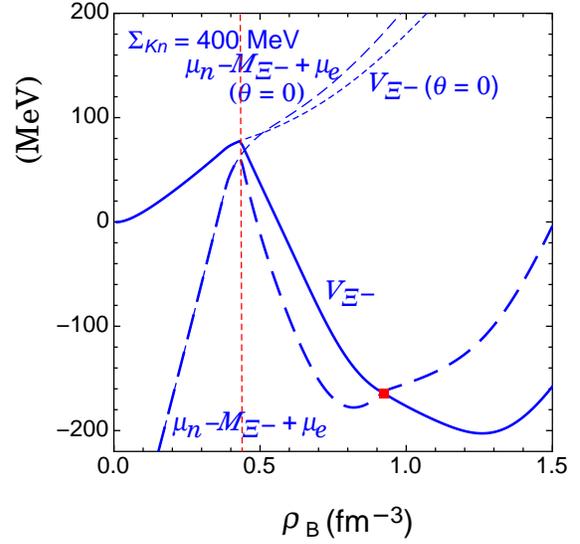}~
\end{center}~
\vspace{1.0cm}~
\caption{The same as in Fig.~\ref{fig12}, but for $\Sigma_{Kn}$ = 400 MeV. }
\label{fig13}
\end{minipage}
\end{figure}

The $V_{\Xi^-}$ is shown as a function of $\rho_{\rm B}$ by the solid line, 
together with the value of ($\mu_n-M_{\Xi^-}+\mu_e$) by the long-dashed line, for $\Sigma_{Kn}$ = 300 MeV in Fig.~\ref{fig12} and for $\Sigma_{Kn}$ = 400 MeV in Fig.~\ref{fig13}. 
The $V_{\Xi^-}$ and the value of ($\mu_n-M_{\Xi^-} +\mu$) for the pure hyperon-mixed matter are also shown by the dotted line and the short dashed line, respectively. 
~
Beyond the onset density $\rho_{\rm B}^c (K^-)$, both the $V_{\Xi^-}$ and $\mu_n$ decrease with density as kaon condensation develops, since the $s$-wave $K$-$\Xi^-$ and $K$-$n$ interactions work attractively. [See Eqs.~(\ref{eq:mub}) and (\ref{eq:vb}).]
However, owing to the decrease in the electron chemical potential $\mu_e$ (=$\mu$) with density, the term ($\mu_n-M_{\Xi^-}+\mu_e$) decreases more rapidly than the $V_{\Xi^-}$. Therefore, the condition for the $\Xi^-$-mixing, $\mu_n-M_{\Xi^-}+\mu_e > V_{\Xi^-}$, is not satisfied, i.e., $\mu_e$ is not large enough to assist the $\Xi^-$-mixing. 
For $\Sigma_{Kn}$ = 300 MeV, although the $\Xi^-$-mixing starts before kaon condensation sets in, but it soon vanishes just after the onset of kaon condensation. At high densities $\rho_{\rm B}\gtrsim$ 1.0 fm$^{-3}$, the $\mu_n$ rises up with density due to the dominant two-body repulsive interaction, and the condition for the $\Xi^-$-mixing gets satisfied again at $\rho_{\rm B}\geq \rho_{\rm B}^c(\Xi^-~{\rm in}~K^-)$ =1.03 fm$^{-3}$ (the second crossing point of the solid line and the long-dashed line in Fig.~\ref{fig12}). 
For $\Sigma_{Kn}$ = 400 MeV, the onset of kaon condensation precedes the $\Xi^-$-mixing, and the $\Xi^-$-mixing does not occur until $\rho_{\rm B}$ exceeds $\rho_{\rm B}^c(\Xi^-~{\rm in}~K^-)$ (= 0.92 fm$^{-3}$) (Fig.~\ref{fig13}). 

It is to be noted that $\Sigma^-$ hyperons do not appear in the ($Y+K$) phase over the relevant densities, as is the case with pure hyperonic matter (Fig.~\ref{fig4}). The total strangeness fraction, $(\rho_{K^-}+\rho_\Lambda+2\rho_{\Xi^-})/\rho_{\rm B}$, increases with $\rho_{\rm B}$ steadily in accordance with the growth of hyperon-mixing and kaon condensates, and it amounts to 0.9 at $\rho_{\rm B}\sim$ 1.5 fm$^{-3}$ for both cases of $\Sigma_{Kn}$ = 300 MeV and 400 MeV. 

\subsection{Self-suppression mechanisms }
\label{subsec:self}

\begin{figure}[h]
\begin{minipage}[l]{0.50\textwidth}~
\vspace{-1.5cm}~
\begin{center}~
\includegraphics[height=.32\textheight]{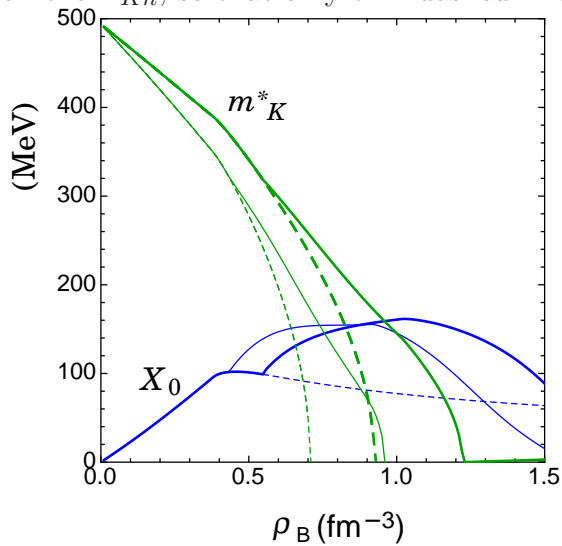}
\end{center}~
\end{minipage}~\vspace{1.0cm}~
\begin{minipage}[r]{0.50\textwidth}
\caption{The density dependence of the effective kaon mass $m_K^\ast$ given by Eq.~(\ref{eq:ekm2}) and $X_0$ by Eq.~(\ref{eq:x0}) in the ($Y$+$K$) phase (solid lines) and in the pure hyperon-mixed matter with $\theta$ being set to be zero (dashed lines). In the case of $m_K^\ast$ and the $X_0$ in the ($Y$+$K$) phase, the bold lines are for $\Sigma_{Kn}$ = 300 MeV and the thin lines for $\Sigma_{Kn}$ = 400 MeV. In the pure hyperonic matter, the $X_0$, responsible for the $s$-wave $K$-$B$ vector attraction, does not depend upon the $\Sigma_{Kn}$, so that only the thin dashed line is depicted. }
\label{fig14}
\end{minipage}
\end{figure}

Here we discuss the relativistic effects of kaon condensates on the $s$-wave $K$-$B$ scalar and vector interactions. 
In Fig.~\ref{fig14}, the density dependence of the effective kaon mass $m_K^\ast$ given by Eq.~(\ref{eq:ekm2}) and that of $X_0$ given by Eq.~(\ref{eq:x0}) are shown by the solid lines for  the ($Y$+$K$) phase. For comparison, those for the pure hyperon-mixed matter with $\theta$ being set to be zero are also shown by the dashed lines. In the case of $m_K^\ast$ and the $X_0$ in the ($Y$+$K$) phase, bold lines are for $\Sigma_{Kn}$ = 300 MeV and thin lines for $\Sigma_{Kn}$ = 400 MeV. In the pure hyperonic matter, the $X_0$, responsible for the $s$-wave $K$-$B$ vector attraction, does not depend upon the $\Sigma_{Kn}$, so that only thin dashed line is depicted.
The difference of $m_K^\ast$ between the solid lines and dashed lines stems from the suppression of the scalar density due to the appearance of kaon condensates. 
Thus one can see the {\it self-suppression mechanism} of the $s$-wave $K$-$B$ scalar interaction in the RMF framework becomes remarkable in the presence of kaon condensates\cite{fmmt96}: As kaon condensation develops with $\rho_{\rm B}$, the effective baryon mass $\widetilde M_b^\ast$ decreases following Eq.~(\ref{eq:effbmci}). The decrease in $\widetilde M_b^\ast$ leads to saturation of the scalar density for baryon, $\rho_b^s$, at higher densities, 
which, in turn, results in the suppression of the $K$-$B$ scalar attraction in the presence of kaon condensates, through the term proportional to $\rho_b^s $ in $m_K^\ast$ [(\ref{eq:ekm2})]. 

On the other hand, $X_0$, representing the $K$-$B$ vector interaction, is enhanced by the appearance of kaon condensates, as is seen from Fig.~\ref{fig14}. The enhancement of $X_0$ results mainly from the increase in proton density $\rho_p$ in response to the growth of kaon condensates. 

In Ref.~\cite{m08}, the EOS of the ($Y+K$) phase is considered based on the effective chiral Lagrangian 
with the nonrelativistic framework for baryons by adding the schematic baryon potential $V_b$ ($b$ = $p$, $n$, $\Lambda$, $\Sigma^-$, $\Xi^-$) parameterized in terms of the number densities of baryons. It has been shown that the energy gain due to the combined effects of both kaon condensates and the hyperon-mixing is so strong that there appears a local minimum of the energy with respect to $\rho_{\rm B}$, leading to self-bound objects with ($Y+K$) phase for $\Sigma_{Kn}$ = 300 MeV\cite{m08}. 
However, in the present result based on the RMF with the self-contained baryon potential $V_b$ [(\ref{eq:vb})], the $s$-wave $K$-$B$ scalar interaction is suppressed at high densities due to the self-suppression mechanism as the relativistic effect. As a result, the energy per baryon monotonically increases with $\rho_{\rm B}$ (Fig.~\ref{fig6}), and the self-bound star formed of the ($Y+K$) phase is unlikely to exist. 

Although the $s$-wave $K$-$B$ scalar attraction is suppressed through the {\it self-suppression mechanism} within the relativistic framework, such suppression is not enough for making the EOS stiff so as to be consistent with the recent observations of massive neutron stars\cite{demo10,f16,ant13,Cromartie2020,Fonseca2021,Romani2021}: The previous result with the coupling constants similar to those in the present paper shows that the maximum gravitational mass is $(1.5-1.6) M_\odot$ for $\Sigma_{Kn}$=(300$ - $400) MeV\cite{mmtt2018}. In order to construct a realistic EOS including the ($Y+K$) phase, being compatible with the recent observations of massive neutron stars, 
some mechanisms to circumvent both the large attraction due to the $s$-wave $K$-$B$ interaction and energy decrease due to the hyperon-mixing effect are still necessary. 

\section{ Circumventing the problems caused by the NLSI term}
\label{sec:discussion}

With regard to repulsive effects for baryons, the three-body $NNN$, $YNN$, $YYN$, $YYY$ forces have been introduced as the extra repulsion in the case of hyperon-mixed matter\cite{nyt02}. It has been shown that the universal three-body repulsion, being derived based on the string junction model by Tamagaki (abbreviated here as UTBR)\cite{t2008}, prevents the EOS from``dramatic softening'' due to the hyperon-mixing and that massive neutron star as high as 2 $M_\odot$ can be obtained\cite{t2008,tnt2008}. 
Recently, we introduced the density-dependent effective two-body potentials for the UTBR based on the string-junction model 2 in \cite{t2008} together with the phenomenological three-nucleon attraction (TNA),
 in addition to the ``minimal'' RMF (abbreviated as MRMF) where baryon interactions are simply composed of the two-body $B$-$B$ interaction mediated by meson-exchange, without recourse to the nonlinear self-interacting $\sigma, \omega$, or $\omega-\rho$ meson-coupling potentials~\cite{mmt2021}. 
In this model (MRMF+UTBR+TNA), the UTBR is supposed to be relevant to the short-range part of the $B$-$B$ interaction, where the quark structure of baryon reveals itself. Therefore, the UTBR has been phenomenologically introduced beyond the RMF picture\cite{mmtt2018}, while baryons can be viewd as point-like within the RMF in the intermediate and long-range part of the interaction. 
 This baryon interaction model can describe the saturation properties of the SNM, following each energy contributions from three-nucleon repulsion and attraction, and two-body parts similar to those obtained by conventional nuclear matter theory~\cite{lp1981}. 
 It is also emphasized that, once the NLSI term is replaced by the UTBR+TNA, the many-body effects arising from the NLSI term in the ME scheme are removed, and that the kaon self-energy in hyperon-mixed matter ($\theta\rightarrow$ 0) with such (MRMF+UTBR+TNA) model is formally equivalent between the CI and ME schemes for the kaon-baryon vertices, even though, in the presence of kaon condensates, there still remains many-body effect (ii) only in  the ME scheme [the fourth (second) term on the r.~h.~s. of Eq.~(\ref{eq:meekm3}) (Eq.~(\ref{eq:mex01}) ) ]  coming from the kaon source terms in the equations of motion for the meson mean fields.  
 
With this baryon interaction model (MRMF+UTBR+TNA) coupled with the effective chiral Lagrangian, we considered the ($Y$+$K$) phase in Ref.~\cite{mmt2021}. It has been shown that softening of the equation of state stemming from both kaon condensation and mixing of hyperons is compensated with the repulsive effect of the UTBR and the relativistic effect for two-body $B$-$B$ interaction. It has also been shown that the EOS and the resulting mass and radius of compact stars accompanying the ($Y$+$K$) phase are consistent with recent observations of massive neutron stars. 

\section{Summary and Concluding Remarks}
\label{sec:summary}

We have compared two coupling schemes, the contact interaction (CI) and  meson-exchange (ME) schemes, concerning the $K$-$B$ and $K$-$K$ interactions in the effective chiral Lagrangian. We have considered how the onset density of kaon condensation realized from hyperon-mixed matter and the EOS of the ($Y+K$) phase are affected in these two schemes. 
The nonlinear self-interacting (NLSI) $\sigma$-meson potential $U_\sigma(\sigma)$ has been commonly introduced in both schemes to reduce the incompressibility at saturation density of symmetric nuclear matter. 
In the ME scheme, there appear many-body effects in the $K$-$B$ interaction through the kaon-multi-$\sigma$-meson coupling stemming from the derivative term, $dU_\sigma/d\sigma$. The kaon-multi-$\sigma$-meson coupling term has a sizable repulsive contribution to the kaon energy $\omega_K$. Hence the onset condition of kaon condensation, $\omega_K=\mu$, is not fulfilled over the relevant baryon densities in the case of the ME scheme unless the $\Sigma_{Kn}$ is taken to be extraordinarily large. In general, the NLSI terms bring about 
extra terms for the kaon self-energy in the ME scheme beyond the scope of chiral symmetry. 

On the other hand, the $K$-$B$ interaction for the kaon self-energy in the CI scheme is specified by chiral symmetry and free from such ambiguity of many-body effects brought about by the NLSI terms.  In this scheme, the onset of kaon condensation occurs at a density $\rho_{\rm B}^c(K^-)=(3-4)\rho_0$ for the standard values of $\Sigma_{Kn}$ = (300 $-$ 400) MeV. 

In the context of stiffening of the EOS at high densities, the NLSI term 
is not relevant to an origin of the extra repulsive energy at high densities leading to a solution to the ``hyperon puzzle'', since the contribution to the repulsive energy gradually decreases with increase in density. 
Actually, 
in the case of the CI scheme, the EOS for the ($Y$+$K$) phase is considerably softened even with the NLSI term  after the appearance of kaon condensates in hyperon-mixed matter. 

As stated in Sec.~\ref{sec:discussion}, the (MRMF+UTBR+TNA) model~\cite{mmt2021} reveals a satisfactory picture for saturation of SNM in view of the standard variational nuclear matter theory with the phenomenological three-nucleon forces (TNR and TNA)~\cite{lp1981,apr1998}. Moreover the UTBR, introduced as the effective two-body baryon potential in the model, has a decisive contribution to stiffening of the EOS with the ($Y$+$K$) phase at high densities as a solution to the ``hyperon puzzle''. In these respects, the (MRMF+UTBR+TNA) model is considered to be more natural and plausible than the (MRMF+NLSI) model elucidated in the present paper. 
As a consequence, the many-body effects (i) originating from the NLSI terms and the resulting extra terms which make difference between the CI and ME schemes in the (MRMF+NLSI) model should not be regarded as solid and universal.
The results on the comparison of the (MRMF+UTBR+TNA) model and the (MRMF+NLSI) model with specific NLSI terms such as $U_\sigma$, quartic terms of the vector $\omega$ meson, and $\omega$-$\rho$ meson coupling terms etc. will be reported in detail elsewhere~\cite{mmt2022}. 

In this paper, we have concentrated specifically on the NLSI terms in the RMF leading to the difference of kaon dynamics in dense matter between the CI and ME schemes. As another prescription for  describing nuclear matter and finite nuclei, density-dependence for meson-baryon coupling constants has been taken into account within the RMF framework to be consistent with the result of the self-energy of Dirac-Brueckner calculation of nuclear matter~\cite{typel1999}. Some authors considered kaon condensation in the ME scheme with the density-dependent meson-baryon coupling strengths in place of the NLSI terms ~\cite{cb2014,ts2020,mbb2021}.
In this approach, the formal expression of the kaon self-energy is essentially the same as in the CI scheme, while density-dependence of the meson-baryon coupling strengths lead to an extra nonlinear density-dependence of the kaon self-energy in addition to the density-dependence of  baryon scalar densities. 

As other possible repulsions between baryons, multi-pomeron exchange potential has been considered as an origin of many-body forces\cite{yfyr2014}.
As another example of many-body forces, the $BMM$, $MMM$-type diagrams, which follow from the specific counting rule of the meson-baryon diagrams, have been considered within the RMF\cite{to12}. 

Anti-symmetrization effect in the Hartree-Fock approximation for baryons is another issue to be elucidated 
 for construction of the realistic EOS of the ($Y$+$K$) phase. In Refs.~\cite{mks12,kms12} the tensor coupling of vector mesons has been introduced in the RMF for the EOS of hyperon-mixed matter. It has been pointed out that the Fock contribution hinders the appearance of hyperons at middle and high densities and also to suppression of hyperon-mixing at high densities. For the ($Y$+$K$) phase, the Fock contribution may have a minor effect on population of kaon condensates, so that kaon condensation may be dominant over hyperons at high densities.

Throughout this paper, hadrons are considered as point-like particles even at high densities, where they are supposed to be overlap with each other and quark degrees of freedom should be explicitly considered. In Ref.~\cite{mht2013,mht2016,k2015}, the hadron phase (hyperon-mixed matter) was connected smoothly to the quark phase in a hadron-quark crossover picture. The resulting EOS has been shown to be stiff enough to have massive stars as much as two solar mass. 
 In this context, kaon condensation may play an important role on both the hadron and quark phases. In particular, kaonic modes may be condensed in the color-flavor locked phase\cite{bs02,kr02,b05,f05}. It is interesting to clarify the relationship between kaon condensation in the hadronic phase and that in the quark phase and to construct a stiff EOS including the hadron-quark crossover, which may be consistent with recent observations of massive neutron stars. 

\section*{Acknowledgement}
This work is supported in part by the Grant-in-Aid for Scientific Research on Innovative Areas "Nuclear Matter in Neutron Stars Investigated by Experiments and Astronomical Observations". One of the authors (T.~Muto) acknowledges the financial support by Chiba Institute of Technology. 

\section{Appendix}
\label{sec:appendix}

\subsection{Estimation of the $KN$ sigma term}
\label{subsec:appendix1}

We estimate the allowable values of the $Kn$ sigma term, $\Sigma_{Kn}$. 
The ``$K$-baryon sigma term'' is defined by 
\begin{equation}
\Sigma_{Kb}=\frac{1}{2}(m_u+m_s)\langle b|(\bar u u +\bar s s) |b\rangle \ , 
\label{eq:kbsigma}
\end{equation}
where $\langle b|\bar q q |b\rangle $ is the quark condensate in the baryon species $b$ ($b$ = $p$, $n$, $\Lambda$, $\Sigma^-$, $\Xi^-$ in this paper). 

In the chiral perturbation theory, one obtains the $\bar q q$ condensates from Eq.~(\ref{eq:fbmass}) 
by the use of the relations, $ \langle b|\bar q q |b\rangle $=$\partial M_b/\partial m_q$:  
\begin{eqnarray}
 \langle p|\bar u u |p\rangle &=&  \langle n|\bar d d |n\rangle = - 2(a_1+a_3) \ , \cr
  \langle p|\bar d d |p\rangle &=&  \langle n|\bar u u |n\rangle = - 2a_3 \ , \cr
    \langle p|\bar s s |p\rangle &=&  \langle n|\bar s s |n\rangle = - 2(a_2+a_3) \ .
    \label{eq:qcondensates}
\end{eqnarray}
Substituting Eq.~(\ref{eq:qcondensates}) into Eq.~(\ref{eq:kbsigma}), one obtains Eq.~(\ref{eq:ckbsigma}) in Sec.~\ref{sec:CI}. For instance, the $K$-neutron $\sigma$ term is given as $\Sigma_{Kn} = -(a_2+2a_3)(m_u+m_s) $.

 The quark masses $m_i$ are chosen to be $m_u$ = 6 MeV, $m_d$ = 12 MeV, and $m_s$ = 240 MeV,  according to Ref.~\cite{kn86}. Further the parameters $a_1$ and $a_2$ are fixed to be $a_1$ = $-$0.28, $a_2$ = 0.56 so as to reproduce the empirical octet baryon mass splittings\cite{kn86}. We fix the remaining parameter $a_3$ with reference to the standard value of the $\pi N$ sigma term, $\Sigma_{\pi N}$ = 45 MeV, which is extracted from the $\pi$-$N$ scattering data\cite{gls91}. By the use of Eq.~(\ref{eq:qcondensates}), $\Sigma_{\pi N}$ is written as 
 \begin{equation}
\Sigma_{\pi N}=\frac{1}{2}(m_u+m_d)\langle N|(\bar u u +\bar d d) |N\rangle = -(a_1+2 a_3)(m_u+m_d) \ , 
\label{eq:piNsigma}
\end{equation}
from which one obtains $a_3=-1.1$. With this value, one obtains $\Sigma_{Kn}$ = 403 MeV and
$y~\equiv~2\langle N|\bar s s |N\rangle/\langle N|(\bar u u +\bar d d) |N\rangle$ = $2(a_2+a_3)/(a_1+2a_3)$=0.44, which implies large $\bar s s$ condensate in the nucleon. 

On the other hand, recent lattice QCD results suggest small $\bar s s $ condensate, $y\simeq$ 0~\cite{ohki08,yt2010,ya2016}, 
for which the expressions (\ref{eq:qcondensates}) followed by Eq.~(\ref{eq:fbmass}) based on lowest-order chiral perturbation theory cannot be applied. In this case, assuming $\langle N|\bar u u|N\rangle$ = $\langle N|\bar d d|N\rangle$, and $\langle N|\bar s s|N\rangle$ = 0, one obtains $\langle N|\bar u u|N\rangle$~=~$\Sigma_{\pi N}/(m_u+m_d)$~=~2.5 with $\Sigma_{\pi N}$=45 MeV, 
and the lower value for $\Sigma_{KN}$ is estimated as $\Sigma_{KN}$= 308 MeV. 

Throughout this paper, we consider two cases for $\Sigma_{Kn}$ = 300 MeV and 400 MeV as the standard values, considering the uncertainty of the $\bar s s$ condensate in the nucleon.

\subsection{$K^-$ optical potential}
\label{aubsec:appendix2}
 
The strengh of the in-medium $K$-$N$ attraction is simulated by the  
the $K^-$ optical potential $U_K$ at $\rho_B$=$\rho_0$ in symmetric nuclear matter. In the CI scheme, it is defined by the use of the $K^-$ self-energy [Eq.~(\ref{eq:selfk})]
\begin{equation}
U_K({\rm CI})\equiv\Pi_K({\rm CI})/(2\omega_K(\rho_B))|_{\rho_p=\rho_n=\rho_0/2}
= -\frac{1}{f^2}\left(\rho_0^s\frac{\Sigma_{Kn}+\Sigma_{Kp}}{4\omega_K(\rho_0)}+\frac{3}{8}\rho_0\right)
\label{eq:ukci}
\end{equation}
with the nuclear scalar density $\rho_0^s$ at $\rho_{\rm B}=\rho_0$ in symmetric nuclear matter. In the ME scheme, one obtains from Eqs.~(\ref{eq:selfkme}) and (\ref{eq:ukci})
\begin{equation}
U_K({\rm ME})\equiv\Pi_K({\rm ME})/(2\omega_K(\rho_B))|_{\rho_B=\rho_0}
\simeq -(g_{\sigma K}\langle\sigma\rangle_0 +g_{\omega K}\langle\omega_0\rangle_0 )=U_K({\rm CI})+\frac{g_{\sigma K}}{m_\sigma^2}\left(\frac{dU_\sigma}{d\sigma}\right)_{\sigma=\langle\sigma\rangle_0} \ , 
\label{eq:ukme}
\end{equation}
where an approximation $\omega_K(\rho_0)\sim m_K$ has been used.


\begin{thebibliography}{9}
\bibitem{kn86} D.~B.~Kaplan and A.~E.~Nelson, 
Phys.~Lett.~{\bf B 175},~57~(1986). 

\bibitem{t88} T.~Tatsumi, Prog.~Theor.~Phys.~{\bf 80},~22~(1988).

\bibitem{mt92} T.~Muto and T.~Tatsumi, Phys.~Lett.~{\bf B 283},~165~(1992).

\bibitem{m93} T.~Muto, Prog.~Theor.~Phys.~{\bf 89}~(1993)~415.

\bibitem{mtt93}
T.~Muto, R.~Tamagaki, and T. Tatsumi, Prog.~Theor.~Phys.~Suppl. {\bf 112},~159~(1993). \\
T.~Muto, T.~Takatsuka, R.~Tamagaki, and T. Tatsumi, Prog.~Theor.~Phys.~Suppl. {\bf 112},~221~(1993).

\bibitem{fmtt94} H.~Fujii, T.~Muto, T.~Tatsumi, R.~Tamagaki, Nucl.~Phys.~{\bf A571},~758~(1994); \\
Phys.~Rev.~{\bf C50},~3140~(1994).

\bibitem{tpl94} V.~Thorsson, M.~Prakash, and J.~M.~Lattimer, 
Nucl.~Phys.~{\bf A 572},~693~(1994); Nucl.~Phys.~{\bf A 574},~851~(1994)~(E). 

\bibitem{kvk95} E.~E.~Kolomeitsev, D.~N.~Voskresensky, 
B.~K{\"a}mpfer, Nucl.~Phys.~{\bf A 588}~(1995)~889. 

\bibitem{lbm95} C.~-H.~Lee, G.~E.~Brown, D.~-P.~Min, and M.~Rho, Nucl.~Phys.~{\bf A 585},~401~(1995). 

\bibitem{fmmt96} H.~Fujii, T.~Maruyama, T.~Muto, and T.~Tatsumi, Nucl.~Phys.~{\bf A 597},~645~(1996).

\bibitem{tstw98} K.~Tsushima, K.~Saito, A.~W.~Thomas, and S.~V.~Wright, 
Phys.~Lett.~{\bf B 429},~239~(1998); 
{\it ibid}. {\bf 436},~453~(1998)~(E).

\bibitem{ty1998} T.~Tatsumi and M.~Yasuhira, Phys.~Lett.~{\bf B 441},~9~(1998); 
Nucl.~Phys.~{\bf A 653},~133~(1999).

\bibitem{g85} N.~K.~Glendenning, 
Astrophys.~J.~{\bf 293},~470~(1985) ; N.~K.~Glendenning, and S.~A.~Moszkowski, 
Phys.~Rev.~Lett.~{\bf 67},~2414~(1991). 

\bibitem{ekp95} P.~J.~Ellis, R.~Knorren, and M.~Prakash, 
Phys.~Lett.~{\bf B349},~11~(1995). 

\bibitem{kpe95} R.~Knorren, M.~Prakash, and P.~J.~Ellis, 
Phys.~Rev.~{\bf C52},~3470~(1995). 

\bibitem{sm96} J.~Schaffner and I.~N.~Mishustin, 
Phys.~Rev.~{\bf C53},~1416~(1996). 

\bibitem{g01} N.~K.~Glendenning and J.~Schaffner-Bielich, Phys.~Rev.~{\bf C 60},~025803~(1999). \\
N.~K.~Glendenning, Phys.~Rep.~{\bf 342},~393~(2001). 

\bibitem{phz99} S.~Pal, M.~Hanauske, I.~Zakout, H.~St{\"o}cker, and 
W.~Greiner, Phys.~Rev.~{\bf C 60},~015802~(1999). 
\bibitem{h00} M.~Hanauske, D.~Zschiesche, S.~Pal, S.~Schramm, H.~St{\"o}cker, and W.~Greiner, Astrophys.~J.~{\bf 537},~958~(2000). 
\bibitem{s00} P.~K.~Sahu, 
Phys.~Rev.~{\bf C 62},~045801~(2000). 

\bibitem{tolos2017} L.~Tolos, M.~Centelles, and A.~Ramos, Astrophys.~J.~{\bf 834}:3~(2017).

\bibitem{bg97} S.~Balberg and A.~Gal, Nucl.~Phys.~{\bf A625},~435~(1997).  

\bibitem{h98} H.~Huber, F.~Weber, M.~K.~Weigel, and Ch.~Schaab, 
Int.~J.~Mod.~Phys.~{\bf E 7},~301~(1998). 

\bibitem{bbs98} M.~Baldo, G.~F.~Burgio, and H.~-J.~Schulze, 
Phys.~Rev.~{\bf C 58},~3688~(1998) ; {\it ibid} {\bf C  61},~055801~(2000).

\bibitem{v00} I.~Vida$\tilde{\rm n}$a, A.~Polls, A.~Ramos, 
M.~Hjorth-Jensen, and V.~G.~J.~Stoks, 
Phys.~Rev.~{\bf C 61},~025802~(2000); 
I.~Vida$\tilde {\rm n}$a, A.~Polls, A.~Ramos, L.~Engvik, and M.~Hjorth-Jensen, 
Phys.~Rev.~{\bf C 62},~035801~(2000). 

\bibitem{nyt02} S.~Nishizaki, Y.~Yamamoto, and T.~Takatsuka,  
Prog.~Theor.~Phys.~{\bf 108},~703~(2002).
 
\bibitem{t04} T.~Takatsuka, Prog.~Theor.~Phys.~{\it Supplement}~{\bf 156}, 84 (2004).

\bibitem{t2016} H.~Togashi, E.~Hiyama, Y.~Yamamoto, and M.~Takano, Phys.~Rev.~{\bf 93}, 035808~(2016).

\bibitem{pplp92} M.~Prakash, M.~Prakash, J.~M.~Lattimer, and C.~J.~Pethick, Astrophys.~J.~{\bf 390}, L77~(1992).

\bibitem{t98} S.~Tsuruta, Phys.~Rep.~{\bf 292}, 1~(1998).

\bibitem{m08} T.~Muto: Phys.~Rev.~{\bf C 77},~015810~(2008), and references cited therein.

\bibitem{demo10} P.~B.~Demorest, T.~Pennucci, S.~M.~Ransom, M.~S.~E.~Roberts, and J.~W.~T.~Hessels, Nature~{\bf 467}~(2010)~1081.

\bibitem{f16} E.~Fonseca, T.~T.~Pennucci et al., Astrophys.~J~{\bf 832}, 167~(2016).

\bibitem{ant13} J.~Antoniadis et al., Science~{\bf 340}~(2013)~6131.

\bibitem{Cromartie2020} H.~T.~Cromartie et al., Nat.~Astron.~{\bf 4},~72~(2020). 

\bibitem{Fonseca2021} E.~Fonseca et al., arXiv:2104.00880~v1~[astroph.HE]. 

\bibitem{Romani2021} R.~W.~Romani et al., Astrophys.~J.~L~{\bf 908},~L46,~(2021). 

\bibitem{bb01} S.~Banik and D.~Bandyopadhyay, Phys.~Rev.~{\bf C 63},~035802~(2001) ; 
{\it ibid}~{\bf C 64},~055805~(2001).

\bibitem{sl2010} G.~Y.~Shao and Y.~X.~Liu, Phys.~Rev.~{\bf C 82},~055801~(2010).

\bibitem{pbg00} S.~Pal, D.~Bandyopadhyay, and W.~Greiner, Nucl.~Phys.~{\bf A 674},~553~(2000).

\bibitem{mk2010} A.~Mishra, A.~Kumar, S.~Sanyal, V.~Dexheimer, and S.~Schramm, Eur.~Phys.~J.~{\bf A 45},~169~(2010).

\bibitem{mpp05} D.~P.~Menezes, P.~K.~Panda, and C.~Providencia, Phys.~Rev.~{\bf C 72},~035802~(2005). 

\bibitem{rhhk07} C.~Y.~Ryu, C.~H.~Hyun, S.~W.~Hong, and B.~T.~Kim, Phys.~Rev.~{\bf C 75},~055804~(2007). 

\bibitem{cb2014} P.~Char and S.~Banik, Phys.~Rev.~{\bf C 90},~015801~(2014). 

\bibitem{ts2020} V.~B.~Thapa and M.~Sinha, Phys.~Rev.~{\bf D 102},~123007~(2020). 

\bibitem{mbb2021} T.~Malik, S.~Banik, and D.~Bandyopadhyay, Astrophys.~J.~{\bf 910},~No.2,~96~(2021). 

\bibitem{a2017} B.~P.~Abott et al., Phys.~Rev.~Lett.~{\bf 119},~161101~(2017).

\bibitem{yy2017} K.~Yagi and N.~Yunes, Phys.~Rep.~{\bf 681},~1~(2017).

\bibitem{a2018} B.~P.~Abbott et al., Phys.~Rev.~Lett.~{\bf 121},~161101~(2018).


\bibitem{Riley2019} T.~E.~Riley et al., Astrophys.~J.~{\bf 887},~L21~(2019). 

\bibitem{Miller2019} M.~C.~Miller et al., Astrophys.~J.~{\bf 887},~L24~(2019). 

\bibitem{Miller2021} M.~C.~Miller et al.,~arXiv : 2105.06979~v1~[astro-ph~HE]. 

\bibitem{Riley2021} T.~E.~Riley et al.,~Astrophys.~J.~L.~{\bf 918}~No.2~(2021), ~arXiv : 2105.06980~[astro-ph~HE]. 

\bibitem{ay02} Y.~Akaishi and T.~Yamazaki, Phys.~Rev.~{\bf C 65},~044005~(2002); 
T.~Yamazaki and Y.~Akaishi, 
Phys.~Lett.~{\bf B 535},~70~(2002). 

\bibitem{yda04} T.~Yamazaki, A.~Dote, and Y.~Akaishi, Phys.~Lett.~{\bf B 587},~167~(2004). 

\bibitem{ynoh2005} J.~Yamagata, H.~Nagahiro, Y.~Okumura, and S.~Hirenzaki, Prog.~Theor.~Phys.~{\bf 114},~301~(2005). \\
J.~Yamagata, H.~Nagahiro, and S.~Hirenzaki, Phys.~Rev.~{\bf C 74}, 014604~(2006).

\bibitem{mmt09} T.~Muto, T.~Maruyama, and T.~Tatsumi, Phys.~Rev.~{\bf C 79},~035207~(2009).

\bibitem{gfgm2009} D.~Gazda, E.~Friedman,  A.~Gal, and J.~Mares,
   Phys.~Rev.~{\bf C 76},~055204~(2007); 
   Phys.~Rev.~{\bf C 77},~045206~(2008); 
   Phys.~Rev.~{\bf C 80},~035205~(2009);
   Nucl.~Phys.~{\bf A 835},~287~(2010).

\bibitem{zs2013} X.~-R.~Zhou and H.~-J.~Schulze, Nucl.~Phys.~{\bf A 914},~332~(2013).

\bibitem{bbg12} E.~Botta, T.~Bressani, and G.~Garbarino, Eur.~Phys.~J.~{\bf A 48}~(2012)~41. 

\bibitem{ghm16} As a review, A.~Gal, E.~V.Hungerford, and D.~J.~Millener, Rev.~Mod.~Phys.~{\bf 88}, 035004~(2016).

\bibitem{hj2016} T.~Hyodo and D.~Jido, Prog.~Part.~Nucl.~Phys.~{\bf 67},~55~(2012).

\bibitem{ichikawa2015} Y.~Ichikawa et al., Prog.~Theor.~Exp.~Phys.~{\bf 2015}, 021D01~(2015).

\bibitem{sada2016} Y.~Sada et al., Prog.~Theor.~Exp.~Phys.~{\bf 2016}, 051D01~(2016).

\bibitem{a2019} S.~Ajimura et al., Phys.~Lett.~{\bf B 789},~620~(2019).
 
\bibitem{yamaga2020} T.~Yamaga et al., Phys.~Rev.~{\bf C102},~044002~(2020). 

\bibitem{mmt14} T.~Muto, T.~Maruyama, and T.~Tatsumi, JPS Conf.~Proc.~{\bf 1},~013081~(2014);
EPJ~Web of Conferences~{\bf 73},~05007~(2014).

\bibitem{mmt2021} T.~Muto, T.~Maruyama, and T.~Tatsumi, Phys.~Lett.~{\bf B 820},~136587~(2021). 

\bibitem{te97} V.~Thorsson and P.~J.~Ellis, Phys.~Rev.~{\bf D 55},~5177~(1997).
   
\bibitem{PDG2020} P.~A.~Zyla et al.~(Particle~Data~Group),~Prog.~Theor.~Exp.~Phys.~{\bf 2020},~083C01~(2020). 
   
\bibitem{mmt2015} T.~Muto, T.~Maruyama, and T.~Tatsumi, JPS Conf.~Proc.~{\bf 17},~102003~(2017).

\bibitem{m81} A.~D.~Martin, Nucl.~Phys.~{\bf A179},~33~(1981).

\bibitem{sdg94} J.~Schaffner, C.~B.~Dover, A.~Gal, C.~Greiner, D.~J.~Millener, and H.~St\"ocker, Ann.~Phys.~{\bf 235},~35~(1994).

\bibitem{mdg88} D.~J.~Millener, C.~B.~Dover, and A.~Gal, Phys.~Rev.~{\bf C 38}, 2700 (1988). 

\bibitem{kf00} M.~Kohno, Y.~Fujiwara, T.~Fujita, C.~Nakamoto, and Y.~Suzuki, 
Nucl.~Phys.~{\bf A 674}, 229  (2000). 

\bibitem{fk01} Y.~Fujiwara, M.~Kohno, C.~Nakamoto, and 
Y.~Suzuki, Phys.~Rev.~{\bf C 64}, 054001 (2001). 

\bibitem{b99} S.~Bart et al., Phys.~Rev.~Lett.~{\bf 83}, 5238 (1999). 
\bibitem{d99} J.~Dabrowski, Phys.~Rev.~{\bf C 60}, 025205  (1999). 
\bibitem{n02} H.~Noumi et al., Phys.~Rev.~Lett.~{\bf 89}, 072301 (2002); 
Phys.~Rev.~Lett.~{\bf 90}, 049902 (2003).
\bibitem{dr04} J.~Dabrowski and J.~Ro\.{z}ynek, Acta~Phys.~Polon.~{\bf B 35}, 2303 (2004). 
\bibitem{hh05} T.~Harada and Y.~Hirabayashi, Nucl.~Phys.~{\bf A 759}, 143 (2005). 
\bibitem{mfgj95} J.~Mares, E.~Friedman, A.~Gal, and B.~K.~Jennings, 
Nucl.~Phys.~{\bf A 594}, 311~(1995). 

\bibitem{f98} T.~Fukuda et al., Phys.~Rev.~{\bf C~58}~(1998)~1306.

\bibitem{k00} P.~Khaustov et al., Phys.~Rev.~{\bf C~61}~(2000)~054603.

\bibitem{h2010} E.~Hiyama, M.~Kamimura, Y.~Yamamoto, and T.~Motoba, Phys.~Rev.~Lett.~{\bf 104}, 212502~(2010).

\bibitem{awt03} G.~Audi, A.~H.~Wapstra and C.~Thibault, Nucl.~Phys.~{\bf A 729}, 337~(2003).

\bibitem{g2016} T.~Gogami et al., Phys.~Rev.~{\bf C~93}, 034314~(2016).

\bibitem{ht01} H.~Takahashi et al., Phys.~Rev.~Lett.~{\bf 87}, 212502~(2001).

\bibitem{a13} J.~K.~Ahn et al., Phys.~Rev.~{\bf C~88}, 014003~(2013).

\bibitem{n15} K.~Nakazawa et al., Prog.~Theor.~Exp.~Phys.~(2015)~033D02. 

\bibitem{sh16} T.~T.~Sun, E.~Hiyama, H.~Sagawa, H.~-J.~Schulze, and J.~Meng, 
Phys.~Rev.~{\bf C~94},~064319~(2016).

\bibitem{hayakawa2021} S.~H.~Hayakawa et al., Phys.~Rev.~Lett.~{\bf 126}, 062501~(2021). 

\bibitem{yoshimoto2021} M.~Yoshimoto et al.,~Prog.~Theor.~Exp.~Phys.~{\bf 2021},~073D02~(2021). 

\bibitem{bc79} G.~Baym and D.~K.~Campbell, in {\it Meson and Nuclei}, 
ed. M.~Rho and D.~H.~Wilkinson, (North Holland, Amsterdam, 1979), 
Vol.~III, p.~1031.

\bibitem{w66} S.~Weinberg, Phys.~Rev.~Lett.~{\bf 17}, 616~(1966).

\bibitem{ichikawa2020} Y.~Ichikawa et al., Prog.~Theor.~Exp.~Phys.~{\bf 2020}, 123D01~(2020). 

\bibitem{st94} Y.~Sugahara and H.~Toki, Nucl.~Phys.~{\bf A 579}, 557~(1994).

\bibitem{Horowitz2001} C.~J.~Horowitz and J.~Piekarewitz, Phys.~Rev.~Lett.~{\bf 86},~5647~(2001). 

\bibitem{fattoyev2020} F.~J~Fattoyev, C.~J.~Horowitz, J.~Piekarewicz, and B.~Reed, Phys.~Rev.~{\bf C 102},~065805~(2020). 

\bibitem{lp1981} I.~E.~Lagaris and V.~R.~Pandharipande, Nucl.~Phys.~{\bf A 359}~,349~(1981). 

\bibitem{apr1998} A.~Akmal, V.~R.~Pandharipande, and D.~G.~Ravenhall, Phys.~Rev.~{\bf C 58},~1804~(1998).

\bibitem{mtvt06} T.~Maruyama, T.~Tatsumi, D.~N.~Voskresensky, and T.~Tanigawa, 
Phys.~Rev.~{\bf C 73}, 035802~(2006).

\bibitem{mmtt2018} T.~Muto, T.~Maruyama, and T.~Tatsumi, and T.~Takatsuka, JPS~Conf.~Proc.~{\bf 20}, 011038~(2018). 

\bibitem{t2008} R.~Tamagaki, Prog.~Theor.~Phys.~{\bf 119},~965~(2008).

\bibitem{tnt2008} T.~Takatsuka, S.~Nishizaki, and R.~Tamagaki, AIP~Conf.~Proc.~{\bf 1011}, 209~(2008).

\bibitem{mmt2022} T.~Muto, T.~Maruyama, and T.~Tatsumi, to be submitted. 

\bibitem{typel1999} S.~Typel and H.~H.~Wolter, Nucl.~Phys.~{\bf A 656},~331~(1999). 

\bibitem{yfyr2014} Y.~Yamamoto, T.~Furumoto, N.~Yasutake, and Th.~A.~Rijken, Phys.~Rev.~{\bf C 90}, 045805~(2014).

\bibitem{to12} K.~Tsubakihara and A.~Ohnishi, Nucl.~Phys.~{\bf A 914},~438~(2013); arXiv:1211.7208.


\bibitem{mks12} T.~Miyatsu, T.~Katayama, and K.~Saito, Phys.~Lett.~{\bf B 709}, 242~(2012).

\bibitem{kms12} T.~Katayama, T.~Miyatsu, and K.~Saito, Astrophys.~J.~Supplement~{\bf 203}, 22~(2012). 

\bibitem{mht2013} K.~Masuda, T.~Hatsuda, and T.~Takatsuka, Astrophys.~J.~{\bf 764},~12~(2013). 

\bibitem{mht2016} K.~Masuda, T.~Hatsuda, and T.~Takatsuka, PTEP~{\bf 2016}. No.~2,~021D01~(2016).

\bibitem{k2015} T.~Kojo, P.~D.~Powel, Y.~Song, and G.~Baym, Phys.~Rev.~{\bf D 91}, 045003~(2015).

\bibitem{bs02} P.~F.~Bedaque and T.~Sch\"afer, Nucl.~Phys.~{\bf A 697},~802~(2002). 
\bibitem{kr02} D.~B.~Kaplan and S.~Reddy, Phys.~Rev.~{\bf D 65},~054042~ (2002). 
\bibitem{b05} M.~Buballa, Phys.~Lett.~{\bf B609},~57~(2005).
\bibitem{f05} M.~M.~Forbes, 
Phys.~Rev.~{\bf D 72},~094032~(2005).
\bibitem{bb03} S.~Banik and D.~Bandyopadhyay, 
Phys.~Rev.~{\bf D~67},~123003~(2003). 

\bibitem{gls91} J.~Gasser, H.~Leutwyler, and M.~E.~Sainio, Phys.~Lett.~{\bf B 253},~252~(1991). 

\bibitem{ohki08} H.~Ohki et al.(JLQCD Collaboration), Phys.~Rev.~{\bf D 78},~054502~(2008).

\bibitem{yt2010} R.~D.~Young and A.~W.~Thomas, Nucl.~Phys.~{\bf A 844}, 266c~(2010); 
Phys.~Rev.~{\bf D 81},~014503~(2010).

\bibitem{ya2016} Y.~B.~Yang A.~Alexandru, T.~Draper, J.~Liang, and K.-F.~Liu ($\chi$QCD Collaboration), 
Phys.~Rev.~{\bf D 94},~054503~(2016).

\end{thebibliography}
\end{document}